\documentclass[
reprint,aip,pop,graphicx,numerical,amsfonts,amsmath,amssymb,twocolumn
]{revtex4-2}

\usepackage{graphicx}
\usepackage{dcolumn}
\usepackage{bm}

\begin{document}


\title{
The Eulerian variational formulation of the gyrokinetic system in general spatial coordinates
}

\author{H. Sugama}
\affiliation{
National Institute for Fusion Science, 
Toki 509-5292, Japan
}
\affiliation{
Department of Fusion Science, SOKENDAI (The Graduate University for Advanced Studies), 
Toki 509-5292, Japan 
}

\author{S. Matsuoka}
\affiliation{
National Institute for Fusion Science, 
Toki 509-5292, Japan
}
\affiliation{
Department of Fusion Science, SOKENDAI (The Graduate University for Advanced Studies), 
Toki 509-5292, Japan 
}

\author{M. Nunami}
\affiliation{
National Institute for Fusion Science, 
Toki 509-5292, Japan
}
\affiliation{
Department of Fusion Science, SOKENDAI (The Graduate University for Advanced Studies), 
Toki 509-5292, Japan 
}

\author{S. Satake}
\affiliation{
National Institute for Fusion Science, 
Toki 509-5292, Japan
}
\affiliation{
Department of Fusion Science, SOKENDAI (The Graduate University for Advanced Studies), 
Toki 509-5292, Japan 
}

\date{\today}

\begin{abstract}
The Eulerian variational formulation of the gyrokinetic system with electrostatic turbulence  is presented in general spatial coordinates by extending our previous work [H. Sugama {\it et al}., Phys.\ Plasmas {\bf 25}, 102506 (2018)]. The invariance of the Lagrangian of the system under an arbitrary spatial coordinate transformation is used to derive the local momentum balance equation satisfied by the gyrocenter distribution functions and the turbulent potential
which are given as solutions of the governing equations. 
In the symmetric background magnetic field,  
the derived local momentum balance equation gives rise to 
the local momentum conservation law in the direction of symmetry. 
This derivation is in contrast with the conventional method using the spatial translation in which the asymmetric canonical pressure tensor generally enters the momentum balance equation. In the present study, the variation of the Lagrangian density with respect to the metric tensor is taken to directly obtain the symmetric pressure tensor which includes the effect of turbulence on the momentum transport. In addition, it is shown in this work how the momentum balance is modified when the collision and/or external source terms are added to the gyrokinetic equation. The results obtained here are considered useful for global gyrokinetic simulations investigating both neoclassical and turbulent transport processes even in general non-axisymmetric toroidal systems. 
\end{abstract}

\pacs{52.25.Dg,52.25.Xz,52.30.Gz,52.35.Ra}

\maketitle


\maketitle 



\section{INTRODUCTION}

Gyrokinetics~\cite{Antonsen,CTB,F-C,B&H,Sugama2000,Schekochihin,Krommes} 
has been used for several decades
as a basic theoretical framework to study microinstabilities and turbulent processes in magnetized plasmas.~\cite{Horton}
A significant number of large-scale simulations are being performed based on  gyrokinetic equations to analyze and predict turbulent transport fluxes of particles, heat, and momentum.~\cite{Idomura,Garbet} 
The gyrokinetic equations derived from the Lagrangian 
and/or Hamiltonian 
describing the gyrocenter motion possess conservation 
properties~\cite{B&H,Sugama2000} 
which are suitable for long-time and global transport 
simulations.
Variational formulations based on the action integral 
of the Lagrangian provide a useful and systematic 
means to obtain governing equations and conservation laws 
of energy and momentum 
for considered systems.~\cite{B&H,Sugama2000,B&T,Fan2019,Hirvijoki2020}  
The variational formulations are also 
applied to derive the useful conservative numerical 
schemes~\cite{Qin,Kraus,Zhou,Bottino,Morrison} 
in plasma physics for solving the guiding center equations and the ideal magnetohydrodynamics equations as well as the Vlasov-Poisson and Vlasov-Maxwell equations.

Because background flow profiles are considered as one of the key factors 
for improving plasma confinement, 
momentum transport processes which determine the flow 
profiles are investigated by large-scale gyrokinetic 
simulations.~\cite{Holod,Wang2009,Sarazin,Abiteboul,Idomura2017} 
Thus, the momentum balance equation satisfied by the gyrokinetic model
attracts our attention as a basis for theoretically and/or numerically 
investigating the physical mechanisms in the formation of the flow 
profiles.~\cite{
Sugama1998,Parra2010,Scott,Sugama2011,Abel,Sugama2017}  
In our previous work,~\cite{Sugama2018} 
the Eulerian formulation,~\cite{Newcomb} which is  
also called  
the Euler-Poincar\'{e} reduction 
procedure,~\cite{B&T,Hirvijoki2020,Cendra,Marsden,Squire}
is applied to derive 
the governing equations of the Vlasov-Poisson-Amp\`{e}re 
(or Vlasov-Darwin) 
system and those of the drift kinetic system 
in the general spatial coordinates, 
and the momentum balance equations for these systems are 
obtained by using the invariance of the action integrals for the systems 
under arbitrary spatial coordinate transformations. 
In this paper, the previous work is extended to present 
the Eulerian variational formulation of the gyrokinetic system with electrostatic turbulence in general spatial coordinates and to derive 
the local momentum balance equation with the symmetric pressure tensor 
including the effects of electrostatic turbulent fluctuations on 
the momentum transport. 

In the present work, 
the governing equations of the gyrokinetic system are represented in general 
spatial coordinates so that they are useful for application to 
systems with complex geometries such as stellarator and heliotron 
plasmas~\cite{Wakatani} 
in which it is convenient to employ flux surface coordinates 
(e.g., Hamada coordinates~\cite{Hamada} 
and Boozer coordinates~\cite{Boozer}) to express these equations 
for analytic and numerical studies. 
In our formulation, the momentum balance  equation is derived using the 
invariance of the Lagrangian under arbitrary spatial coordinate transformations, 
which is analogous to the derivation of 
energy-momentum conservation laws from 
the invariance of the action integral 
under arbitrary transformations of spatiotemporal 
coordinates in the theory of general relativity.~\cite{Landau} 
In the same way as in the previous work,~\cite{Sugama2018} 
the symmetric pressure tensor entering 
our momentum balance equation
is directly obtained by taking the variational derivative of  
the Lagrangian density with respect to the metric tensor, 
which is in contrast to the conventional technique using 
the spatial translation transformation for the derivation of
the canonical momentum balance including 
the asymmetric pressure tensor in the presence of the 
magnetic field. 
The canonical pressure tensor 
is asymmetric and  gauge-dependent because of the vector potential 
included in the canonical momentum. 
In order to obtain the symmetric pressure tensor from 
the asymmetric canonical pressure tensor, 
additional complicated procedures of the 
Belinfante-Rosenfeld type 
using the angular momentum conservation law derived from the rotational symmetry 
are required.~\cite{Sugama2013,Dixon}
On the other hand, our method can more directly derive 
the symmetric pressure tensor which is clearly shown in this work to 
describe both neoclassical~\cite{H&S,Helander} and turbulent transport of the momentum 
in the gyrokinetic system. 

We should note here that while employing the Hamiltonian gyrocenter motion 
equations, an irreversible collision term must be included into the 
gyrokinetic equation in order to treat neoclassical and turbulent transport  
simultaneously.~\cite{Sugama2015,Burby}
It is shown in our formulation how 
the momentum balance equations are modified when 
the collision and/or external source terms are 
added into the gyrokinetic equation.
This is possible because the momentum balance equations are derived from 
the invariance of the Lagrangian under the general spatial coordinate transformations,
with the help of the gyrocenter motion equations and the gyrokinetic Poisson's equation 
while we have freedom in choosing the governing equation for the gyrocenter distribution function. 
The momentum balance equations including the effects of collisions and
external sources are useful for checking and analyzing results of global 
gyrokinetic simulations using the Lagrangian and Hamiltonian equations to 
investigate neoclassical and turbulent transport in 
plasmas with particle, momentum, and/or heat sources. 

It is valuable to make comparisons of the present work 
with recent works~\cite{Brizard2019,Fan2020}
on similar subjects. 
In Ref.~\cite{Brizard2019}, 
a constrained Eulerian variational principle presented by 
Brizard~\cite{Brizard2000} 
is used to derive the governing equations and 
local energy-momentum conservation laws of 
the electromagnetic gyrokinetic Vlasov-Maxwell system, for which 
the gyrocenter motion equations are expressed only 
in terms of the perturbed electric and magnetic fields by including 
the perturbed fields in the Poisson-bracket structure.
In Ref.~\cite{Fan2020}, the field theory presented by 
Qin {\it et al}.~\cite{Qin2014}  
for particle-field systems on heterogenous manifolds is applied to 
derive local energy and momentum conservation laws 
in the electromagnetic gyrokinetic system.
     Except for dropping magnetic fluctuations and some of the second or higher order 
terms with respect to the small perturbation amplitude, 
the gyrokinetic system treated here is basically the same as seen in 
other earlier works~\cite{B&H,Sugama2000}, and it contains full 
finite gyroradius effects due to electrostatic fluctuations with high wavenumbers 
which are not included in Ref.~~\cite{Fan2020}. 
     Also, in the same way as in our previous work~\cite{Sugama2018}, 
the governing equations for the gyrokinetic system are derived 
in this paper based on  
the Eulerian (or the Euler-Poincar\'{e}) formulation 
which is historically older than the methods employed 
in Refs.~\cite{Brizard2019,Fan2020}. 
     The Euler-Poincar\'{e} formulation was used in the pioneering work by Newcomb~\cite{Newcomb} to derive the ideal MHD equations, and later   
it was applied to the derivation of the Vlasov-Maxwell equations, 
the guiding-center (or drift kinetic) system as well as the gyrokinetic system 
as shown in Refs.~\cite{Sugama2018,B&T,Hirvijoki2020,Cendra,Marsden,Squire}. 
     Here, the Eulerian formulation implies that, for the 
present gyrokinetic case, 
the phase-space velocity (or the temporal change rate of the 
gyrocenter coordinates in the phase space) of the gyrocenter 
is regarded as a field function of time and phase-space coordinates of 
the gyrocenter at the time when the gyrocenter passes 
through the considered point. 
  In the formulation presented in Refs.~\cite{Fan2020,Qin2014}, 
the gyrocenter (or particle) phase-space velocity is described by not 
the Eulerian but Lagrangian view point and is coupled with the Eulerian 
description of the electromagnetic fields. 
     In Refs.~\cite{Brizard2019,Fan2020}, 
not general but isometric transformations such as spatial translation and rotation 
are considered to derive local conservation laws of canonical linear and 
angular momentum in collisionless systems. 
In the present work, recognizing that 
the Lagrangian and the governing equations of the gyrokinetic system 
can be expressed in the invariant form under arbitrary spatial coordinate 
transformations in the Eulerian framework,   
this invariance property is used to derive the local momentum balance 
equation even in the presence of collisions and external sources, 
and the derived balance equation is shown to give the
local momentum conservation law in the direction of symmetry 
for collisionless systems without external sources.  

The rest of this paper is organized as follows. 
In Sec.~II, 
the Lagrangian for describing the single-particle gyrocenter motion 
is given, in which the electrostatic potential fluctuation with the 
wavelength of the order of the gyroradius is included. 
Next, in Sec.~III, we use the general spatial coordinates and 
define the Lagrangian of the whole system 
including particles of all species and turbulent electrostatic fields 
to present the Eulerian variational principle, from which 
the collisionless gyrokinetic equations for the gyrocenter 
distribution functions and the gyrokinetic Poisson's equation 
for the electrostatic potential are derived. 
In Sec.~IV, we make use of the invariance of the Lagrangians 
for the single-species and whole systems 
under arbitrary spatial coordinate transformations to 
derive the momentum balance equations  
for both systems
while allowing the collision and/or external 
source terms to be included in the gyrokinetic equations. 
Then, the symmetric pressure tensors, which enter these 
momentum balance equations, are obtained from the variational 
derivatives of the Lagrangians with respect to the metric tensor 
components and they are verified to describe both neoclassical 
and turbulent transport of the momentum. 
Finally, conclusions are given in Sec.~V. 
In addition, Appendix~A is given to briefly explain 
covariant derivatives and Christoffel symbols, 
which are used in the general spatial coordinates. 
Variations in the functional forms of vector and tensor fields 
under the infinitesimal transformation of the spatial coordinates 
and variational derivative with respect to 
the metric tensor components are described in Appendices~B and C, 
respectively. 
In Appendix~D,  
the WKB representation~\cite{WKB} is used to express the 
turbulent part of the pressure tensor in the form 
comparable with the one obtained in the past work 
on the momentum transport.  
Another derivation of the momentum balance equations, 
in which the asymmetric canonical pressure tensors appear, 
is shown in Appendix~E and the energy balance equations are derived 
in Appendix~F. 
The case of the symmetric background magnetic field is considered 
in Appendix~G where it is shown how
the local conservation law of the canonical momentum 
in the direction of symmetry is derived in the present formulation.

\section{THE GYROCENTER LAGRANGIAN}

The Lagrangian for describing the gyrocenter motion of the particle with 
mass $m$ and charge $e$ is given by~\cite{B&H}
\begin{eqnarray}
\label{LGY}
L_{GY} 
& \equiv & 
\left( \frac{e}{c} {\bf A} ({\bf X}, t ) + 
m v_\parallel {\bf b}({\bf X}, t ) \right) \cdot \dot{\bf X}
+ \frac{m c}{e} \mu \dot{\vartheta}  
\nonumber \\
& & \mbox{}
- H_{GY}
,
\end{eqnarray}
where the gyrocenter phase-space coordinates ${\bf X}$, $v_\parallel$, 
$\mu \equiv m v_\perp^2/(2B)$, 
and $\vartheta$ 
denote the gyrocenter position, the velocity component parallel to the magnetic field, the magnetic moment, and the gyrophase angle, respectively, and 
$\dot{} \equiv d/dt$ represents the time derivative along the trajectory in the phase space. 
The vector potential and the unit vector parallel to the magnetic field ${\bf B}$ 
are written by ${\bf A}$ and ${\bf b} \equiv {\bf B}/B$, respectively. 
Here, we suppose that ${\bf A}$ can weakly depend on time $t$ and accordingly 
the background magnetic field ${\bf B} = \nabla \times {\bf A}$ 
is allowed to slowly vary in time. 
Thus we can treat the inductive electric field 
$- c^{-1} \partial {\bf A}/\partial t$ which drives the ohmic current in tokamaks. 

The gyrocenter Hamiltonian 
which appears on the right-hand side of Eq.~(\ref{LGY}) is defined by 
\begin{equation}
\label{HGY}
 H_{GY} \equiv \frac{1}{2} m v_\parallel^2 + \mu B + e \Psi
,
\end{equation}
and its electrostatic fluctuation part is written as     
\begin{equation}
\label{Psi}
e \Psi  \equiv 
e \langle \phi ({\bf X} + 
\mbox{\boldmath$\rho$},t)  \rangle_\vartheta
- \frac{e^2}{2 B}
\frac{\partial}{\partial \mu}
\langle (\widetilde{\phi})^2 \rangle_\vartheta
,
\end{equation}
where 
the gyroradius vector is denoted by 
$\mbox{\boldmath$\rho$}\equiv {\bf b} \times {\bf v} / \Omega$,  
${\bf v}$ is the particle's velocity, $\Omega \equiv e B/(mc)$ is the gyrofrequency, 
and 
the average of the electrostatic potential $\phi$ at the particle 
position ${\bf X} + \mbox{\boldmath$\rho$}$ over the gyrophase $\vartheta$ 
is given by 
\begin{equation}
\label{phiav}
\langle \phi ({\bf X} + \mbox{\boldmath$\rho$}) \rangle_\vartheta
\equiv
\oint \frac{d\vartheta}{2\pi} \phi ({\bf X} + \mbox{\boldmath$\rho$})
,
\end{equation}
and the gyrophase-dependent part is denoted by 
\begin{equation}
\label{phitilde}
\widetilde{\phi} 
\equiv
\phi ({\bf X} + \mbox{\boldmath$\rho$}) 
- 
\langle \phi ({\bf X} + \mbox{\boldmath$\rho$}) \rangle_\vartheta
. 
\end{equation}
%
The last term on the right-hand side of Eq.~(\ref{Psi}) is of the second order in 
the gyrokinetic ordering parameter $\epsilon$ where 
$\epsilon \sim e \widetilde{\phi}/ (m |{\bf v}|^2) \sim \rho / L$ with $L$ 
representing the equilibrium gradient scale length is assumed. 
This second-order term is retained because it is 
necessary for deriving the gyrokinetic Poisson's equation correctly 
including the polarization effect as shown in Sec.~III.C. 
However, other second-order terms shown in 
Ref.~\cite{P&C}
are neglected in the gyrocenter Hamiltonian given by 
Eqs.~(\ref{HGY}) and (\ref{Psi}).
Therefore, rigorously speaking, the accuracy of the present model 
is up to the first order. 
The turbulent fluctuations are assumed to have 
the characteristic wavelength 
$\sim (k_\perp)^{-1} \sim \rho$. 
Then, 
the fluctuation terms in Eq.(\ref{Psi}) are considered to contain 
all-order terms in $k_\perp \rho (\sim 1)$
even though small amplitude terms of higher order in  
$\epsilon \sim \rho/L (\ll 1)$ are neglected.
In Eqs.~(\ref{phiav}) and (\ref{phitilde}) as well as in the equations shown below, 
the time variable $t$ on which $\phi$ depends is omitted for simplicity.

In this section, the Cartesian spatial coordinates are used and 
three-dimensional vectors are represented in terms of boldface letters. 
Then, the electrostatic potential 
$\phi ({\bf X} + \mbox{\boldmath$\rho$} )$ is Taylor expanded about 
the gyrocenter position ${\bf X}$ as 
\begin{equation}
\label{phixr}
\phi ({\bf X} + \mbox{\boldmath$\rho$}) 
= \sum_{n=0}^\infty  \frac{1}{n!}
\sum_{j_1 = 1}^3 \cdots \sum_{j_n = 1}^3
\rho^{j_1} \cdots \rho^{j_n}
\frac{\partial^n \phi ({\bf X})}{\partial X^{j_1} \cdots \partial X^{j_n}}
,
\end{equation}
where $X^j$ and $\rho^j$ $(j=1,2,3)$ and 
the Cartesian spatial coordinates of the gyrocenter position vector  
${\bf X}$ and the gyroradius vector 
$\mbox{\boldmath$\rho$}$, 
respectively. 
Substituting Eq.~(\ref{phixr}) into Eq.~(\ref{phiav}), we obtain 
\begin{equation}
\label{phixrav}
\langle \phi ({\bf X} + \mbox{\boldmath$\rho$}) \rangle_\vartheta
= \sum_{n=0}^\infty  
\sum_{j_1 = 1}^3 \cdots \sum_{j_n = 1}^3
\frac{\alpha^{j_1 \cdots j_n}}{n!}
\frac{\partial^n \phi ({\bf X})}{\partial X^{j_1} \cdots \partial X^{j_n}}
, 
\end{equation}
where the gyrophase average of a product of $n$ gyroradius 
vector components is denoted by 
\begin{equation}
\label{alpha}
\alpha^{j_1 \cdots j_n}
\equiv
\langle \rho^{j_1} \cdots \rho^{j_n}
\rangle_\vartheta
. 
\end{equation}
Obviously, $\alpha^{j_1 \cdots j_n}$ is symmetric with respect to 
arbitrary permutations of the indices $j_1, \cdots, j_n$. 
It can be shown that 
\begin{equation}
\label{alphaodd}
\alpha^{j_1 \cdots j_{n}}
= 0
\hspace*{5mm}
\mbox{for odd $n$}
,
\end{equation}
and 
\begin{equation}
\label{alphaeven}
\alpha^{j_1 \cdots j_{2l}}
= 
\frac{1}{(2 l)!}
\sum_{\sigma \in \mathfrak{S}_{2l}}
\eta^{j_{\sigma (1)}  \cdots j_{\sigma (2l)}}
, 
\end{equation}
where 
$\mathfrak{S}_{2l}$ is the symmetric group of 
permutations of the set $\{1, 2, \cdots, 2l \}$ 
and 
$\eta^{j_1 \cdots j_{2l}}$ is defined by 
\begin{equation}
\label{eta}
\eta^{j_1 \cdots j_{2l}}
= 
\frac{(2l)!}{(l !)^2}
\left( \frac{\rho}{2} \right)^{2l}
h^{j_1 j_2} h^{j_3 j_4} \cdots h^{j_{2l-1} j_{2l}} 
,
\end{equation}
with 
\begin{equation}
\label{rho}
\rho
\equiv 
\frac{c }{e}
\sqrt{\frac{2m\mu}{B}}
, 
\end{equation}
and 
\begin{equation}
\label{hij}
h^{ij}
\equiv 
\delta^{ij} - b^i b^j
. 
\end{equation}
Here, $b^i$ is the $i$th contravariant component of 
${\bf b} \equiv {\bf B}/B$ and 
$\delta^{ij}$ represents the Kronecker delta; 
$\delta^{ij} = 1$ (for $i=j$), 0 (for $i\neq j$). 
In Eq.~(\ref{Psi}), 
we also find that 
\begin{eqnarray}
\label{phitilde2}
\langle (\widetilde{\phi})^2 \rangle_\vartheta
& = & 
\sum_{m=1}^\infty  \sum_{n=1}^\infty  
\sum_{i_1 = 1}^3 \cdots \sum_{i_m = 1}^3
\sum_{j_1 = 1}^3 \cdots \sum_{j_n = 1}^3
\frac{\beta^{i_1 \cdots i_m ;  j_1 \cdots j_n}}{m! \; n!}
\nonumber \\ & & \mbox{} \times 
\frac{\partial^n \phi ({\bf X})}{\partial X^{i_1} \cdots \partial X^{i_m}}
\; 
\frac{\partial^n \phi ({\bf X})}{\partial X^{j_1} \cdots \partial X^{j_n}}
,
\end{eqnarray}
where 
\begin{equation}
\label{beta}
\beta^{i_1 \cdots i_m ;  j_1 \cdots j_n}
\equiv
\alpha^{i_1 \cdots i_m   j_1 \cdots j_n}
-
\alpha^{i_1 \cdots i_m}
\alpha^{j_1 \cdots j_n}
\end{equation}
is defined and it satisfies
\begin{equation}
\label{beta0}
\beta^{i_1 \cdots i_m ;  j_1 \cdots j_n}
= 0
\hspace*{5mm}
\mbox{for odd $(m+n)$}
. 
\end{equation}
The expressions given in 
Eqs.~(\ref{phixrav}),  (\ref{hij}), and (\ref{phitilde2}) are valid in the 
Cartesian spatial coordinates although they can be easily transformed 
into those in general spatial coordinates as shown in Sec.~III.

\section{EULERIAN VARIATIONAL PRINCIPLE FOR DERIVATION OF THE COLLISIONLESS GYROKINETIC SYSTEM OF EQUATIONS IN GENERAL SPATIAL COORDINATES}

In this section, the governing equations for the distribution functions and the turbulent electrostatic potential in the collisionless gyrokinetic system are derived 
from the Eulerian variational principle using the Lagrangian density represented 
in general spatial coordinates. 

\subsection{The Lagrangian density represented in general spatial coordinates}

  The action integral $I_{GKF}$ for the Eulerian variational principle to derive 
all of the governing equations of 
the collisionless electrostatic gyrokinetic turbulent system 
is written as 
\begin{equation}
\label{IGKF}
I_{GKF}  \equiv  \int_{t_1}^{t_2} dt \; L_{GKF} 
\equiv \int_{t_1}^{t_2}  dt \int_V d^3 x\; {\cal L}_{GKF} 
,
\end{equation}
where the Lagrangian density ${\cal L}_{GKF}$ is given by 
\begin{eqnarray}
\label{LGKF}
{\cal L}_{GKF} 
& \equiv & 
{\cal L}_{GK} + {\cal L}_{F} 
\nonumber \\ 
& \equiv & 
\sum_a
\int d^3 v
\; F_a (x, v, t) L_{GYa}  (x, v, t) 
\nonumber \\ 
& & \mbox{}
+ \frac{\sqrt{g}}{8 \pi} g^{ij}(x) (E_L)_i (x, t) (E_L)_j (x, t)
, 
\end{eqnarray}
Here, $d^3 x \equiv dx^1 dx^2 dx^3$
and $d^3 v \equiv d v_\parallel d\mu d\vartheta$ are used and 
the subscript $a$ represents the particle species 
with mass $m_a$ and charge $e_a$.  
We use $x \equiv (x^i)_{i=1,2,3}$ and  
$v \equiv (v_\parallel, \mu, \vartheta)$  
as the gyrocenter phase-space coordinates. 
It should be emphasized that, in this section,  
$x \equiv (x^i)_{i=1,2,3}$ represent general spatial coordinates of the 
gyrocenter position which can be either Cartesian or any other curved coordinates. 
However, here we assume that 
the spatial position vector ${\bf r} = {\bf r} (x)$ 
is a function of only the spatial coordinates 
$x \equiv (x^i)_{i=1,2,3}$ and it is independent of time $t$. 
The gyrocenter distribution function on 
the $(x, v)$-space 
is denoted by $F_a$, and 
the number of particles of species $a$ in 
the phase-space volume element 
$d^3 x d^3 v \equiv dx^1 dx^2 dx^3 d v_\parallel d\mu d\vartheta$ 
at time $t$ 
is given by $F_a (x, v, t) d^3 x d^3 v$. 
The field part ${\cal L}_F$ of the Lagrangian density 
${\cal L}_{GKF}$ in 
Eq.~(\ref{LGKF}) is written in terms of the covariant 
components of the electric field due to the electrostatic potential, 
\begin{equation}
\label{EL}
(E_L)_i
\equiv 
- \frac{\partial \phi (x, t)}{\partial x^i}
. 
\end{equation}

In this paper, we employ the summation convention that 
the same symbol used for a pair of upper and lower indices  
within a term 
[such as seen in Eq.~(\ref{LGKF}) as well as 
in the equations shown below]  
indicates summation over the range $\{ 1, 2, 3 \}$ 
of the symbol index.
The contravariant metric tensor components $g^{ij}$
in the general spatial coordinates $x \equiv (x^i)$ are
related to the covariant components $g_{ij}$ by 
$
g^{ik} g_{kj} = \delta^i_j
$,
where $\delta^i_j$ represents the Kronecker delta. 
The determinant of the covariant metric tensor matrix 
is denoted by  
\begin{equation}
\label{detg}
g (x) \equiv  \det [ g_{ij} (x) ]
. 
\end{equation}
Note that $g_{ij}(x)$,  $g^{ij}(x)$, and $g(x)$ are 
all independent of time $t$ 
because the spatial position vector ${\bf r}$ 
is assumed to be given by  
a function of only the spatial coordinates 
$x \equiv (x^i)_{i=1,2,3}$, independently of time $t$ 
as mentioned earlier.

The gyrocenter Lagrangian for species $a$, 
which is multiplied by $F_a$ to define 
${\cal L}_{GK}$ in Eq.~(\ref{LGKF}), 
is represented in the Eulerian picture by 
\begin{eqnarray}
\label{LGYa}
L_{GYa} 
& \equiv & 
\left( \frac{e_a}{c} A_j (x, t)+ m_a v_\parallel b^i (x, t) g_{ij}(x) \right) 
u_{ax}^j (x, v, t)
\nonumber \\ & & \mbox{}
+ \frac{m_a c}{e_a} \mu u_{a\vartheta} (x, v, t) 
- H_{GYa} (x, v_\parallel, \mu, t)
,
\end{eqnarray}
where 
$A_j$ is 
the $j$th covariant component of the vector potential,  
$b^i \equiv B^i/B$ is the $i$th contravariant component 
of the unit vector parallel to the background magnetic field, and
the magnetic field strength is given by 
\begin{equation}
\label{Bmag}
B (x, t)
\equiv
\sqrt{ g_{ij}(x) B^i (x, t) B^j (x, t)}
. 
\end{equation}
   The contravariant components $(B^i)_{i=1,2,3}$  
of the magnetic field 
are expressed in terms of the covariant components 
$(A_i)_{i=1,2,3}$  of the vector potential as 
\begin{equation}
\label{Bi}
B^i (x, t) = \frac{\epsilon^{ijk}}{\sqrt{g(x) }} 
\frac{\partial A_k (x, t)}{\partial x^j}
,
\end{equation}
where the Levi-Civita symbol is denoted by  
\begin{eqnarray}
\label{eijk}
& & 
\epsilon^{ijk} 
 \equiv   \epsilon_{ijk}
\nonumber \\
&  & 
\equiv
\left\{
\begin{array}{cl}
1 
&
\mbox{($(i, j, k) = (1,2,3), (2,3,1), (3,1,2)$)} 
\\
-1
& 
\mbox{($(i, j, k) = (1,3,2), (2,1,3), (3,2,1)$)} 
\\
0 
& 
\mbox{(otherwise)}. 
\end{array}
\right. 
\end{eqnarray}

The gyrocenter Hamiltonian is written here as  
\begin{equation}
\label{HGYa}
 H_{GYa} (x, v_\parallel, \mu, t) 
\equiv \frac{1}{2} m_a v_\parallel^2 + \mu B (x, t)+ e_a \Psi_a (x, \mu, t)
,
\end{equation}
and the electrostatic fluctuation part is given by    
\begin{equation}
\label{Psia}
\Psi_a (x, \mu, t)\equiv 
\phi (x, t) + \Psi_{E1a} (x, \mu, t) + \Psi_{E2a} (x, \mu, t)
, 
\end{equation}
where 
\begin{equation}
\label{PsiE1a}
\Psi_{E1a} (x, \mu, t) 
\equiv \sum_{n=1}^\infty  
\frac{\alpha_a^{j_1 \cdots j_n}(x, \mu, t)}{n!}
\nabla_{j_1} \cdots \nabla_{j_n} \phi (x, t)
,
\end{equation}
and 
\begin{eqnarray}
\label{PsiE2a}
\Psi_{E2a} (x, \mu, t) 
& \equiv & 
- e_a
\sum_{m=1}^\infty  \sum_{n=1}^\infty  
\frac{(m+n)}{m! \; n!} 
\frac{\beta_a^{i_1 \cdots i_m ;  j_1 \cdots j_n}(x, \mu, t) 
}{4 \mu B(x,t)}
\nonumber \\ & & \mbox{} 
\hspace*{-8mm} \times 
( \nabla_{i_1} \cdots \nabla_{i_m} \phi (x, t) )
\, 
( \nabla_{j_1} \cdots \nabla_{j_n} \phi (x, t) )
. 
\hspace*{8mm}
\end{eqnarray}
Here, $\alpha_a^{j_1 \cdots j_n}(x, \mu, t)$ and 
$\beta_a^{i_1 \cdots i_m ;  j_1 \cdots j_n}(x, \mu, t)$ 
are defined  by  Eqs.~(\ref{alpha})--(\ref{rho}) and (\ref{beta}) in which 
$h^{ij}$ is represented in the general spatial 
coordinates $x \equiv (x^i)_{i=1,2,3}$ 
by 
\begin{equation}
\label{hijg}
h^{ij}
\equiv 
g^{ij} - b^i b^j
,
\end{equation}
and the covariant derivative $\nabla_j$ is defined in 
Appendix~A. 

The Eulerian representation of 
the gyrocenter Lagrangian $L_{GYa}$ shown in Eq.~(\ref{LGYa}) contains 
the temporal change rates of the gyrocenter position 
coordinates, parallel velocity, magnetic moment, and 
gyrophase at time $t$
which are represented by the functions 
$u_{ax}^i(x, v, t)$, 
$u_{av_\parallel}(x, v, t)$, 
$u_{a\mu}(x, v, t)$, and 
$u_{a\vartheta}(x, v, t)$, respectively, 
in the same way as in Ref.~\cite{Sugama2018}. 
Then, the distribution function
$F_a (x, v, t)$ satisfies 
\begin{eqnarray}
\label{GKE0}
& & \frac{\partial F_a}{\partial t} 
+ \frac{\partial}{\partial x^j}
( F_a u_{ax}^j  )
+ \frac{\partial}{\partial v_\parallel}
( F_a u_{a v_\parallel}  )
+ \frac{\partial}{\partial \mu}
( F_a u_{a \mu} )
\nonumber \\ 
& & 
\hspace*{6mm} \mbox{}
+ \frac{\partial}{\partial \vartheta}
( F_a u_{a \vartheta}  )
= 0
. 
\end{eqnarray}

We find here that 
the gyrocenter Hamiltonian $H_{GYa}$ given in Eq.~(\ref{HGYa}) takes a functional form, 
\begin{equation}
\label{HGYaf}
H_{GYa} 
=
H_{GYa} 
 \left[ v,   
\partial_j A_i (x,t), 
\{ \partial_J \phi (x,t) \},  
 \{ \partial_J g_{ij} (x) \}   \right]
, 
\end{equation}
which depends on the velocity space coordinates
(except for $\vartheta$) as well as the 
general spatial coordinates $x \equiv (x^i)_{i=1,2,3}$ 
through the field variables 
$[\partial_j A_i (x,t),  \{ \partial_J \phi (x,t) \},  
\{ \partial_J g_{ij} (x) \} ]$. 
Here, 
we use the notation
$J \equiv (j_1, j_2, \cdots, j_n )$ 
(for $n = 0, 1, 2, \cdots$ and $j_1, j_2, \cdots, j_n = 1, 2, 3$) 
to write 
\begin{equation}
\label{pJQ}
\partial_J {\cal F} \equiv 
\left\{
\begin{array}{lc}
 {\cal F}
&
(n=0)
\\
\partial_{j_1 j_2 \cdots j_n} {\cal F}
\;
\equiv 
\;
\partial^n {\cal F}/
\partial x^{j_1} \partial x^{j_2} \cdots \partial x^{j_n}
&
(n \geq 1)
\end{array}
\right.
\end{equation}
where 
 ${\cal F}$ is an arbitrary function of $x = (x^i)_{i=1,2,3}$. 
Then, the compact notations 
$\{ \partial_J \phi (x,t) \}$ and 
 $\{ \partial_J g_{ij} (x) \}$ in Eq.~(\ref{HGYaf}) imply
\begin{equation}
\label{phiJ}
\{ \partial_J \phi \} \equiv 
\{ \phi, \partial_j \phi, \partial_{jk} \phi, 
\partial_{jkl} \phi, \cdots \}
,
\end{equation}
and 
\begin{equation}
\label{gijJ}
 \{ \partial_J g_{ij} \}   \equiv 
\{ g_{ij}, \partial_k g_{ij}, \partial_{kl} g_{ij}, 
\partial_{klm} g_{ij}, \cdots \}
,
\end{equation}
respectively. 
Note that the high-order spatial derivatives included in 
Eqs.~(\ref{phiJ}) and (\ref{gijJ}) 
enter the gyrocenter Hamiltonian $H_{GYa}$ 
due to finite gyroradius as seen in Eqs.~(\ref{PsiE1a}) 
and Eqs.~(\ref{PsiE2a}) where the covariant 
derivatives contain the spatial derivatives of $g_{ij}$ through the 
Christoffel symbols defined by Eq.~(\ref{Christoffel}) in Appendix~A. 

In the same way as in Eq.~(\ref{HGYaf}), the functional form of the 
gyrocenter Lagrangian $L_{GYa}$ is written as 
\begin{eqnarray}
\label{LGYaf}
L_{GYa} 
& = & 
L_{GYa} 
 \left[ v,   u_{ax}^i (x, v, t),  u_{a\vartheta}(x, v, t),  
A_i (x, t), 
\right. 
\nonumber \\ & & 
\left. 
\mbox{} \hspace*{6mm}
\partial_j A_i (x,t), 
\{ \partial_J \phi (x) \},   \{ \partial_J g_{ij} (x) \}   \right]
, 
\end{eqnarray}
where the Eulerian representations of the temporal change rates of the gyrocenter 
position and the gyrophase, $u_{ax}^i (x, v, t)$ and $u_{a\vartheta}(x, v, t)$, 
are additionally included. 
In Eq.~(\ref{LGYaf}), 
$u_{ax}(x, v, t)$, $u_{a\vartheta}(x, v, t)$, and $\phi(x,t)$ are 
the functions, the governing equations of which are derived 
from the variation principle in Sec.~III.C  
while the dependence of $L_{GYa}$ on $A_i (x, t)$, $\partial_j A_i (x, t)$, 
and $\partial_J g_{ij} (x, t)$ is also explicitly shown 
because their variations need to be taken into account to evaluate 
the variation of $L_{GYa}$ in Sec.~IV where, in order to derive 
the local momentum balance,  
we consider the general spatial coordinate transformation which causes 
the variations in the functional forms of both  
$(u_{ax}, u_{a\vartheta}, \phi)$ and 
$(A_i, g_{ij})$. 
Here,  using Eq.~(\ref{EL}) and (\ref{phiJ}), we note that 
$\{ \partial_J \phi \}$ can be replaced by 
\begin{equation}
\label{phiELJ}
\{ \phi, \{ \partial_J (E_L)_i \}  \}
,
\end{equation}
where
\begin{equation}
\label{ELJ}
\{ \partial_J (E_L)_i \} \equiv   
\{ (E_L)_i,  \partial_j (E_L)_i, \partial_{jk} (E_L)_i, 
\partial_{jkl} (E_L)_i, \cdots   \}
. 
\end{equation}

\subsection{The Lagrangian density associated with polarization}

We find from Eqs.~(\ref{LGYa}) and (\ref{HGYa})--(\ref{PsiE2a}) that 
the part of the Lagrangian density $L_{GYa}$ 
which includes the perturbed potential 
is given by 
\begin{equation}
\label{LPsia}
{\cal L}_{\Psi a} 
\equiv  
-
\int d^3 v \, F_a  
\, e_a \Psi_a 
=
- e_a N_a^{(g)}  \phi 
+ {\cal L}_{E 1 a} + {\cal L}_{E 2 a} 
, 
\end{equation}
where 
the gyrocenter density $N_a^{(g)}$ is defined by 
\begin{equation}
\label{Nag}
N_a^{(g)} 
\equiv
\int d^3 v \, F_a 
,
\end{equation}
${\cal L}_{E1a}$ is given in the linear form with respect to 
the longitudinal electric field $(E_L)_i  \equiv - \partial \phi/\partial x^i$ and
its covariant derivatives, 
\begin{eqnarray}
\label{LE1a}
 {\cal L}_{E1a} 
& \equiv  & 
-
\int d^3 v
\, F_a 
\,  e_a
\Psi_{E1a}
\nonumber \\ 
& = & 
\sum_{k=1}^\infty 
Q_{0a}^{j_1 \cdots j_{2k}} 
\nabla_{j_1} \cdots  \nabla_{j_{2k-1}}
 (E_L)_{j_{2k}}
,
\end{eqnarray}
and ${\cal L}_{E2a}$ 
is given in the quadratic form, 
\begin{eqnarray}
\label{LE2a}
{\cal L}_{E2a} 
& \equiv  & 
-
\int d^3 v
\, F_a 
\,  e_a
\Psi_{E2a}
\nonumber \\ 
&  =   & 
\frac{1}{2} \sum_{m=1}^\infty
 Q_{Ea}^{i_1 \cdots i_m}
\nabla_{i_1} \cdots  \nabla_{i_{m-1}} (E_L)_{i_m}
\nonumber \\ 
&  =   & 
\frac{1}{2} \sum_{m=1}^\infty  \sum_{n=1}^\infty  
\chi_a^{i_1 \cdots i_m ;  j_1 \cdots j_n}
\nabla_{i_1} \cdots  \nabla_{i_{m-1}} (E_L)_{i_m}
\nonumber \\ 
& & 
\hspace*{5mm}
\mbox{} 
\times
\nabla_{j_1} \cdots  \nabla_{j_{n-1}} (E_L)_{j_n}
. 
\end{eqnarray}
In this paper, the longitudinal (irrotational) and transverse (solenoidal) parts 
of any vector field are represented by the subscripts $L$ and $T$, 
respectively.~\cite{Jackson} 
In Eq.~(\ref{LE1a}), there appear 
the multipole moments $Q_{0a}^{j_1 \cdots j_{2k}}$ 
of the electric charge distribution~\cite{Jackson} 
 of species $a$ induced by finite gyroradius, 
\begin{equation}
\label{Q0a}
Q_{0a}^{j_1 \cdots j_{2k}} 
\equiv
e_a
\int d^3 v
\; F_a 
\frac{\alpha_a^{j_1 \cdots j_{2k}}}{(2k)!}
,
\end{equation}
which can exist even without the electric field $(E_L)_i$. 
The multipole moments $Q_{Ea}^{i_1 \cdots i_m}$ of 
the electric charge distribution of species $a$ induced by 
the longitudinal electric field  $(E_L)_i$ is shown in Eq.~(\ref{LE2a}), 
and they are given in the linear form  
with respect to  $(E_L)_i$ and its covariant derivatives
as 
\begin{equation}
\label{QEa}
 Q_{Ea}^{i_1 \cdots i_m}
\equiv 
\sum_{n=1}^\infty
\chi_a^{i_1 \cdots i_m ;  j_1 \cdots j_n}
\nabla_{j_1} \cdots  \nabla_{j_{n-1}} (E_L)_{j_n}
.
\end{equation}
Here, the coefficients 
$\chi_a^{i_1 \cdots i_m ;  j_1 \cdots j_n} $ 
are defined by 
\begin{equation}
\label{chia}
\chi_a^{i_1 \cdots i_m ;  j_1 \cdots j_n} 
\equiv
e_a^2
\int d^3 v
\frac{F_a}{2 \mu B}
\frac{(m+n) \beta_a^{i_1 \cdots i_m ;  j_1 \cdots j_n}}{m! \; n!}
,
\end{equation}
which are regarded as the generalized electric susceptibility 
produced by finite gyroradius. 
It is remarked that  $Q_{0a}^{i_1 \cdots i_m}$ and $Q_{Ea}^{i_1 \cdots i_m}$ 
are both symmetric with respect to arbitrary permutations of the 
indices $i_1,  \cdots,  i_m$ 
because $\alpha_a^{i_1 \cdots i_m}$ has the same 
symmetry. 

Taking the summation of Eq.~(\ref{LPsia}) over species $a$, 
we obtain  
\begin{eqnarray}
\label{LPsi}
& & 
{\cal L}_\Psi 
 \equiv  
\sum_a 
{\cal L}_{\Psi a} 
=
- \rho^{(g)} \phi   + {\cal L}_ {E 1}  
+ {\cal L}_{E 2} 
, 
\end{eqnarray}
where the gyrocenter charge density $\rho^{(g)}$ 
is given by 
\begin{equation}
\label{rhog}
\rho^{(g)} \equiv 
\sum_a e_a N_a^{(g)} 
, 
\end{equation}
and the parts of the Lagrangian density associated with polarization  
are represented by ${\cal L}_ {E 1}$ and ${\cal L}_ {E 2}$. 
Here, ${\cal L}_ {E 1}$ is defined by 
\begin{equation}
\label{LE1}
{\cal L}_ {E 1} 
 \equiv   
\sum_a {\cal L}_ {E 1 a} 
=
\sum_{k=1}^\infty 
Q_0^{j_1 \cdots j_{2k}} 
\nabla_{j_1} \cdots \nabla_{j_{2k-1}} (E_L)_{j_{2k}}
,
\end{equation}
with the multiple moments induced by finite gyroradius, 
\begin{equation}
\label{Q0}
Q_0^{j_1 \cdots j_{2k}} 
\equiv
\sum_a 
Q_{0a}^{j_1 \cdots j_{2k}} 
, 
\end{equation}
and ${\cal L}_ {E 2}$ is defined by 
\begin{eqnarray}
\label{LE2}
{\cal L}_ {E2} 
& \equiv  &
\sum_a {\cal L}_ {E 2 a} 
= \frac{1}{2} \sum_{m=1}^\infty
 Q_E^{i_1 \cdots i_m}
\nabla_{i_1} \cdots  \nabla_{i_{m-1}} (E_L)_{i_m}
\nonumber \\
& = & 
\frac{1}{2} \sum_{m=1}^\infty  \sum_{n=1}^\infty  
\chi^{i_1 \cdots i_m ;  j_1 \cdots j_n}
\nabla_{i_1} \cdots  \nabla_{i_{m-1}} (E_L)_{i_m}
\nonumber \\ 
& & 
\hspace*{5mm}
\mbox{} 
\times
\nabla_{j_1} \cdots  \nabla_{j_{n-1}} (E_L)_{j_n}
,
\end{eqnarray}
with the multiple moments induced by the longitudinal electric field 
$(E_L)_i$ and its covariant derivatives, 
\begin{eqnarray}
\label{QE}
 Q_E^{i_1 \cdots i_m}
& \equiv & 
\sum_a  Q_{Ea}^{i_1 \cdots i_m}
\nonumber \\
& = &
\sum_{n=1}^\infty
\chi^{i_1 \cdots i_m ;  j_1 \cdots j_n}
\nabla_{j_1} \cdots  \nabla_{j_{n-1}} (E_L)_{j_n}
,
\end{eqnarray}
and the generalized electric susceptibility due to finite gyroradius, 
\begin{equation}
\label{chi}
\chi^{i_1 \cdots i_m ;  j_1 \cdots j_n}  
\equiv
\sum_a 
\chi_a^{i_1 \cdots i_m ;  j_1 \cdots j_n}   
. 
\end{equation}

\subsection{Derivation of the governing equations of 
the collisionless electrostatic gyrokinetic turbulent system}

Here we virtually allow the phase-space trajectories for all species and 
the electrostatic potential to vary infinitesimally. 
Following the same procedure as in Ref.~\cite{Sugama2018}, 
the variations in the gyrocenter position,  parallel velocity,  
magnetic moment, and gyrophase of the phase-space trajectory  
are represented in the Eulerian picture by
$\delta x_{aE}^i (x, v, t)$, 
$\delta  v_{a\parallel E} (x, v, t)$, 
$\delta \mu_{aE} (x, v, t)$, 
and 
$\delta \vartheta_{aE} (x, v, t)$, 
respectively. 
We also denote the variation in the electrostatic potential 
by $\delta \phi$. 
   Then, in terms of  
$\delta x_E^i$, $\delta  v_{\parallel E}$, $\delta \mu_E$, 
and $\delta \vartheta_E$, 
the variations in the functional forms of $u_{ax}^i$, 
$u_{a v_\parallel}$, $u_{a \mu}$, and $u_{a \vartheta}$ 
due to the virtual displacement are written as 
\begin{eqnarray}
\label{duvmt}
& & 
\hspace*{-3mm}
\delta u_{ax}^i
= 
\left( \frac{\partial }{\partial t}
+ u_{ax}^j \frac{\partial }{\partial x^j}
+ u_{a v_\parallel}  \frac{\partial }{\partial v_\parallel}
+ u_{a \mu}  \frac{\partial }{\partial \mu}
+ u_{a \vartheta} \frac{\partial }{\partial \vartheta}
\right) \delta x_{aE}^i
\nonumber \\ 
& & \mbox{}
- \left( 
\delta x_{aE}^j \frac{\partial }{\partial x^j}
+ \delta v_{a\parallel E} \frac{\partial }{\partial v_\parallel}
+  \delta \mu_{aE} \frac{\partial }{\partial \mu}
+ \delta \vartheta_{aE} \frac{\partial }{\partial \vartheta}
\right) u_{ax}^i
,
\nonumber \\ 
& & 
\hspace*{-3mm}
\delta u_{a v_\parallel} 
= 
\left( \frac{\partial }{\partial t}
+ u_{ax}^j \frac{\partial }{\partial x^j}
+ u_{a v_\parallel}  \frac{\partial }{\partial v_\parallel}
+ u_{a \mu}  \frac{\partial }{\partial \mu}
+ u_{a \vartheta} \frac{\partial }{\partial \vartheta}
\right) \delta v_{a\parallel E}
\nonumber \\ 
& & \mbox{}
- \left( 
\delta x_{aE}^j \frac{\partial }{\partial x^j}
+ \delta v_{a\parallel E} \frac{\partial }{\partial v_\parallel}
+  \delta \mu_{aE} \frac{\partial }{\partial \mu}
+ \delta \vartheta_{aE} \frac{\partial }{\partial \vartheta}
\right) u_{a v_\parallel}
,
\nonumber \\ 
& & 
\hspace*{-3mm}
\delta u_{a \mu}
=
\left( \frac{\partial }{\partial t}
+ u_{ax}^j \frac{\partial }{\partial x^j}
+ u_{a v_\parallel}  \frac{\partial }{\partial v_\parallel}
+ u_{a \mu}  \frac{\partial }{\partial \mu}
+ u_{a \vartheta} \frac{\partial }{\partial \vartheta}
\right) \delta \mu_{aE}
\nonumber \\ 
& & \mbox{}
- \left( 
\delta x_{aE}^j \frac{\partial }{\partial x^j}
+ \delta v_{a\parallel E} \frac{\partial }{\partial v_\parallel}
+  \delta \mu_{aE} \frac{\partial }{\partial \mu}
+ \delta \vartheta_{aE} \frac{\partial }{\partial \vartheta}
\right) u_{a\mu}
,
\nonumber \\ 
& & 
\hspace*{-3mm}
\delta u_{a \vartheta}
=
\left( \frac{\partial }{\partial t}
+ u_{ax}^j \frac{\partial }{\partial x^j}
+ u_{a v_\parallel}  \frac{\partial }{\partial v_\parallel}
+ u_{a \mu}  \frac{\partial }{\partial \mu}
+ u_{a \vartheta} \frac{\partial }{\partial \vartheta}
\right) \delta \vartheta_{aE}
\nonumber \\ 
& & \mbox{}
- \left( 
\delta x_{aE}^j \frac{\partial }{\partial x^j}
+ \delta v_{a\parallel E} \frac{\partial }{\partial v_\parallel}
+  \delta \mu_{aE} \frac{\partial }{\partial \mu}
+ \delta \vartheta_{aE} \frac{\partial }{\partial \vartheta}
\right) u_{a \vartheta}
, 
\nonumber \\ 
& & \mbox{}
\end{eqnarray}
and the variation in the distribution function $F_a$ is given by 
\begin{eqnarray}
\label{dFGK}
\delta F_a
& = & 
- \frac{\partial}{\partial x^j}
( F_a \delta x_{aE}^j  )
- \frac{\partial}{\partial v_{a\parallel}}
( F_a \delta v_{a\parallel E})
- \frac{\partial}{\partial \mu}
( F_a \delta \mu_{aE})
\nonumber \\ & & \mbox{}
- \frac{\partial}{\partial \vartheta}
( F_a \delta \vartheta_{aE})
. 
\end{eqnarray}
Using Eqs.~(\ref{IGKF}), (\ref{LGKF}), (\ref{LPsi}), (\ref{duvmt}) and (\ref{dFGK}), 
the variation in the action integral $I_{GKF}$
is expressed as 
\begin{eqnarray}
\label{dIGKF}
& & \delta I_{GKF}
= 
\sum_a
\int_{t_1}^{t_2}  dt \int_V d^3 x
\int d^3 v
\; F_a
 \nonumber \\
& & 
\hspace*{2mm}
\mbox{} \times 
\left[
\left\{ 
\left( \frac{\partial L_{GYa}}{\partial x^i} \right)_u
-
\left(\frac{d}{d t}\right)_a
\left( \frac{\partial L_{GYa}}{\partial u_{ax}^i} \right)
\right\} \delta x_{aE}^i 
\right. 
\nonumber \\
& & 
\hspace*{2mm}
\mbox{} + 
\left( \frac{\partial L_{GYa}}{\partial v_\parallel} \right)_u \delta v_{\parallel a E}
+ 
\left( \frac{\partial L_{GYa}}{\partial \mu } \right)_u \delta \mu_{aE}
\nonumber \\
& & 
\left. 
\hspace*{2mm}
+ 
  \left\{ 
\left( \frac{\partial L_{GYa}}{\partial \vartheta} \right)_u
-    \left(\frac{d}{d t}\right)_a 
\left( \frac{\partial L_{GYa}}{\partial u_{a\vartheta}} \right)
\right\} 
\delta \vartheta_{aE}
\right]
\nonumber \\
& & 
\hspace*{2mm} \mbox{} 
- 
\int_{t_1}^{t_2} dt \int_V d^3 x   
\left(
\rho^{(g)}  - \frac{1}{4\pi} \frac{\partial ( \sqrt{g} D^i) }{\partial x^i}
\right) \delta \phi 
+ \mbox{B.T.}
, 
\hspace*{8mm} 
\end{eqnarray}
where $D^i$ represents the electric displacement vector defined later in Eq.~(\ref{Di}) 
and  B.T.\ represents boundary terms that appear due to partial integrals. 
Here, 
$( \partial L_{GYa}/\partial x^i )_u$, 
$( \partial L_{GYa}/\partial v_\parallel )_u$, 
$( \partial L_{GYa}/\partial \mu )_u$, 
 and 
$( \partial L_{GYa}/\partial \vartheta )_u$ denote  
the derivatives of $L_{GYa}$ in $x^i$, $v_\parallel$, $\mu$, 
and $\vartheta$, respectively, 
with $(u_{ax}^i, u_{a\vartheta})$ kept fixed in $L_{GYa}$, and 
the time derivative along the phase-space trajectory 
is represented by 
\begin{equation}
\label{ddt2}
\left( 
\frac{d}{dt}
\right)_a
 \equiv  
\frac{\partial}{\partial t } 
+  u_{ax}^k \frac{\partial}{\partial x^k} 
+ u_{av_\parallel} \frac{\partial}{\partial v_\parallel} 
+  u_{a\mu} \frac{\partial}{\partial \mu} 
+  u_{a\vartheta} \frac{\partial}{\partial \vartheta} 
. 
\end{equation}

We now employ the Eulerian variation principle which implies that 
the collisionless gyrokinetic equations for the distribution functions of all species and 
the gyrokinetic Poisson's equation for the electrostatic potential 
can be derived from the condition that $\delta I_{GKF} = 0$ 
for arbitrary variations 
$\delta x_{aE}^i$, $\delta v_{a\parallel E}$, 
$\delta \mu_{aE}$, 
$\delta \vartheta_{aE}$, and $\delta \phi$ 
which vanish on the boundaries of the integral region 
to make B.T.\ disappear in Eq.~(\ref{dIGKF}). 

  We first use 
$\delta I_{GKF}/\delta x_{aE}^i =0$
to obtain 
\begin{equation}
\label{dpidt}
\left( 
\frac{d}{dt}
\right)_a 
p_{ai}
=
\left( \frac{\partial L_{GYa}}{\partial x^i} \right)_u
, 
\end{equation}
where $p_{ai}$ represents the covariant vector component 
of the canonical momentum defined by
\begin{equation}
\label{pai}
p_{ai}
\equiv 
\frac{\partial L_{GYa}}{\partial u_{ax}^i} 
= 
\frac{e_a}{c} A_i (x, t) + m_a v_\parallel b_i (x, t)
\equiv \frac{e_a}{c} A^*_{ai} (x, v_\parallel, t) 
. 
\end{equation}
We should note that 
the distribution function $F_a$ is included as a factor 
in  $\delta I_{GKF}/\delta x_{aE}^i =0$ 
although it is omitted from Eq.~(\ref{dpidt}) for simplicity. 
This omission of $F_a$ is also performed in the other equations obtained below from 
$\delta I_{GKF}/\delta v_{a\parallel E} =0$, 
$\delta I_{GKF}/\delta \mu_{aE}=0$, 
and $\delta I_{GKF}/\delta \vartheta_{aE}=0$
although it does not make a difference in deriving 
the resultant collisionless gyrokinetic equation in Eq.~(\ref{GKE}). 
   We can rewrite  Eq.~(\ref{dpidt}) as 
\begin{eqnarray}
\label{mupb}
m_a u_{av_\parallel} b_i  & = & 
 e_a \left( - \frac{\partial \Psi_a}{\partial x^i}
-   \frac{1}{c}
\frac{\partial  A^*_{ai}}{\partial t}
 + \frac{1}{c} \sqrt{g} \epsilon_{ijk} u_x^j 
B^{*k} \right)
\nonumber \\ & & \mbox{}
- \mu \frac{\partial B}{\partial x^i}
, 
\end{eqnarray}
where the modified magnetic field is defined by 
\begin{equation}
\label{B*i}
B^{*i}_a  \equiv
\frac{\epsilon^{ijk}}{\sqrt{g}} \frac{\partial A^*_{ak}}{\partial x^j}
,
\end{equation}
respectively. 

Next, 
$\delta I_{GKF}/\delta v_{a\parallel E} =0$
is used to obtain 
\begin{equation}
\label{dLdv}
\left( \frac{\partial L_{GYa}}{\partial v_\parallel} \right)_u
= m_a \left( 
u_{ax}^i b_i  - v_\parallel 
\right)
= 0 
, 
\end{equation}
from which we have 
\begin{equation}
\label{ubvp}
u_{ax}^i b_i  
=
 v_\parallel 
.
\end{equation}
   Furthermore,  
$\delta I_{GKF}/\delta \mu_{aE} =0$
and $\delta I_{GKF}/\delta \vartheta_{aE} =0$ 
yield 
\begin{equation}
\label{dIdmu}
\left( \frac{\partial L_{GYa}}{\partial \mu} \right)_u
= 
\frac{m_a c}{e_a} u_{a\vartheta} - B 
- e_a \frac{\partial \Psi_a}{\partial \mu}
= 0 
,
\end{equation}
and 
\begin{equation}
\label{dIdth}
\left( \frac{d}{d t } \right)_a
\left( \frac{\partial L_{GYa}}{\partial u_{a\vartheta}} \right)
= 
\frac{m_a c}{e_a} u_{a\mu}
=
\left( \frac{\partial L_{GYa}}{\partial \vartheta} \right)_u
=
0  
, 
\end{equation}
respectively. 

Equations~(\ref{mupb}), (\ref{ubvp}), (\ref{dIdmu}), and 
(\ref{dIdth}) are rewritten as 
\begin{equation}
\label{uxieq}
u_{ax}^i
= 
\frac{1}{B^*_{a\parallel}}
\left[ v_\parallel B^{*i}_a
+
c \frac{\epsilon^{ijk}}{\sqrt{g}} b_j
\left( 
 \frac{\mu}{e_a} \frac{\partial B}{\partial x^k}
+ \frac{\partial \Psi_a}{\partial x^k}
+ \frac{1}{c}
\frac{\partial  A^*_{ak}}{\partial t}
\right)
\right]
,
\end{equation}
\begin{equation}
\label{uvpeq}
m_a u_{av_\parallel} 
 =  
- \frac{B^{*i}_a}{B^*_{a\parallel}}
 \left[
 \mu \frac{\partial B}{\partial x^i}
+
e_a   \left( \frac{\partial \Psi_a}{\partial x^i}
+ \frac{1}{c}
\frac{\partial  A^*_{ai}}{\partial t} 
\right)
\right]
,
\end{equation}
\begin{equation}
\label{umueq}
u_{a\mu}
= 
0
,
\end{equation}
   and 
\begin{equation}
\label{uvteq}
u_{a\vartheta} 
= 
\Omega_a
+ \frac{e_a^2}{m_a c} 
\frac{\partial \Psi_a}{\partial \mu}
, 
\end{equation}
 where 
$\Omega_a \equiv e_a B / (m_a c)$
and 
\begin{equation}
\label{B*para}
B^*_{a\parallel}
\equiv
B^{*i}_a b_i
. 
\end{equation}
Equations~(\ref{uxieq}) and (\ref{uvpeq}) are obtained by 
taking the vector and scalar products between the magnetic 
field and Eq.~(\ref{mupb}), respectively. 
We can verify that the right-hand sides of Eqs.~(\ref{uxieq})--(\ref{uvteq}) 
are all independent of $\vartheta$ and that the magnetic moment 
$\mu$ is an invariant of motion as seen from Eq.~(\ref{umueq}). 

Substituting Eqs.~(\ref{uxieq})--(\ref{uvteq}) into 
Eqs.~(\ref{GKE0}) and taking its average with respect to 
the gyrophase $\vartheta$, 
the collisionless gyrokinetic equation is derived as 
\begin{eqnarray}
\label{GKE}
& & 
\frac{\partial \overline{F}_a}{\partial t} 
+ \frac{\partial}{\partial x^i}
\left[\overline{F}_a \frac{1}{B^*_{a\parallel}}
\left\{ v_\parallel B^{*i}_a
+
c \frac{\epsilon^{ijk}}{\sqrt{g}} b_j
\right.  \right. 
\nonumber \\ 
& & \mbox{}
\left. \left. 
\hspace*{30mm}
\times
\left( 
 \frac{\mu}{e_a} \frac{\partial B}{\partial x^k}
+ \frac{\partial \Psi_a}{\partial x^k}
+ \frac{1}{c}
\frac{\partial  A^*_{ak}}{\partial t}
\right)
\right\}
 \right]
\nonumber \\ 
& & \mbox{}
+ \frac{\partial}{\partial v_\parallel}
\left[ 
\overline{F}_a 
\frac{B^{*i}_a}{m_a B^*_{a\parallel}}
 \left\{
- e_a   \left( \frac{\partial \Psi_a}{\partial x^i}
+ \frac{1}{c}
\frac{\partial  A^*_{ai}}{\partial t} 
\right)
- \mu \frac{\partial B}{\partial x^i}
\right\}
 \right]
\nonumber \\ 
& & \mbox{}
 = 0
, 
\end{eqnarray}
where $\overline{F}_a$ denotes the gyrophase-averaged 
distribution function, 
\begin{equation}
\label{Fbar}
\overline{F}_a
\equiv 
\langle F_a \rangle_\vartheta
\equiv
\oint \frac{d\vartheta}{2\pi} F_a
. 
\end{equation}

The remaining governing equation of the system, namely, 
the gyrokinetic Poisson's equation
is derived from the condition that 
the variational derivative of the action integral  $I_{GKF}$ 
with respect to 
the electrostatic potential $\phi$ vanishes, 
$\delta I_{GKF} / \delta \phi =  0$.
Since the time derivative of $\phi$ never appears in 
the Lagrangian density ${\cal L}_{GKF}$, 
the above-mentioned condition can be replaced 
using the Lagrangian $L_{GKF}$ instead of $I_{GKF}$
by 
\begin{equation}
\label{dLdphi}
\frac{\delta L_{GKF} [\phi]}{\delta \phi} (x)
\equiv 
\left. 
\frac{d}{d\epsilon} L_{GKF} [\phi + \epsilon \delta^3_x]
\right|_{\epsilon=0} 
= 0
,
\end{equation}
where $\delta^3_x$ with the subscript $x = (x^i)_{i=1,2,3}$ represents the function 
that takes a value 
$\delta^3_x (y) = 
\delta (y^1-x^1) \delta (y^2-x^2) \delta (y^3-x^3)$
at $y= (y^i)_{i=1,2,3}$. 
Equation~(\ref{dLdphi}) gives the gyrokinetic Poisson's equation, 
\begin{equation}
\label{GKP}
\frac{\partial ( \sqrt{g} D^i) }{\partial x^i}
= 4\pi \rho^{(g)} 
, 
\end{equation}
where the electric displacement vector $D^i$ is 
written as 
\begin{equation}
\label{Di}
\sqrt{g} D^i \equiv 
\sqrt{g} E_L^i + 4\pi 
P_G^i
. 
\end{equation}
Here, the generalized polarization vector density $P_G^i$ is defined by
\begin{eqnarray}
\label{PG}
& & P_G^i
 \equiv 
\frac{\delta L_{GK}}{\delta (E_L)_i}
 \equiv 
\left. 
\frac{d}{d\epsilon} L_{GK} [(E_L)_i + \epsilon \delta^3_x]
\right|_{\epsilon=0} 
\nonumber \\ 
& &  =  
\sum_{m=1}^\infty  \sum_{n=1}^\infty  
(-1)^{m-1}
\nabla_{i_1} \cdots  \nabla_{i_{m-1}} 
Q^{i \, i_1 \cdots i_{m-1} }
\nonumber \\ 
& & 
= -
\sum_{k=1}^\infty  
\nabla_{j_1} \cdots \nabla_{j_{2k-1}} 
Q_0^{j_1 \cdots j_{2k-1} i}
+
\sum_{m=1}^\infty  \sum_{n=1}^\infty  
(-1)^{m-1}
\nonumber \\
& & \mbox{} 
\hspace*{1mm}
\times
\nabla_{i_1} \cdots  \nabla_{i_{m-1}} 
[
\chi^{i \, i_1 \cdots i_{m-1} ;  j_1 \cdots j_n}
\nabla_{j_1} \cdots  \nabla_{j_{n-1}} (E_L)_{j_n}
]
,
\hspace*{7mm}
\end{eqnarray}
in which 
not only the dipole moment but also other multipole moments~\cite{Jackson} 
occurring due to finite gyroradius are included. 
The multipole moments $Q^{i_1 \cdots i_m }$ in Eq.~(\ref{PG}) 
are generally given by the sum of the two parts, 
\begin{equation}
\label{Q}
Q^{i_1 \cdots i_m }
\equiv
Q_0^{i_1 \cdots i_m }
+ 
Q_E^{i_1 \cdots i_m }
, 
\end{equation}
where $Q_0^{i_1 \cdots i_m }$ and $Q_E^{i_1 \cdots i_m }$ are defined by 
Eqs.~(\ref{Q0}) and (\ref{QE}), respectively. 
It should be noted 
that $Q_0^{i_1 \cdots i_m }$ vanishes unless $m$ is an even integer. 
Taking only the contribution of $m=1$ to the summation 
over $m$ in Eq.~(\ref{Q}), 
we obtain the dipole moment density which is denoted by 
\begin{equation}
\label{PD}
P_D^i
\equiv
Q_E^i
\equiv 
\frac{\partial {\cal L}_{GK}}{\partial (E_L)_i}
.
\end{equation}
%

The presence of the infinite series due to 
the finite gyroradius 
in Eq.~(\ref{PG}) and other places 
has an analogy with that for the case of 
macroscopic electromagnetism described in detail 
in Ref.~\cite{Jackson} where 
the macroscopic charge density is evaluated for the system consisting 
of molecules. 
The contribution of each molecule to the macroscopic charge density 
is calculated by spatially averaging the microscopic charge density 
(given by the point charges constituting the molecule)
around the center of mass of the molecule. 
Then, 
the resultant expression of the macroscopic charge density 
is given by the series expansion associated with the multipole moments
due to the finite distance of each point charge from the 
center of mass of the molecule. 
  The local spatial average of the microscopic charge density 
in the system of molecules is replaced by 
the phase-space integration in the present case of 
the gyrokinetic system
to represent the macroscopic charge density as    
\begin{eqnarray}
\label{macroscopic_cd}
& & \rho^{(g)} ({\bf x}, t) - \nabla \cdot {\bf P}_G ({\bf x}, t)
\nonumber \\
& = & 
 \frac{\delta}{\delta \phi({\bf x})}
\left[
 \sum_a e_a \int d^3 x'  \int d^3 v \, 
F_a ({\bf x}', v, t) \Psi_a ({\bf x}', v, t)
\right]
\nonumber \\
& = & 
\sum_a e_a
\int d^3 x'
\int_{-\infty}^\infty  dv_\parallel 
\int_0^\infty  d\mu  
\int_0^{2\pi} d\vartheta \, \overline{F}_a ({\bf x}', v_\parallel, \mu )
\nonumber \\
&  &  
\hspace*{-5mm} 
\mbox{} \times
\left[ 
\delta ( {\bf x}' + \mbox{\boldmath$\rho$}_a - {\bf x} )
- 
\frac{e_a}{B} \frac{\partial}{\partial \mu} 
\left\{
\widetilde{\phi} ({\bf x}', \mu, \vartheta )
\delta ( {\bf x}' + \mbox{\boldmath$\rho$}_a - {\bf x} )
\right\}
\right]
,
\nonumber \\
&  &  
\end{eqnarray}
which contains the polarization effect due to the finite 
gyroradius and the microscopic electrostatic fluctuations.
In Eq.~(\ref{macroscopic_cd}), 
the Cartesian spatial coordinates are used and three-dimensional vectors 
are represented in terms of boldface letters.
In the square brackets on the right-hand side of Eq.~(\ref{macroscopic_cd}), 
the first and second terms give the point charge density of the particle 
at ${\bf x} = {\bf x}' + \mbox{\boldmath$\rho$}_a$  
and its correction due to 
the electrostatic fluctuation, respectively. 
We can see that the effect of the finite gyroradius (or the finite distance between 
the particle and gyrocenter positions) is included in the delta functions,  
which cause the infinite series expansions to appear in Eq.~(\ref{PG}) and 
other equations related to the polarization (or dipole and multipole moments). 

\section{DERIVATION OF THE MOMENTUM BALANCE}

In this section, we use the invariance of the Lagrangian 
under arbitrary infinitesimal transformations of spatial coordinates 
to derive the momentum balance equation for the single-particle-species 
system and that for the whole system including all species 
and the field.

\subsection{Invariance of the Lagrangian under arbitrary 
spatial coordinate transformations}

We now consider the infinitesimal transformation of 
the spatial coordinates from $x = (x^i)_{i=1,2,3}$ to $x' = (x'^i)_{i=1,2,3}$,  
which is written as  
\begin{equation}
\label{xprime}
 x'^i 
=
x^i + \xi^i (x)
. 
\end{equation}
Here, the infinitesimal variation in the spatial coordinate $x^i$ 
is denoted by $\xi^i (x)$
which is an arbitrary function of 
only the spatial coordinates $x = (x^i)_{i=1,2,3}$ 
and independent of time $t$. 

From Eq.~(\ref{LGKF}), 
the gyrokinetic part of the Lagrangian for species $a$ 
is found to be given by 
\begin{equation}
\label{LGKa}
L_{GKa}  
\equiv 
\int_V d^3 x  \, 
{\cal L}_{GKa}
\equiv 
\int_V d^3 x  
\int d^3 v\, 
F_a L_{GYa}
,  
\end{equation}
where the gyrocenter Lagrangian $L_{GYa}$ is defined in Eq.~(\ref{LGYa}). 
Since $L_{GKa}$ shown in Eq.~(\ref{LGKa}) is a scalar constant 
which is invariant under arbitrary spatial coordinate transformations 
including the infinitesimal transformation given by Eq.~(\ref{xprime}),  we have
\begin{equation}
\label{dLGKa}
\overline{\delta} L_{GKa}  
\equiv 
\int_V d^3 x  
\left(
\frac{\partial ( \xi^i {\cal L}_{GKa} )}{\partial x^i} 
+ 
\overline{\delta}
{\cal L}_{GKa}
\right)
= 0. 
\end{equation}
Note that here and hereafter  
we use $\overline{\delta} \cdots$ 
 to represent the variation associated with 
the infinitesimal spatial coordinate transformation 
which should be distinguished from 
the variation $\delta  \cdots$ due to the virtual 
displacement treated in Sec.~III. 
The divergence term in the integrand of Eq.~(\ref{dLGKa}) appears due to 
the difference between the domains of integrations in 
$x = (x^i)_{i=1,2,3}$ and $x' = (x'^i)_{i=1,2,3}$
while $\overline{\delta}{\cal L}_{GKa}$ in the integrand 
represents the variation in the spatial functional form of 
${\cal L}_{GKa}$ due to the infinitesimal spatial 
coordinate transformation.  

As shown in Appendix~B, 
the variation in the spatial 
functional form of an arbitrary tensor field 
(as well as an arbitrary tensor field density) 
due to the infinitesimal spatial coordinate transformation given in 
Eq.~(\ref{xprime}) 
is written as the opposite sign of its Lie derivative~\cite{Marsden2} 
with respect to the generating vector field $\xi^i$, and it is represented by 
$\overline{\delta} = - L_\xi$.   
It is also noted that  
since the Lagrangian density behaves as a scalar field density under 
arbitrary spatial coordinate transformations, 
the Lie derivative of ${\cal L}_{GKa}$ is given by 
\begin{equation}
\label{LxLGKa}
L_\xi  {\cal L}_{GKa}
=
\frac{\partial ( \xi^i {\cal L}_{GKa} )}{\partial x^i}
.
\end{equation}
Therefore, the integrand in Eq.~(\ref{dLGKa}) is simply rewritten as 
$L_\xi  {\cal L}_{GKa} - L_\xi  {\cal L}_{GKa} (= 0)$, 
from which its spatial integral $\overline{\delta} L_{GKa}$ 
is naturally found to vanish as shown in Eq.~(\ref{dLGKa}). 

We now use Eq.~(\ref{LGKa}) and the Leibniz rule for the derivative operation by  
$\overline{\delta} = - L_\xi$ to write the variation in the 
spatial functional form of  the Lagrangian density ${\cal L}_{GKa}$ as 
\begin{eqnarray}
\label{dLGKa2}
\overline{\delta}  {\cal L}_{GKa}
& =  & 
 \int d^3 v \, \overline{\delta} (F_a  L_{GYa})
\nonumber \\ 
& =  & 
 \int d^3 v  \, ( \overline{\delta} F_a  \cdot L_{GYa} 
+ F_a \cdot \overline{\delta}   L_{GYa}
)
.
\end{eqnarray}
Then, 
using Eqs.~(\ref{LxLGKa})--(\ref{dLGKa2}) and $\overline{\delta} = - L_\xi$,  
Eq.~(\ref{dLGKa}) is rewritten as 
\begin{eqnarray}
\label{dLGKa3}
\overline{\delta}
L_{GKa}
& = &  
\int_V d^3 x \int d^3 v \, 
[ -  \overline{\delta} ( F_a  L_{GYa} )
+ \overline{\delta} ( F_a  L_{GYa} )
]
\nonumber \\
& = & 
\int_V d^3 x \int d^3 v \, F_a 
(
-  \overline{\delta}  L_{GYa} 
+ \overline{\delta}  L_{GYa} 
)
\nonumber \\
& = & 
 \int_V d^3 x \int d^3 v \, 
 F_a \left( \xi^i \frac{\partial L_{GYa}}{\partial x^i}  + 
\frac{\partial L_{GYa}}{\partial u_{ax}^i} \overline{\delta} u_{ax}^i
\right.  
\nonumber \\ & & \mbox{} \hspace*{0mm}
+ \frac{\partial L_{GYa}}{\partial u_{a\vartheta}^i} 
\overline{\delta} u_{a\vartheta}^i
+ \sum_J  \frac{\partial L_{GYa}}{\partial (\partial_J A_i)} 
\overline{\delta} (\partial_J A_i)
\nonumber \\ & & \mbox{} \hspace*{0mm}
\left. 
+ \sum_J  \frac{\partial L_{GYa}}{\partial (\partial_J \phi)} 
\overline{\delta} (\partial_J \phi)
+ \sum_J  \frac{\partial L_{GYa}}{\partial (\partial_J g_{ij})} 
\overline{\delta} (\partial_J g_{ij})
\right)
\nonumber \\
& = & 
0
.
\hspace*{8mm}
\end{eqnarray}
Equation~(\ref{dLGKa3}) is found to hold by noting that 
$ L_\xi L_{GYa} = \xi^i \partial L_{GYa}/\partial x^i$ 
(because $L_{GYa}$ behaves as a scalar field under arbitrary spatial coordinate 
transformations) and that the chain rule 
is applied to the derivative operation $\overline{\delta} = - L_\xi$ on 
$L_{GYa}[u_{ax}^i, u_{a\vartheta}^i, 
\{ \partial_J A_i \}, \{ \partial_J \phi \}, \{ \partial_J g_{ij} \} ]$ 
as 
\begin{eqnarray}
\label{dLGYa}
& & 
 \overline{\delta} L_{GYa} 
= 
-  \xi^i \frac{\partial L_{GYa}}{\partial x^i} 
\nonumber \\
& & 
=
\frac{\partial L_{GYa}}{\partial u_{ax}^i} \overline{\delta} u_{ax}^i 
+ \frac{\partial L_{GYa}}{\partial u_{a\vartheta}^i} 
\overline{\delta} u_{a\vartheta}^i
+ \sum_J  \frac{\partial L_{GYa}}{\partial (\partial_J A_i)} 
\overline{\delta} (\partial_J A_i)
\nonumber \\ & & \mbox{} \hspace*{3mm}
+ \sum_J  \frac{\partial L_{GYa}}{\partial (\partial_J \phi)} 
\overline{\delta} (\partial_J \phi)
+ \sum_J  \frac{\partial L_{GYa}}{\partial (\partial_J g_{ij})} 
\overline{\delta} (\partial_J g_{ij})
,
\hspace*{5mm}
\end{eqnarray}
where 
$\partial L_{GYa}/\partial (\partial_J A_i) = 0$ 
when the order of $J$ is greater than or equal to two [see Eq.~(\ref{LGYaf})]. 

We next consider the invariance of the Lagrangian $L_{GKF}$ of the whole system 
under the infinitesimal spatial coordinate transformation, which is written 
in the same way as in Eq.~(\ref{dLGKa}) by 
\begin{equation}
\label{dLGKF}
\overline{\delta} L_{GKF}  
\equiv 
\int_V d^3 x  
\left(
\frac{\partial ( \xi^i {\cal L}_{GKF} )}{\partial x^i} 
+ 
\overline{\delta}
{\cal L}_{GKF}
\right)
= 0
,
\end{equation}
where $L_{GKF}$ consists of the gyrokinetic and field parts 
as shown in Eqs.~(\ref{IGKF}) and (\ref{LGKF}). 
Procedures similar to those leading to Eq.~(\ref{dLGKa3}) can be made 
to obtain 
\begin{eqnarray}
\label{dLGKF2}
\overline{\delta}
& & L_{GKF}
=  
 \int_V d^3 x \left[ \sum_a \int d^3 v \,
 F_a \left( \xi^i \frac{\partial L_{GYa}}{\partial x^i}  + 
\frac{\partial L_{GYa}}{\partial u_{ax}^i} \overline{\delta} u_{ax}^i
\right.  
\right.
\nonumber \\ & & \mbox{} \hspace*{7mm}
+ \frac{\partial L_{GYa}}{\partial u_{a\vartheta}^i} 
\overline{\delta} u_{a\vartheta}^i
+ \sum_J  \frac{\partial L_{GYa}}{\partial (\partial_J A_i)} 
\overline{\delta} (\partial_J A_i)
\nonumber \\ & & \mbox{} \hspace*{7mm}
\left. 
+ \sum_J  \frac{\partial L_{GYa}}{\partial (\partial_J \phi)} 
\overline{\delta} (\partial_J \phi)
+ \sum_J  \frac{\partial L_{GYa}}{\partial (\partial_J g_{ij})} 
\overline{\delta} (\partial_J g_{ij})
\right)
\nonumber \\ & & \mbox{} \hspace*{7mm}
\left. 
\mbox{}
+  \frac{\partial ( \xi^i {\cal L}_F)}{\partial x^i}
+  \frac{\partial {\cal L}_F}{\partial (\partial \phi/\partial x^i)} 
\overline{\delta} 
\left( \frac{\partial \phi}{\partial x^i}
\right)
+  \frac{\partial {\cal L}_F}{\partial g_{ij}} 
\overline{\delta} g_{ij} 
\right]
\nonumber \\
&  & \hspace*{3mm}
= 0
.
\end{eqnarray}
The invariance shown in Eq.~(\ref{dLGKF2}) can also be confirmed using 
Eq.~(\ref{dLGYa}) and the following formula for the derivative of 
${\cal L}_F[\partial \phi/\partial x^i, g_{ij} ]$ 
with respect to 
$\overline{\delta} = - L_\xi$,  
\begin{eqnarray}
\label{dLF}
\overline{\delta} {\cal L}_F
& = & 
- \frac{\partial ( \xi^i {\cal L}_F)}{\partial x^i}
\nonumber \\
& = & 
\frac{\partial {\cal L}_F}{\partial (\partial \phi/\partial x^i)} 
\overline{\delta} 
\left( \frac{\partial \phi}{\partial x^i}
\right)
+  \frac{\partial {\cal L}_F}{\partial g_{ij}} 
\overline{\delta} g_{ij} 
.
\end{eqnarray}
It is summarized from Eqs.~(\ref{dLGKa3}), 
(\ref{dLGYa}), (\ref{dLGKF2}), and (\ref{dLF}) that  
the invariance of the scalar constants of $L_{GKa}$ and $L_{GKF}$ 
under the infinitesimal spatial transformation can be verified using 
the chain rule formulas for the derivative operation 
$\overline{\delta} = - L_\xi$  on the scalar field $L_{GYa}$ and the 
scalar field density ${\cal L}_F$. 
The invariance formulas  shown in 
Eqs.~(\ref{dLGKa3}) and (\ref{dLGKF2}) are used to 
derive the momentum balance equation for the single-species 
system in Sec.IV.B and that for the whole system including all species 
and the field in Sec.~IV.C, respectively.

\subsection{Derivation of the momentum balance for a single particle species}

We now use the Euler-Lagrange equations for gyrocenter motion 
[Eqs.~(\ref{dpidt}), (\ref{dLdv}), (\ref{dIdmu}), and (\ref{dIdth})] and 
perform partial integrals to rewrite (\ref{dLGKa3}) as 
\begin{eqnarray}
\label{dLGKa4}
\overline{\delta}
L_{GKa}
& = & 
 \int_V d^3 x \, 
\left[ 
\xi^i \int d^3 v \left\{
\frac{\partial}{\partial t}
\left( 
 F_a 
\frac{\partial L_{GYa}}{\partial u_{ax}^i} 
\right)
\right.  \right. 
\nonumber \\ & & \mbox{} 
\left. 
-
{\cal K}_a
\frac{\partial L_{GYa}}{\partial u_{ax}^i} 
\right\}
+ \frac{\delta L_{GKa}}{\delta A_i} 
\overline{\delta} A_i
+ \frac{\delta L_{GKa}}{\delta \phi} 
\overline{\delta} \phi
\nonumber \\ & & \mbox{} 
\left. 
+ \frac{\delta L_{GKa}}{\delta g_{ij}} 
\overline{\delta} g_{ij}
\right]
+  \mbox{B.T.}
\nonumber \\ 
& = & 0
, 
\end{eqnarray}
where, instead of using Eq.~(\ref{GKE0}), 
the gyrocenter distribution function $F_a$ is assumed to satisfy 
\begin{eqnarray}
\label{GKB}
& & \frac{\partial F_a}{\partial t} 
+ \frac{\partial}{\partial x^j}
( F_a u_{ax}^j  )
+ \frac{\partial}{\partial v_\parallel}
( F_a u_{a v_\parallel}  )
+ \frac{\partial}{\partial \mu}
( F_a u_{a \mu} )
\nonumber \\ 
& & 
\hspace*{6mm} \mbox{}
+ \frac{\partial}{\partial \vartheta}
( F_a u_{a \vartheta}  )
= {\cal K}_a
. 
\end{eqnarray}
Here, ${\cal K}_a$ represents the rate of temporal 
change in $F_a$ due to collisions and/or external sources for the species $a$. 
In the present work, we assume that  
\begin{equation}
\label{eK}
\sum_a e_a \int d^3 v \,{\cal K}_a = 0
\end{equation}
is satisfied by ${\cal K}_a$. 
In fact, it is shown in Ref.~\cite{Sugama2015} that 
Eq.~(\ref{eK}) corresponds to the intrinsic ambipolarity of the
classical particle fluxes when ${\cal K}_a$ represents the collision 
operator in the gyrocenter coordinates. 
The variational derivative of $L_{GKa}$ with respect to $A_i$ is given by
\begin{eqnarray}
\label{dLGKadA}
& & 
 \frac{\delta L_{GKa}}{\delta A_i} 
=
\sum_J (-1)^{\#J}
\partial_J
\left(
\frac{\partial {\cal L}_{GKa}}{\partial (\partial_J A_i)}  
\right) 
\nonumber \\ 
& & 
=
 \frac{\partial {\cal L}_{GKa}}{\partial A_i} 
- \frac{\partial}{\partial x^j}
\left(
\frac{\partial {\cal L}_{GKa}}{\partial (\partial A_i /\partial x^j)}  
\right) 
=
\frac{e_a}{c} \Gamma_a^i
,
\end{eqnarray}
where the particle flux of species $a$ is represented by 
\begin{eqnarray}
\label{NV}
& & 
 \Gamma_a^k
\equiv 
\int d^3 v \, F_a u_{ax}^k
+ \frac{c}{e_a} \epsilon^{kij}
\frac{\partial }{\partial x^i}
\left(
\int d^3 v \, 
\frac{F_a}{\sqrt{g}}
\right.
\nonumber \\ & & 
\left.
\times \left[
- \mu b_j +\frac{m_a v_\parallel}{B}
\left\{
(u_{ax})_j - (u_{ax})_i b^i b_j
\right\}
- e_a \frac{\partial \Psi_a}{\partial B^j} 
\right]
\right)
.
\hspace*{8mm}
\end{eqnarray}

The variational derivatives 
$\delta L_{GKa}/\delta \phi$ and 
$\delta L_{GKa}/\delta g_{ij}$ are 
written as
\begin{equation}
\label{dLGKadphi}
 \frac{\delta L_{GKa}}{\delta \phi} 
= 
\sum_J (-1)^{\#J}
\partial_J
\left(
\frac{\partial {\cal L}_{GKa}}{\partial (\partial_J \phi)}  
\right) 
=
- e_a N_a^{(p)} 
\end{equation}
and
\begin{equation}
\label{dLGKadg}
\frac{\delta L_{GKa}}{\delta g_{ij}} 
= 
\sum_J (-1)^{\#J}
\partial_J
\left(
\frac{\partial {\cal L}_{GKa}}{\partial (\partial_J g_{ij})}  
\right) 
=
\frac{1}{2} P_a^{ij}
,
\end{equation}
respectively, where 
the particle density $N_a^{(p)}$ and 
the symmetric pressure tensor density $P_a^{ij}$ of species $a$ 
are defined by 
\begin{equation}
\label{Np}
e_a N_a^{(p)} 
\equiv
e_a N_a^{(g)} - \nabla_i P_{Ga}^i
\end{equation}
and 
\begin{equation}
\label{Paij}
P_a^{ij}
 \equiv 
2 
\sum_J (-1)^{\#J}
\int d^3 v \; F_a \frac{\partial L_{GYa}}{\partial (\partial_J g_{ij})}
=
P_{{\rm CGL}a}^{ij}
+ \pi_{\land a}^{ij}
+ P_{\Psi a}^{ij}
, 
\end{equation}
respectively, and 
$\#J = n$  represents the order of 
$J \equiv (j_1, j_2, \cdots, j_n )$. 

On the right-hand side of Eq.~(\ref{Np}), 
$P_{Ga}^i$ represents the contribution of species $a$ to the 
generalized polarization vector density $P_G^i$ defined in Eq.~(\ref{PG}) 
and it is written as 
\begin{eqnarray}
\label{PGa}
& & 
P_{Ga}^i
 \equiv 
\frac{\delta L_{GKa}}{\delta (E_L)_i}
 \nonumber \\ 
& & = 
-
\sum_{k=1}^\infty  
\nabla_{j_1} \cdots \nabla_{j_{2k-1}} 
Q_{0a}^{j_1 \cdots j_{2k-1} i}
+
\sum_{m=1}^\infty  \sum_{n=1}^\infty  
(-1)^{m-1}
\nonumber \\
& & \mbox{} 
\hspace*{2mm}
\times
\nabla_{i_1} \cdots  \nabla_{i_{m-1}} 
[
\chi_a^{i \, i_1 \cdots i_{m-1} ;  j_1 \cdots j_n}
\nabla_{j_1} \cdots  \nabla_{j_{n-1}} (E_L)_{j_n}
]
.
\hspace*{7mm}
\end{eqnarray}
On the right-hand side of Eq.~(\ref{Paij}), 
$P_{{\rm CGL}a}^{ij}$ is given in 
the Chew-Goldberger-Low (CGL) form,~\cite{H&S}  
\begin{equation}
\label{CGL0}
P_{{\rm CGL}a}^{ij}
=
\int d^3 v \, F_a
[ m_a v_\parallel^2 b^i b^j + \mu B ( g^{ij} - b^i b^j ) ]
, 
\end{equation}
and $\pi_{\land a}^{ij}$ is defined by   
\begin{equation}
\label{non-CGL0}
\pi_{\land a}^{ij}
\equiv 
 \int d^3 v \, F
m_a v_\parallel 
[ b^i ( u_x )_\perp^j  + ( u_x )_\perp^i b^j ]
.
\end{equation}
Here, the perpendicular component of the 
gyrocenter velocity is represented by 
$
( u_{ax} )_\perp^i 
\equiv 
 u_{ax}^i  - u_{ax}^k b_k b^i
$. 
We find that Eqs.~(\ref{CGL0}) and (\ref{non-CGL0}) 
agree with those given by Eqs.~(124) and (125) in Ref.~\cite{Sugama2018} 
except for the effects of the electrostatic fluctuations included in 
$F_a$ and $( u_{ax} )_\perp^i$ on the right-hand side of 
Eqs.~(\ref{CGL0}) and (\ref{non-CGL0}). 
We also note that Eq.~(\ref{non-CGL0}) contains the product of the fluctuation 
parts of $F_a$ and $( u_{ax} )_\perp^i$, the ensemble average of which 
does not disappear but contributes to the turbulent momentum transport. 
In the neoclassical transport theory,~\cite{H&S,Helander} 
it is considered that 
the CGL pressure tensor shown in Eq.~(\ref{CGL0}) contains 
the scalar (or isotropic) part, which represents background pressure, 
and the anisotropic part, the magnitude of which is 
smaller than the background pressure 
by the factor $\sim \rho/L$ where $\rho$ and $L$ represent 
the gyroradius and the equilibrium gradient scale length, respectively. 
The anisotropic part of the CGL pressure tensor causes the viscous force which 
plays an essential role in the neoclassical transport processes 
when the distribution function deviates 
from the local Maxwellian under the influence of collisions.~\cite{H&S,Helander}  
The magnitude of $\pi_{\land a}^{ij}$ defined in Eq.~(\ref{non-CGL0})
is regarded as $\sim (\rho/L)^2$. 

The last term on the right-hand side of Eq.~(\ref{Paij}) is given by 
\begin{equation}
\label{PPsia}
P_{\Psi_a}^{ij}
 \equiv 
- 2  
\sum_J (-1)^{\#J}
 \int d^3 v \; F_a e_a \frac{\partial \Psi_a}{\partial (\partial_J g_{ij})}
=
P_ {E1a}^{ij} + P_ {E2a}^{ij}
, 
\end{equation}
where $P_ {E1a}^{ij}$ and  $P_ {E2a}^{ij}$ are defined 
in Eqs.~(\ref{PE1a}) and (\ref{PE2a}) of Appendix~C, respectively. 
We note that the effects of the turbulent electrostatic potential are 
included in the definitions of $P_{\Psi_a}^{ij}$ explicitly 
as well as in $\pi_{\land a}^{ij}$ through the turbulent drift 
velocity part of  $( u_{ax} )_\perp^i$. 
Then, the nonlinear interaction of the turbulent potential and 
the fluctuation part of the gyrocenter distribution function 
included in $P_{\Psi_a}^{ij}$ and $\pi_{\land a}^{ij}$
causes the turbulent momentum transport. 
In Appendix~D, the ensemble-averaged pressure tensor describing 
the turbulent momentum transport is given by the WKB representation. 

Performing further partial integrals in Eq.~(\ref{dLGKa4}) finally gives 
\begin{equation}
\label{dLGKa5}
\overline{\delta} L_{GKa}  
=
\int_V d^3 x \,
\xi_j  J_{GKa}^j 
+ \mbox{B.T.} 
= 0
, 
\end{equation}
where
\begin{eqnarray}
\label{JGKaj}
J_{GKa}^j
& \equiv & 
\frac{\partial}{\partial t} 
\left(  m_a N_a^{(g)} V_{a g \parallel} b^j \right)
- \int d^3 v \, {\cal K}_a m_a v_\parallel b^j
\nonumber \\
& & 
\mbox{}
+
e_a \left(
N_a^{(p)} g^{jk} \frac{\partial \phi}{\partial x^k}
+ 
\frac{1}{c} N_a^{(g)} \frac{\partial A^j}{\partial t} 
\right. 
\nonumber \\
& & 
\mbox{}
 \left. 
 - \frac{1}{c} \frac{\epsilon^{jkl}}{\sqrt{g}} \Gamma_{ak} B_l 
\right)
+ \nabla_i P_a^{ij}
\end{eqnarray}
and 
\begin{equation}
\label{NVagpara}
 N_a^{(g)} V_{ag \parallel}
\equiv 
\int d^3 v \, F_a v_\parallel
. 
\end{equation}
%
In deriving Eq.~(\ref{dLGKa5})--(\ref{NVagpara}), we have deformed 
Eq.~(\ref{dLGKa4}) using
Eqs.~(\ref{GKB}), (\ref{dLGKadA}), (\ref{dLGKadphi}), (\ref{dLGKadg}), 
$\overline{\delta} A_i = - \xi^j (\partial_j A_i) - 
(\partial_i \xi^j) A_j$ [see Eq.~(\ref{dEgg})], 
$\overline{\delta} \phi = - \xi^j (\partial_j \phi)$ [see Eq.~(\ref{dS})], 
$\overline{\delta} g_{ij} = - \nabla_i \xi_j - \nabla_i \xi_i $ 
[see Eq.~(\ref{dg})]
and partial integration.
Especially, the term $\nabla_i P_a^{ij}$ in Eq.~(\ref{JGKaj}) is derived from 
the term $(\delta L_{GKa} / \delta g_{ij}) \overline{\delta} g_{ij}$ 
which is rewritten as
$
(\delta L_{GKa}/ \delta g_{ij}) \overline{\delta} g_{ij}
= -\frac{1}{2} P_a^{ij} (\nabla_i \xi_j + \nabla_i \xi_i)
= \xi_j (\nabla_i P_a^{ij} ) - \nabla_i ( \xi_j P_a^{ij} )
$,
where $P_a^{ij} = P_a^{ji}$ is also used and the spatial integral 
of the last term $\nabla_i ( \xi_j P_a^{ij} )$ becomes one of the 
boundary terms. 
We now recall that an arbitrary infinitesimal vector field can be employed as $\xi^i$. 
Then, in order for Eq.~(\ref{dLGKa5}) to hold for any $\xi^i$, 
we need to have 
$
J_{GKa}^j = 0, 
$
which gives 
the momentum balance equation as 
\begin{eqnarray}
\label{mombal}
& & 
\frac{\partial}{\partial t} 
\left(  m_a N_a^{(g)} V_{a g \parallel} b^j \right)
- \int d^3 v \, {\cal K}_a m_a v_\parallel b^j
\nonumber \\
& & 
= 
e_a \left(
- N_a^{(p)} g^{jk} \frac{\partial \phi}{\partial x^k}
  - \frac{1}{c} N_a^{(g)} \frac{\partial A^j}{\partial t}  + \frac{1}{c} 
\frac{\epsilon^{jkl}}{\sqrt{g}} \Gamma_{ak} B_l  \right) 
\nonumber \\
& & 
\hspace*{5mm}
\mbox{}
- \nabla_i P_a^{ij}
. 
\end{eqnarray}
As seen in Eqs.~(126) and (152) of Ref.~\cite{Sugama2018}, 
the relation between 
the symmetric and canonical pressure tensors is obtained from the other condition 
for the sum of the boundary terms (B.T.) in Eq.~(\ref{dLGKa5}) to vanish 
although its complicated expression is not shown here. 
We see that the inertia term in the momentum balance equation, Eq.~(\ref{mombal}), 
contains only the parallel momentum component while
 the electric current $e_a \Gamma_a^k$ in the Lorentz force term 
consists of the guiding-center current and  
the magnetization current as shown in 
Eq.~(\ref{NV}) [see also Eq.~(\ref{magnetization})]. 
For comparison with Eq.(\ref{mombal}), 
the canonical momentum balance equation 
derived by the conventional method is shown in 
Eq.~(\ref{canmomba}) of Appendix~E where the 
divergence of the asymmetric canonical pressure tensor 
appears. 
In addition, 
the energy balance equation for the single-species 
gyrokinetic system is given in Eq.~(\ref{energy_balance})
of Appendix~F.

\subsection{Derivation of the momentum balance for the whole system}

In the same way as in deriving Eq.~(\ref{dLGKa4}), 
performing partial integrals in Eq.~(\ref{dLGKF2}) 
and using Eqs.~(\ref{GKB}), 
we have 
\begin{eqnarray}
\label{dLGKF3}
\overline{\delta}
L_{GKF}
& = & 
 \int_V d^3 x \, 
\left[ 
\xi^i  \sum_a \int d^3 v \left\{
\frac{\partial}{\partial t}
\left( 
 F_a 
\frac{\partial L_{GYa}}{\partial u_{ax}^i} 
\right)
\right.  \right. 
\nonumber \\ & & \mbox{} 
\left. 
-
{\cal K}_a
\frac{\partial L_{GYa}}{\partial u_{ax}^i} 
\right\}
+ \frac{\delta L_{GKF}}{\delta A_i} 
\overline{\delta} A_i
+ \frac{\delta L_{GKF}}{\delta \phi} 
\overline{\delta} \phi
\nonumber \\ & & \mbox{} 
\left. 
+ \frac{\delta L_{GKF}}{\delta g_{ij}} 
\overline{\delta} g_{ij}
\right]
+  \mbox{B.T.}
\end{eqnarray}
%
%
%
Now, substituting Eq.~(\ref{dLGKadA}) into 
$\delta L_{GKF}/\delta A_i = \sum_a \delta L_{GKa}/\delta A_i $ 
and 
using the gyrokinetic Poisson's equation, 
$\delta L_{GKF} / \delta \phi = 0$, and Eq.~(\ref{dLGKadg}), 
Eq.~(\ref{dLGKF3}) is finally rewritten as
\begin{equation}
\label{dLGKF4}
\overline{\delta}
L_{GKF}
=
\int_V d^3 x \, 
\xi_j  J_{GKF}^j 
+ \mbox{B.T.}
= 0
,
\end{equation}
where 
\begin{eqnarray}
\label{JGKFj}
J_{GKF}^j
& \equiv & 
\sum_a 
\left[
\frac{\partial}{\partial t} 
\left(  m_a N_a^{(g)} V_{a g \parallel} b^j \right)
- \int d^3 v \, {\cal K}_a m_a v_\parallel b^j
\right]
\nonumber \\
& & 
\mbox{}
+ 
\sum_a  \frac{e_a }{c} \left( 
N_a^{(g)} 
\frac{\partial A^j}{\partial t}
- \frac{\epsilon^{jkl}}{\sqrt{g}} \Gamma_{ak} B_l  \right) 
+ \nabla_i \Theta^{ij}
. 
\hspace*{10mm}
\end{eqnarray}
Here, the symmetric pressure tensor $\Theta^{ij}$ is defined by 
\begin{eqnarray}
\label{Theta}
& &  \Theta^{ij}
\equiv 
2 \frac{\delta L_{GKF}}{\delta g_{ij}} 
\equiv 
2 \left[
\sum_J (-1)^{\#J}
\int d^3 v \; F_a \frac{\partial L_{GYa}}{\partial (\partial_J g_{ij})}
+ 
\frac{\partial {\cal L}_F}{\partial g_{ij}} 
\right]
\nonumber \\
& & 
\hspace*{5mm}
=
P_{\rm CGL}^{ij}
+ \pi_\land^{ij} 
+ P_\Psi^{ij}
\nonumber \\
& & \mbox{}
\hspace*{10mm}
+
\frac{\sqrt{g}}{4\pi} 
\left( \frac{(E_L)^k (E_L)_k}{2} g^{ij}
-
(E_L)^i (E_L)^j 
\right)
,
\hspace*{10mm}
\end{eqnarray}
where the last group of terms including $(E_L)_i$ 
represents the Maxwell stress tensor due to the 
electrostatic field with the opposite sign. 
Taking the summation of 
Eqs.~(\ref{CGL0}), (\ref{non-CGL0}), and 
(\ref{PPsia})
over species defines 
$P_{\rm CGL}^{ij}$, $\pi_\land^{ij}$, and $P_\Psi^{ij}$ on 
the right-hand side of Eq.~(\ref{Theta}) as 
\begin{equation}
\label{CGL}
P_{\rm CGL}^{ij}
=
\sum_a \int d^3 v \, F_a
[ m_a v_\parallel^2 b^i b^j + \mu B ( g^{ij} - b^i b^j ) ]
, 
\end{equation}
\begin{equation}
\label{non-CGL}
\pi_\land^{ij}
\equiv 
\sum_a \int d^3 v \, F_a
m_a v_\parallel 
[ b^i ( u_{ax} )_\perp^j  + ( u_{ax} )_\perp^i b^j ]
, 
\end{equation}
and
\begin{equation}
\label{PPsi}
P_\Psi^{ij}
\equiv 
\sum_a
P_{\Psi_a}^{ij}
=
\sum_a ( P_ {E1a}^{ij} + P_ {E2a}^{ij} )
, 
\end{equation}
respectively. 
As mentioned after Eq.~(\ref{PPsia}) as well as in Appendix~F, 
the turbulent momentum transport caused by 
the nonlinear interaction of the turbulent potential and 
the fluctuation part of the gyrocenter distribution function 
is included in $\pi_\land^{ij}$ and $P_\Psi^{ij}$. 

From Eqs.~(\ref{dLGKF4}) and (\ref{JGKFj}), 
we obtain $J_{DKF}^j = 0$ which represents 
the momentum balance equation for the whole system, 
\begin{eqnarray}
\label{totmomb0}
& & 
\frac{\partial}{\partial t}
\left( 
\sum_a \int d^3 v \, F_a 
m_a v_\parallel b^j 
\right) 
- \sum_a \int d^3 v \, {\cal K}_a m_a v_\parallel b^j
+
\nabla_i \Theta^{ij} 
\nonumber \\ 
& & = 
\sum_a   \frac{e_a }{c} \left( 
- N_a^{(g)}  \frac{\partial A^j}{\partial t}
+ \frac{\epsilon^{jkl}}{\sqrt{g}} \Gamma_{ak} B_l  \right) 
. 
\end{eqnarray}
For comparison with Eq.(\ref{totmomb0}), 
the canonical momentum balance equation 
derived by the conventional method is shown in 
Eq.~(\ref{totcanmomb}) of Appendix~E where the 
divergence of the asymmetric pressure tensor 
appears. 
The condition for the sum of the boundary terms (B.T.) in Eq.~(\ref{dLGKF4}) 
to vanish results in a complicated expression which is not shown here 
while it gives the relation between the symmetric pressure tensor 
$\Theta^{ij}$ and the asymmetric canonical pressure tensor 
in Eq.~(\ref{totcanmomb}). 
Here, using Eq.~(\ref{NV}), the electric current density can be written as 
\begin{equation}
\label{current}
J^k
\equiv 
\sum_a 
e_a
 \Gamma_a^k
=
\sum_a e_a
\int d^3 v \, F_a u_{ax}^k
+
c \,\epsilon^{kij}
\frac{\partial  M_j}{\partial x^i}
,
\end{equation}
where the covariant components of 
the magnetization vector are defined by 
\begin{eqnarray}
\label{magnetization}
M_j
& \equiv & 
\sum_a 
\int d^3 v \, 
\frac{F_a}{\sqrt{g}}
\left[
- \mu b_j +\frac{m_a v_\parallel}{B}
\right.
\nonumber \\ & & 
\left.
\times
\left\{
(u_{ax})_j - (u_{ax})_i b^i b_j
\right\}
- e_a \frac{\partial \Psi_a}{\partial B^j} 
\right]
.
\end{eqnarray}
Using Eqs.~(\ref{GKB}) and (\ref{eK}), 
we have  
\begin{equation}
\label{charge_conservation}
\frac{\partial \rho^{(g)}}{\partial t} + \frac{\partial J^i}{\partial x^i} = 0, 
\end{equation}
where $\rho^{(g)}$ is defined in Eq.~(\ref{rhog}). 
Combining Eq.~(\ref{GKP}) and (\ref{charge_conservation}), 
we obtain  
\begin{equation}
\label{JL}
 J_L^i + \frac{1}{4\pi} 
\frac{\partial ( \sqrt{g}D_L^i )}{\partial t}= 0, 
\end{equation}
where the subscript $L$ represents the longitudinal part. 
It should be noted that
the polarization current 
$4\pi \partial P_G^i / \partial t$
 is included not in $J^i$ but 
in $\partial (\sqrt{g} D^i )/ \partial t$. 

It can be shown by performing further vector calculations that 
Eq.~(\ref{totmomb0}) is rewritten as 
\begin{eqnarray}
\label{totmomb}
& & 
\frac{\partial}{\partial t}
\left( 
\sum_a \int d^3 v \, F_a 
m_a v_\parallel {\bf b} 
+ \frac{1}{4\pi c} ( {\bf D}_L \times {\bf B} )
\right) 
\nonumber \\ 
& & 
\mbox{} 
+
\nabla \cdot 
\left(
{\bf P}_{\rm CGL}
+ \mbox{\boldmath$\pi$}_\land
+ {\bf P}_\Psi
\right)
+
\nabla \left( 
\frac{|{\bf E}_L|^2}{8 \pi} + \frac{{\bf E}_T \cdot {\bf D}_L }{4 \pi}  
\right)
\nonumber \\ 
& & 
\mbox{} 
-  \nabla \cdot  \left( 
\frac{{\bf E}_L {\bf E}_L + {\bf D}_L {\bf E}_T + {\bf E}_T {\bf D}_L}{4 \pi}
\right) 
\nonumber \\ 
& & 
\mbox{} 
+
\nabla \left( 
\frac{B^2}{8 \pi}  
\right)
-  \nabla \cdot  \left( \frac{{\bf B} {\bf B}}{4 \pi} \right) 
\nonumber \\ 
& & = 
 \sum_a \int d^3 v \, {\cal K}_a m_a v_\parallel {\bf b} 
+ 
\left(
\frac{\nabla \times {\bf B}}{4\pi}  - 
\frac{{\bf J}_T }{c} 
\right) \times {\bf B}
, 
\end{eqnarray}
where the Cartesian spatial coordinates are used 
and three-dimensional vectors are represented in terms of boldface letters. 
The longitudinal part of the electric displacement vector defined in Eq.~(\ref{Di}) 
is represented by ${\bf D}_L$. 
The longitudinal part of the electric field is written as 
${\bf E}_L \equiv - \nabla \phi$ while 
${\bf E}_T \equiv - c^{-1} \partial {\bf A}/\partial t$ 
gives the transverse part under the Coulomb gauge condition 
$\nabla \cdot {\bf A} = 0$. 
We see that 
the time derivative of 
the momentum density $( {\bf D}_L \times {\bf B} ) / (4\pi c)$ 
due to the electromagnetic field and the spatial divergences of 
the pressure tensors produced by both the 
electric and magnetic fields 
(the opposite sign of the Maxwell stress tensor) appear in the momentum balance 
equation in Eq.~(\ref{totmomb}) where the effects of the polarization ${\bf P}_G$ 
[see Eq.~(\ref{PG})] are included through ${\bf D}_L$. 

Here, recall that $F_a (x, v, t)$ represents 
the gyrocenter distribution function 
of the gyrocenter (not particle) position coordinates 
$x \equiv (x^i)_{i=1,2,3}$, $v \equiv (v^i)_{i=1,2,3} \equiv 
(v_\parallel, \mu, \vartheta)$ and time $t$ 
[see descriptions after Eq.(\ref{LGKF})]. 
Then, from typical gyrokinetic codes 
using the gyrocenter position coordinates as 
independent variables for distribution functions, 
we can directly evaluate the term 
$
\int d^3 v \, F_a m_a v_\parallel {\bf b} \equiv \int_{-\infty}^\infty  dv_\parallel 
\int_0^\infty  d\mu  
\int_0^{2\pi} d\vartheta \, F_a (x, v_\parallel, \mu, \vartheta ) 
m_a v_\parallel {\bf b} (x, t)
$
in Eq.~(\ref{totmomb}), 
for which we do not need to specify 
the transformation from the gyrocenter position coordinates    
to the particle coordinates. 
As explained at the end of Sec.~III.C, 
the difference between the gyrocenter and particle positions 
is taken into account in Eq.~(\ref{totmomb}) through 
the terms related to the polarization due to the effects of 
the finite gyroradius and the electrostatic fluctuation.
Such polarization terms also 
appear in the energy balance equations 
as seen in Eqs.~(\ref{totenergyb1}) and (\ref{totenergyb2}) in Appendix~F.

When the polarization ${\bf P}_G$ is eliminated from Eq.~(\ref{totmomb}), 
the terms including the electric field are given by
$\partial [( {\bf E}_L \times {\bf B} ) / (4\pi c) ]/ \partial t$
and 
$
\nabla [ |{\bf E}_L|^2/(8 \pi) + ({\bf E}_T \cdot {\bf E}_L)/(4 \pi) ]
-  \nabla \cdot  [
({\bf E}_L {\bf E}_L + {\bf E}_L {\bf E}_T + {\bf E}_T {\bf E}_L)/(4 \pi ) ]
$
which are verified to be the same 
as given in the momentum conservation law 
for the Vlasov-Poisson-Amp\`{e}re (or Vlasov-Darwin) system 
[see Eq.~(32) of Ref.~\cite{Sugama2013}]. 
In the case using the quasineutrality condition and the self-consistent magnetic field 
given 
by $\nabla \times {\bf B} = (4\pi/c) {\bf J}$ (${\bf J} = {\bf J}_T$ from 
$\nabla \cdot {\bf J} = 0$ due to the quasineutrality) 
as well as removing polarization effects and ${\cal K}_a$,  
we find that $( {\bf D}_L \times {\bf B} ) / (4\pi c)$ and 
the part of the pressure tensor caused by the electric field 
disappear from Eq.~(\ref{totmomb}) and 
that Eq.~(\ref{totmomb}) agrees with 
the momentum conservation law for the drift kinetic 
system shown in Eq.~(151) of Ref.~\cite{Sugama2018}. 

The energy balance equation for the whole system is derived  
in Appendix~F [see Eqs.~(\ref{totenergyb1}) and (\ref{totenergyb2})] 
where, in the same way as seen in Eq.~(\ref{totmomb}), 
we can confirm the consistency of the derived energy balance 
with the energy conservation laws obtained for 
the Vlasov-Poisson-Amp\`{e}re (or Vlasov-Darwin) 
system~\cite{Sugama2013} and the drift kinetic 
sytem~\cite{Sugama2018}.

We note that, if the Lagrangian ${L}_{GKF}$ for 
the gyrokinetic system 
defined by Eqs.~(\ref{IGKF}) and (\ref{LGKF}) 
is modified to 
$L_{GKF*} \equiv L_{GKF} - \int_V d^3 x \, \sqrt{g} B^2/(8\pi)$, 
the variational equation $\delta L_{GKF*}/ \delta {\bf A} = 0$  
yields  
$
\nabla \times {\bf B} = (4\pi /c ) {\bf J}_T  
$
which makes Eq.~(\ref{totmomb}) take the form of 
the total momentum conservation law in the case of 
$\sum_a \int d^3 v \, {\cal K}_a m_a v_\parallel = 0$. 
Nevertheless, in the gyrokinetic turbulent system, 
this condition is not generally imposed on the given 
equilibrium magnetic field ${\bf B}$. 
It is because ${\bf B}$ is considered not to contain
the fluctuation part while ${\bf J}_T$ can have fluctuations.  
However, when the background magnetic field satisfies  
spatial translation, rotation, or helical symmetry and the effect of
${\cal K}_a$ is neglected, 
the local conservation law of the canonical momentum in the 
direction of symmetry can be derived from $\overline{\delta} L_{GKF} = 0$  
with Eq.~(\ref{dLGKF3}) as shown in Appendix~G. 

\section{CONCLUSIONS}

In this paper, 
the governing equations of the gyrokinetic system with electrostatic turbulence 
are derived in the general spatial coordinates based on the Eulerian variational principle. 
The local momentum balance equation for each particle species and that for the 
whole system, which the gyrocenter distribution functions and 
the potential field satisfy, are obtained from 
the invariance of the Lagrangians of these systems under arbitrary spatial coordinate transformations. 

It is shown that, when the background magnetic 
field satisfies the consistency condition that 
its rotation is given by the solenoidal part of the current density as in the Darwin model, 
the momentum and energy balance equations for the whole system are 
rewritten in the complete conservative forms where contributions of the turbulent electric field and the background magnetic field are clearly given  
in the expressions of the momentum and energy densities,  the pressure tensor, and 
the energy flux. 
The effects of the collision and/or external source terms added into the gyrokinetic equation on the momentum and energy balance equations are clarified as well. 

The symmetric pressure tensor is directly obtained by the variational derivative of the Lagrangian with respect to the metric tensor components and it  
is shown to contain the CGL part representing the neoclassical viscosity 
as well as the turbulent momentum transport part, the ensemble average of which 
is confirmed to agree with the previous result obtained from the gyrokinetic theory 
using the WKB representation. 

The representation in terms of the general spatial coordinates is useful in 
treating complex toroidal plasmas in suitable coordinates such as 
the flux coordinates. 
The momentum and energy balance equations obtained here are applicable 
as a reference for verification of long-time global gyrokinetic simulations 
based on the Lagrangian and Hamiltonian formulations to study neoclassical and 
turbulent transport in plasmas with external sources. 
    It may seem troublesome for global simulation codes to 
treat the finite gyroradius effect represented by the 
infinite series expansions appearing in Eq.~(\ref{PsiE1a}) and other places. 
However, for those simulations in which the expansions are truncated or 
approximated by other simpler expressions such that the Lagrangian 
corresponding to the reduced model given by the truncation or approximation is clearly defined, 
the same technique as shown in this work can be applied to that Lagrangian 
to derive the local energy and momentum balance equations 
which can be compared 
with those simulation results. 
Such applications of the present work to the global simulations based on 
the reduced gyrokinetic model are considered as future works. 
   For comparison with local flux tube gyrokinetic 
simulations~\cite{Dimits,GENE,GYRO,GKV,GKW}  
treating the full finite gyroradius effect, 
useful informations such as expressions of local 
turbulent momentum transport can be obtained from this work 
using the WKB approximation as shown in Appendix~D. 
The extension of the present work to the case with magnetic 
microturbulence also 
remains as a future study. 

\begin{acknowledgments}
This work is supported in part by 
JSPS Grants-in-Aid for Scientific Research Grant No.~19H01879 
 and in part by the NIFS Collaborative Research Program NIFS20KNTT055. 
\end{acknowledgments}

\section*{DATA AVAILABILITY}
Data sharing is not applicable to this article as no new data were created or analyzed in this study.

\appendix

\section{COVARIANT DERIVATIVES AND CHRISTOFFEL SYMBOLS}

This Appendix briefly describes definitions of covariant derivatives and 
Christoffel symbols which are used in the main text of the present paper.  
In the general spatial coordinates $x = (x^i)_{i=1,2,3}$, 
the components $\nabla_i S$  $(i =1,2,3)$ of the covariant derivative 
of an arbitrary scalar field $S(x)$ is given by
\begin{equation}
\nabla_i S  = 
\frac{\partial S}{\partial x^i}
, 
\end{equation}
while those of the covariant derivatives of arbitrary 
contravariant and covariant vector fields, 
$V^j$ and $W_j$ $(j=1,2,3)$,  are written as 
\begin{equation}
\nabla_i V^j   
 = 
\frac{\partial V^j}{\partial x^i}
+ \Gamma_{ik}^j V^k
,
\end{equation}
and 
\begin{equation}
\nabla_i W_j    
=  
\frac{\partial W_j}{\partial x^i}
- \Gamma_{ij}^k W_k
, 
\end{equation}
respectively. 
Here, the Christoffel symbols $\Gamma_{ij}^k$ $(i,j,k=1,2,3)$ are defined 
by~\cite{Schutz}
\begin{eqnarray}
\label{Christoffel}
& & \Gamma_{ij}^k (x)    \equiv  
 g^{kl} (x) \Gamma_{l, ij}(x)  
\nonumber \\ 
& & \equiv 
\frac{1}{2} g^{kl}(x) 
\left[ 
\frac{\partial g_{jl} (x) }{\partial x^i}
+ \frac{\partial g_{li} (x) }{\partial x^j}
- \frac{\partial g_{ij} (x) }{\partial x^l}
\right]
. 
\hspace*{5mm}
\end{eqnarray}
The covariant and contravariant components of the metric tensor 
are denoted by $g_{ij}$ and $g^{ij}$, respectively, 
and they satisfy 
\begin{equation}
g^{ik} g_{kj} = \delta^i_j
, 
\end{equation}
where $\delta^i_j$ represents the Kronecker delta defined by 
\begin{equation}
\delta^i_j
\equiv 
\left\{
\begin{array}{cc}
1 & (i = j) \\
0 & (i \neq j) .
\end{array}
\right. 
\end{equation}

   In general, the components of the covariant derivative
of an arbitrary 
mixed tensor field  $T^{j_1 \cdots j_r}_{k_1 \cdots k_s}$ of type $(r, s)$ 
are defined by
\begin{eqnarray}
\label{nablaT}
\nabla_i T^{j_1 \cdots j_r}_{k_1 \cdots k_s} 
&  = & 
\frac{\partial T^{j_1 \cdots j_r}_{k_1 \cdots k_s} }{\partial x^i}
+ \Gamma_{i l}^{j_1} T^{l j_2 \cdots j_r}_{k_1 \cdots k_s} 
+ \cdots +
 \Gamma_{i l}^{j_r}  T^{j_1 \cdots j_{r-1} l}_{k_1 \cdots k_s} 
\nonumber \\ & & \mbox{}
- \Gamma_{i k_1}^{l} T^{j_1 \cdots j_r}_{l k_2 \cdots k_s}
- \cdots -
 \Gamma_{i k_s}^{l}  T^{j_1 \cdots j_r}_{k_1 \cdots k_{s-1} l} 
.
\end{eqnarray}
Here, it is also noted that 
the covariant derivatives of 
$g_{jk}$, $g^{jk}$, and $g\equiv  \det ( g_{ij} )$ 
all vanish, 
\begin{equation}
\label{nablag}
\nabla_i g_{jk} =0,
\hspace*{5mm}
\nabla_i g^{jk} =0,
\hspace*{5mm}
\mbox{and}
\hspace*{5mm}
\nabla_i  g  =0. 
\end{equation}
We also note that 
when the covariant derivatives act on arbitrary tensors, 
the commutative property, 
\begin{equation}
\label{comm}
\nabla_i \nabla_j
= \nabla_j \nabla_i
, 
\end{equation}
holds for the case of the present paper where
the considered real space is a flat one 
with no Riemann curvature.

\section{VARIATIONS IN THE FUNCTIONAL FORMS OF VECTOR AND TENSOR FIELDS 
UNDER THE INFINITESIMAL TRANSFORMATION OF SPATIAL COORDINATES}

It is shown in this Appendix how to represent variations in the functional forms of 
vector and tensor fields under the infinitesimal spatial coordinate transformation.  
The infinitesimal transformation from 
$x= (x^i)_{i=1,2,3}$ to $x'= (x'^i)_{i=1,2,3}$ is given by Eq.~(\ref{xprime})   
with the infinitesimal variation $\xi^i (x)$ in the spatial coordinate $x^i$.  
Here, $( \xi^i  (x) )_{i=1,2,3}$ can be regarded as components of 
a vector field $\xi (x)$. 

We first consider a scalar field $S(x)$ which is invariant under 
the spatial coordinate transformation, 
\begin{equation}
\label{phi}
S' (x') 
=
S(x) 
, 
\end{equation}
and define  the variation $\overline{\delta} S$ 
in the functional form of a scalar field 
$S(x)$ under the infinitesimal spatial coordinate transformation by 
\begin{equation}
\label{dbphi}
\overline{\delta} S(x) 
\equiv 
S' (x) - S (x).  
\end{equation}
Note that the spatial arguments of $S'$ and $S$ are the same as 
each other on the right-hand side of Eq.~(\ref{dbphi}) 
while they are different in Eq.~(\ref{phi}). 
Next, substituting  
$S' (x') \simeq S' (x) 
+ \xi^i (x) \partial  S' (x) /\partial x^i
\simeq S' (x, t) 
+ \xi^i (x) \partial  S (x) /\partial x^i$ 
into Eq.~(\ref{phi}) and 
using Eq.~(\ref{dbphi}), 
we obtain 
\begin{equation}
\label{dS}
 \overline{\delta} S (x)
=
-  \xi^i (x)\frac{\partial S (x)}{\partial x^i} 
\equiv 
- ( L_\xi  
S
) (x)
,
\end{equation}
where $L_\xi$ denotes the Lie derivative~\cite{Marsden2} 
with respect to the vector field $\xi$ with the components  
$(\xi^i)$. 
 
Under the general spatial coordinate transformation, 
the components of a contravariant vector field $V^i (x)$
are transformed as 
\begin{equation}
\label{Vprime}
V'^i(x') 
=
\frac{\partial x'^i}{\partial x^j}
V^j(x) 
.  
\end{equation}
  In the same way as in Eq.~(\ref{dbphi}), 
we define the variation $\overline{\delta} V^i (x)$  
in the functional form of 
$V^i (x)$ under the infinitesimal spatial coordinate transformation by 
\begin{equation}
\label{dbV}
\overline{\delta} V^i (x) \equiv 
V'^i (x)  -  V^i (x)
.
\end{equation}
Substituting the formulas 
$V'^i (x') \simeq  V'^i (x, t)  + \xi^j (x) \partial V^i (x)/\partial x^j$ 
and 
$\partial x'^i /\partial x^j\simeq 
\delta^i_j  + \partial \xi^i (x) /\partial x^j$ 
into Eq.~(\ref{Vprime}) 
and using Eq.~(\ref{dbV}), 
we obtain  
\begin{eqnarray}
\label{deltaV}
\overline{\delta} V^i (x)
& = & 
 - \xi^j (x) \frac{\partial V^i (x)}{\partial x^j} 
+ \frac{\partial \xi^i (x)}{\partial x^j}
V^j (x) 
\nonumber \\ 
& \equiv & 
- ( L_\xi V^i ) (x)
, 
\end{eqnarray}
where we see that 
the Lie derivative $L_\xi$  
can be used again to represent $\overline{\delta} V^i (x)$.  
The components of a covariant vector field $W_i (x)$ 
are transformed as 
\begin{equation}
\label{Egg}
W'_i (x')
=
\frac{\partial x^j }{\partial x'^i}
W_j (x) 
.
\end{equation}
Next, following the procedure similar to those used in deriving 
Eqs.~(\ref{dS}) and (\ref{deltaV}), 
the variation in the functional form of $W_i (x)$
under the infinitesimal spatial coordinate transformation is  
derived as
\begin{equation}
\label{dEgg}
\overline{\delta} W_i (x)
=
- ( L_\xi W_i ) (x)
= 
- \xi^j \frac{\partial W_i }{\partial x^j}
- \frac{\partial \xi^j }{\partial x^i}
W_j 
.
\end{equation}

It is shown in the same way as shown above that 
the variation in the functional form of 
a mixed tensor field $T^{j_1 \cdots j_r}_{k_1 \cdots k_s} (x)$ 
is given by the opposite sign of its Lie derivative as 
\begin{eqnarray}
& & \overline{\delta} T^{j_1 \cdots j_r}_{k_1 \cdots k_s} (x)
=
- (L_\xi T^{j_1 \cdots j_r}_{k_1 \cdots k_s}) (x)
\nonumber \\ & & \mbox{}
= 
- \xi^i \frac{\partial T^{j_1 \cdots j_r}_{k_1 \cdots k_s} }{\partial x^i}
+ \frac{\partial \xi^{j_1}}{\partial x^l} 
T^{l j_2 \cdots j_r}_{k_1 \cdots k_s} 
+ \cdots + \frac{\partial \xi^{j_r}}{\partial x^l} 
T^{j_1 \cdots j_{r-1} l}_{k_1 \cdots k_s} 
\nonumber \\ & & \mbox{}
\hspace*{5mm}
- \frac{\partial \xi^l}{\partial x^{k_1}} 
T^{j_1 \cdots j_r}_{l k_2 \cdots k_s}
- \cdots - \frac{\partial \xi^l}{\partial x^{k_s}}
T^{j_1 \cdots j_r}_{k_1 \cdots k_{s-1} l} 
.
\end{eqnarray}
It is also shown that the variations in the functional forms of 
$g_{ij}$ and $g^{ij}$ under the infinitesimal 
spatial coordinate transformation are expressed as 
\begin{eqnarray}
\label{dg}
& & \overline{\delta} g_{ij}
=
- L_\xi g_{ij}
= 
- \xi^k \frac{\partial g_{ij} }{\partial x^k}
- \frac{\partial \xi^k}{\partial x^i} g_{kj}
- \frac{\partial \xi^k}{\partial x^j} g_{ik}
\nonumber \\
& & 
\hspace*{7mm}
=
 - \nabla_i \xi_j - \nabla_j \xi_i
, 
\nonumber \\
& & \overline{\delta} g^{ij}
=
- L_\xi g^{ij}
= 
- \xi^k \frac{\partial g^{ij} }{\partial x^k}
+ \frac{\partial \xi^i}{\partial x^k} g^{kj}
+ \frac{\partial \xi^j}{\partial x^k} g^{ik}
\nonumber \\
& & 
\hspace*{7mm}
= 
\nabla^i \xi^j + \nabla^j \xi^i
. 
\end{eqnarray}

We now apply the chain rule to the derivative operation  
$\overline{\delta} = - L_\xi$ 
on $\sqrt{g} = \sqrt{\det ( g_{ij} )}$ 
and use Eqs.~(\ref{nablag}) and (\ref{dg}) to obtain 
\begin{eqnarray}
& & 
\overline{\delta} \sqrt{g} 
 = - L_\xi \sqrt{g}
=  \frac{\overline{\delta} g }{2\sqrt{g} } 
=  \frac{ \sqrt{g} }{2 } g^{ij} \overline{\delta} g_{ij}
\nonumber \\ 
& & 
=
 -  \sqrt{g} g^{ij} \nabla_i  \xi_j
=
- \sqrt{g} \nabla_i  \xi^i 
= 
 - \frac{\partial  (\sqrt{g} \xi^i)}{\partial x^i } 
.
\end{eqnarray}
We here note that 
an arbitrary scalar density ${\cal S}$ is transformed in the same manner 
as $\sqrt{g}$ 
under arbitrary spatial coordinate transformations, 
\begin{equation}
{\cal S}' (x')
= 
\left| \det \left( \frac{\partial x^i}{\partial x'^j} \right) \right| 
{\cal S} (x)
,
\end{equation}
and that ${\cal S} (x)/ \sqrt{g}(x)$ is regarded as a scalar field. 
Then, the variation $\overline{\delta} {\cal S} (x)$  
in the functional form of ${\cal S} (x)$ under the 
infinitesimal spatial coordinate transformation is given by 
\begin{eqnarray}
\label{dcalS}
& & 
\overline{\delta} {\cal S} (x)
=
- ( L_\xi {\cal S} ) (x)
= ( \overline{\delta} \sqrt{g})
\frac{\cal S}{ \sqrt{g}} +  \sqrt{g} \, \overline{\delta} 
\left( \frac{\cal S}{ \sqrt{g}} \right)
\nonumber \\ 
& & 
=
 - \frac{\partial  ( \sqrt{g} \xi^i )}{\partial x^i } \frac{\cal S}{ \sqrt{g}} 
- \sqrt{g} \xi^i\frac{\partial}{\partial x^i }
\left( \frac{\cal S}{ \sqrt{g}} \right)
=
 - \frac{\partial  ( \xi^i {\cal S} )}{\partial x^i } 
. 
\hspace*{8mm}
\end{eqnarray}
Since the distribution function $F(x, v, t)$ behaves as a scalar density 
under arbitrary spatial coordinate transformations, 
the variation $\overline{\delta} F$ in its spatial functional form 
due to the infinitesimal spatial coordinate transformation 
is written using Eq.~(\ref{dcalS}) as 
\begin{equation}
\label{dFDK3}
\overline{\delta} F
= 
- L_\xi F
= 
 - \frac{\partial}{\partial x^j}
( F \xi^j  )
. 
\end{equation}

Note that  $u_{ax}^i$ are the contravariant  components of 
the gyrocenter velocity vector field while $u_{av_\parallel}$, 
$u_{a\mu}$, and $u_{a\vartheta}$, which represent 
the temporal change rates of the  parallel velocity, magnetic moment, and 
gyrophase, respectively, behave as scalar fields under arbitrary 
spatial coordinate transformations. 
Then, we find from Eqs.~(\ref{deltaV}) and (\ref{dS}) that 
the variations in the functional forms of $u_{ax}^i$, $u_{av_\parallel}$, 
$u_{a\mu}$, and $u_{a\vartheta}$ under  
the infinitesimal spatial coordinate transformation
are given by 
\begin{eqnarray}
\label{dupAg}
 \overline{\delta} u_{ax}^i
& = & -  L_\xi  u_{ax}^i  
\equiv  
- \xi^j  \frac{\partial u_{ax}^i}{\partial x^j} 
+  u_{ax}^j \frac{\partial \xi^i}{\partial x^j}
,
\nonumber \\ 
\overline{\delta} u_{av_\parallel}
& = & -  L_\xi  u_{av_\parallel}  
\equiv  
- \xi^j  \frac{\partial u_{av_\parallel}}{\partial x^j} 
,
\nonumber \\ 
\overline{\delta} u_{a\mu}
& = & -  L_\xi  u_{a\mu}  
\equiv  
- \xi^j  \frac{\partial u_{a\mu}}{\partial x^j} 
,
\nonumber \\ 
\overline{\delta} u_{a\vartheta}
& = & -  L_\xi  u_{a\vartheta}  
\equiv  
- \xi^j  \frac{\partial u_{a\vartheta}}{\partial x^j} 
. 
\end{eqnarray}

It is also useful to know that even though 
the Christoffel symbols are not regarded as tensor components, 
their derivatives with respect to $\overline{\delta}$ 
can be defined by the variations in their functional forms 
under the infinitesimal spatial coordinate transformation 
and written as
\begin{equation}
\label{dGamma}
 \overline{\delta} \Gamma_{ij}^k
= 
\frac{g^{kn}}{2}
\left(
\nabla_i \overline{\delta} g_{nj} + 
 \nabla_j \overline{\delta}  g_{ni} -  
\nabla_n \overline{\delta}  g_{ij}
\right)
. 
\end{equation}
Equation~(\ref{dGamma}) can be derived 
from Eqs.~(\ref{Christoffel}), (\ref{nablaT}), (\ref{dg}), 
and the commutative property, 
\begin{equation}
\label{comm2}
\frac{\partial}{\partial x^i} \overline{\delta}
=
\overline{\delta} \frac{\partial}{\partial x^i} 
. 
\end{equation}

\section{VARIATIONAL DERIVATIVE WITH RESPECT TO METRIC TENSOR COMPONENTS}

We here consider the parts of the Lagrangian including the polarization effects  
denoted by $L_ {E 1a}$ and $L_ {E 2a}$, 
\begin{equation}
[ L_ {E 1a} , L_ {E 2a} ]
\equiv   
\int_V d^3 x \;
[ {\cal L}_ {E1a} , {\cal L}_ {E2a} ]
, 
\end{equation}
where the 
${\cal L}_ {E1a}$ and  ${\cal L}_ {E2a}$ 
defined in Eqs.~(\ref{LE1a}) and (\ref{LE2a}) 
appear due to finite gyroradius for species $a$ and they are given, respectively,  
in the linear and quadratic forms of the longitudinal electrostatic field and 
its spatial gradients. 
For the purpose of deriving the pressure tensor $P_{\Psi a}^{ij}$ 
due to electrostatic gyrokinetic
turbulence, we need to evaluate the variational derivatives of 
${\cal L}_ {E1a}$ and  ${\cal L}_ {E2a}$ with respect to 
the metric tensor components $g_{ij}$. 
It should be noted here that   
partial derivatives with respect to $g_{ij}$ 
need to be carefully performed because 
$3\times 3$ metric tensor components $g_{ij}$
are not completely independent of each other 
due to the constraint $g_{ij} = g_{ji}$. 
Here, for an arbitrary function $f$ of $g_{ij}$, 
the notation  $\partial f/ \partial g_{ij}$ 
is defined such that the infinitesimal variations $\delta g_{ij}$ 
in $g_{ij}$ give rise to the variation 
$\delta f = (\partial f/ \partial g_{ij}) \delta g_{ij}$ in $f$
where  both $\delta g_{ij}$ and  $\partial f/ \partial g_{ij}$ 
must be symmetric under exchange of the 
indices $i$ and $j$.~\cite{Landau} 
For example,  
we have  
$\partial g_{kl}/ \partial g_{ij} = \frac{1}{2}
( \delta^i_k \delta^j_l + \delta^j_k \delta^i_l )$
according to the above-mentioned definition. 
In the same manner, 
derivatives with respective to $\partial g_{ij}/\partial x^k$  
are defined taking into account 
the symmetry under exchange of the indices $i$ and $j$. 

The variation in $L_ {E1a}$ caused by 
the variation in the metric tensor with keeping
the gyrocenter distribution function $F_a$ fixed in  $L_ {E1a}$ 
is written as 
\begin{eqnarray}
\label{dLE1a}
& & 
(\delta_g L_ {E1a} )_F
\equiv
\left.
\left(
\frac{d}{d\epsilon}
L_ {E1a} [g_{ij} + \epsilon \delta g_{ij}]
\right|_{\epsilon = 0}
\right)_F
\nonumber \\ 
& & 
= \int_V d^3 x \;
( \delta_g {\cal L}_ {E1a} )_F
\nonumber \\ 
& & 
= 
\int_V d^3 x \; \sum_{k=1}^\infty 
\left[
 (\delta_g  Q_{0a}^{j_1 \cdots j_{2k}}  )_F
 \nabla_{j_1} \cdots  \nabla_{j_{2k-1}} 
 (E_L)_{j_{2k}}
\right. 
\nonumber \\ & & 
 \left.  \mbox{} 
\hspace*{10mm} + 
 Q_{0a}^{j_1 \cdots j_{2k}} 
\delta_g  
 ( \nabla_{j_1} \cdots  \nabla_{j_{2k-1}} 
 (E_L)_{j_{2k}} )
\right]
\nonumber \\ & & 
=
\int_V  d^3 x 
\left(\frac{\delta L_ {E1a} }{\delta g_{ij}}\right)_F
\delta g_{ij}
+ \mbox{B.T.}
,
\end{eqnarray}
where 
integration by parts is repeatedly performed 
to finally derive $(\delta L_ {E1a}/\delta g_{ij})_F$. 
Here, $(\cdots)_F$ implies that the gyrocenter function $F_a$ 
is fixed when taking the variation with respect to the metric tensor. 
From Eqs.~(\ref{nablaT}), we have 
\begin{eqnarray}
\label{dnablaE}
& & \delta_g (
\nabla_{j_1} \cdots  \nabla_{j_{n-1}} (E_L)_{j_n} )
\nonumber \\
&  & = 
\sum_{l=1}^{n-1}
\nabla_{j_1} \cdots  \nabla_{j_{l-1}} 
(\delta_g \nabla_{j_l})
 \nabla_{j_{l+1}} \cdots \nabla_{j_{n-1}} (E_L)_{j_n} 
,
\hspace*{5mm}
\end{eqnarray}
where  
\begin{eqnarray}
\label{dnablajk}
& &
(\delta_g \nabla_{j_l})
 \nabla_{j_{l+1}} \cdots \nabla_{j_{n-1}} (E_L)_{j_n} 
\nonumber \\ 
&  & \equiv 
- (n - l) ( \delta_g \Gamma^p_{j_l j_n} )
 \nabla_{j_{l+1}} \cdots \nabla_{j_{n-1}} (E_L)_p 
\nonumber \\ 
&  &  
=
- (n - l) \frac{g^{pq}}{2}
( \nabla_{j_l} \delta g_{j_n q}
+ \nabla_{j_n} \delta g_{j_l q}
-  \nabla_{q} \delta g_{j_l j_n} )
\nonumber \\ 
&  &  
\hspace*{5mm}
\mbox{} \times
 \nabla_{j_{l+1}} \cdots \nabla_{j_{n-1}} (E_L)_p 
.
\hspace*{5mm}
\end{eqnarray}
We note that, in Eq.~(\ref{dnablajk}),  
$\delta_g \Gamma^p_{j_l j_n}$ is expressed in the same form as in 
Eq.~(\ref{dGamma}). 
Now, the pressure tensor $P_ {E1a}^{ij}$ is given by
two times of $(\delta L_ {E1a}/\delta g_{ij})_F$ as  
\begin{eqnarray}
\label{PE1a}
& & 
P_ {E1a}^{ij}
\equiv
2 
\left(
\frac{\delta L_ {E1a} }{\delta g_{ij}}
\right)_F
\nonumber \\ & & 
=
\sum_{k=1}^\infty 
\left[ 2
\left(
\frac{\partial Q_{0a}^{j_1 \cdots j_{2k}} }{\partial  g_{ij}}
\right)_F
 \nabla_{j_1} \cdots  \nabla_{j_{2k-1}} 
 (E_L)_{j_{2k}}
\right. \nonumber \\ & & 
\hspace*{5mm}
\mbox{}
+ 
\sum_{l=1}^{2k-1}
(-1)^{l-1} 
(2k - l)
\left\{
(  \delta^i_{j_{2k}} g^{j p} + \delta^j_{j_{2k}} g^{i p} )
\nabla_{j_l} 
\right.
\nonumber \\ & & 
\hspace*{5mm} \left. \mbox{}
-    \delta^i_{j_l}  \delta^j_{j_{2k}} \nabla^p
\right\}
\left\{
(  \nabla_{j_1} \cdots \nabla_{j_{l-1}}
Q_{0a}^{j_1 \cdots j_{2k}}) 
\right.
\nonumber   \\ & & 
\hspace*{5mm} \left. \left. \mbox{}
\times (
 \nabla_{j_{l+1}} \cdots  \nabla_{j_{2k-1}} 
 (E_L)_p )
\right\}
\right]
,
\end{eqnarray}
where Eqs.~(\ref{dLE1a})--(\ref{dnablajk}) are used and 
$(\partial Q_{0a}^{j_1 \cdots j_{2k}}  / \partial  g_{ij})_F$ is 
given by using  Eq.~(\ref{Q0a}) and 
\begin{eqnarray}
\label{dadg}
& & 
\frac{\partial \alpha_a^{j_1 \cdots j_{2k}}}{\partial g_{mn}}
=
k \left[
\frac{h^{mn}}{2}
\alpha_a^{j_1 \cdots j_{2k}}
+ \frac{1}{(k!)^2}
\left( \frac{\rho}{2} \right)^{2k}
\right.
\nonumber \\ & & 
\hspace*{5mm}
\left. \mbox{}
\times 
\sum_{\sigma  \in \mathfrak{S}_{2k}}
\frac{\partial h^{j_{\sigma(1)} j_{\sigma(2)}}}{\partial g_{mn}}
h^{j_{\sigma(3)} j_{\sigma(4)}}
\cdots 
h^{j_{\sigma(2k-1)} j_{\sigma(2k)}}
\right]
.
\hspace*{10mm}
\end{eqnarray}

Next, the variation in $L_ {E2a}$ caused by 
the variation in the metric tensor with keeping 
the gyrocenter distribution function $F_a$ fixed is written as 
\begin{eqnarray}
\label{dLE2a}
& & 
(\delta_g L_ {E2a} )_F
= \int_V d^3 x \;
( \delta_g {\cal L}_ {E2a} )_F
\nonumber \\ 
& & 
= 
\int_V d^3 x \; 
 \sum_{m=1}^\infty  \sum_{n=1}^\infty  
\left[
\frac{1}{2}
(\delta_g 
\chi^{i_1 \cdots i_m ;  j_1 \cdots j_n} )_F
\nabla_{i_1} \cdots  \nabla_{i_{m-1}} (E_L)_{i_m}
\right. \nonumber \\ & & 
\hspace*{0mm}
\mbox{} 
\times
\nabla_{j_1} \cdots  \nabla_{j_{n-1}} (E_L)_{j_n}
+ Q_{Ea}^{j_1 \cdots j_n}
\delta_g (
\nabla_{j_1} \cdots  \nabla_{j_{n-1}} (E_L)_{j_n} )
\nonumber \\ & & 
=
\int_V  d^3 x 
\left(
\frac{\delta L_ {E2a} }{\delta g_{ij}}
\right)_F
\delta g_{ij}
+ \mbox{B.T.}
\end{eqnarray}
Then, using Eqs.~(\ref{dnablaE}), (\ref{dnablajk}), 
and (\ref{dLE2a}), we obtain 
the pressure tensor $P_ {E2a}^{ij}$ which is given by
two times of $(\delta L_ {E2a}/\delta g_{ij})_F$ as  
\begin{eqnarray}
\label{PE2a}
& & 
P_ {E2a}^{ij}
\equiv
2 
\left(
\frac{\delta L_ {E2a} }{\delta g_{ij}}
\right)_F
\nonumber \\ & & 
=
\sum_{m=1}^\infty \sum_{n=1}^\infty 
\left[
\left(
\frac{\partial \chi_a^{i_1 \cdots i_m ;  j_1 \cdots j_n} }{\partial  g_{ij}}
\right)_F
\nabla_{i_1} \cdots  \nabla_{i_{m-1}} (E_L)_{i_m}
\right. \nonumber \\ & & 
\hspace*{5mm}
\mbox{} 
\times
\nabla_{j_1} \cdots  \nabla_{j_{n-1}} (E_L)_{j_n}
\nonumber \\ & & 
\hspace*{5mm}
\mbox{}
+ 
\sum_{l=1}^{n-1}
(-1)^{l-1}
(n - l)
\left\{
( \delta^i_{j_n} g^{j p} + \delta^j_{j_n} g^{i p} )
\nabla_{j_l} 
\right.
\nonumber \\ & &  
\hspace*{5mm}
\left. \mbox{}
 -    \delta^i_{j_l}  \delta^j_{j_n} \nabla^k
\right\} 
\left\{
(  \nabla_{j_1} \cdots \nabla_{j_{l-1}}
Q_{Ea}^{j_1 \cdots j_n} ) 
\right.
\nonumber   \\ & & 
\hspace*{5mm}
\left. \left. \mbox{}
\times (
 \nabla_{j_{l+1}} \cdots  \nabla_{j_{n-1}} 
 (E_L)_p )
\right\}
\right]
, 
\end{eqnarray}
where
$(\partial \chi_a^{i_1 \cdots i_m ;  j_1 \cdots j_n} / \partial  g_{ij})_F$ 
is given by using Eqs.~(\ref{beta}), (\ref{chia}), (\ref{dadg}), and 
\begin{equation}
\frac{\partial}{\partial g_{ij}}
\left( \frac{1}{B} \right)
= 
-  \frac{1}{B^2} 
\frac{\partial B}{\partial g_{ij}}
=
\frac{h^{ij}}{2B}
.
\end{equation}

\section{WKB REPRESENTATION}

Here, we use the WKB (or ballooning) representation~\cite{WKB}
for turbulent fluctuations which 
have small wavelengths of the order of the gyroradius $\rho$ in the directions  
perpendicular to the background magnetic field. 
Such rapid spatial variations are represented using  
the perpendicular wavenumber vector ${\bf k}_\perp$. 
We assume that $k_\perp \rho = {\cal O}(1)$ and 
$\rho/L \ll 1$ where $L$ is the gradient scale length of equilibrium variables. 

The gyrocenter distribution function $F_a$ for species $a$ is 
given by the sum of the zeroth  and first-order parts in $\rho/L$  as
$
F_a (x, v) 
=
F_{a0} + F_{a1} + \hat{F}_{a1}
$
where the zeroth-order part $F_{a0}$ is the local Maxwellian equilibrium 
distribution function, and the first-order part representing 
the deviation from the local Maxwellian consists of 
the non-turbulent and turbulent functions denoted by 
$F_{a1}$ and $\hat{F}_{a1}$, respectively.  
Then, the turbulent gyrocenter distribution function $\hat{F}_{a1}$ is 
expanded as 
$
\hat{F}_{a1}
= 
\sum_{{\bf k}_\perp} \hat{F}_{a1{\bf k}_\perp}
\exp ( i {\bf k}_\perp \cdot {\bf x} )
$,
where ${\bf x}$ is the gyrocenter position vector 
and the ${\bf k}_\perp$-component $F_{a1{\bf k}_\perp}$ 
is given by the sum of 
the adiabatic and nonadiabatic parts as~\cite{Sugama2009} 
\begin{equation}
\hat{F}_{a1{\bf k}_\perp}
=
- F_{a0} \frac{e_a}{T_a}
J_0(k_\perp \rho_a) \hat{\phi}_{{\bf k}_\perp}
+ \hat{h}_{a{\bf k}_\perp}
.
\end{equation}
Here, $\hat{\phi}_{{\bf k}_\perp}$ and $\hat{h}_{a{\bf k}_\perp}$
are the ${\bf k}_\perp$-components of the turbulent 
electrostatic potential and the nonadiabatic part of the turbulent 
distribution function, respectively, 
$J_0$ is the zeroth-order Bessel function, and $T_a$ is 
the background temperature of the species $a$. 

We now follow the conventional assumption that 
$\langle \hat{\cal F}_{{\bf k}_\perp} \rangle_{\rm ens} = 0$
and 
$\langle \hat{\cal F}  \hat{\cal F}' \rangle_{\rm ens}
= \sum_{{\bf k}_\perp}
\langle \hat{\cal F}_{{\bf k}_\perp}^* \hat{\cal F}'_{{\bf k}_\perp} \rangle_{\rm ens}$ 
are satisfied by 
arbitrary real-valued turbulent fluctuations $\hat{\cal F}$ and $\hat{\cal F}'$ 
where $(\cdots)^*$ and $\langle  \cdots  \rangle_{\rm ens}$ represents 
the complex conjugate and the ensemble average, respectively. 
Note that the ensemble-averaged quantities are smooth spatial functions with 
the gradient scale length $L$ and 
that $\langle \hat{\cal F}_{{\bf k}_\perp}^* 
\hat{\cal F}'_{{\bf k}'_\perp} \rangle_{\rm ens} = 0$ for 
${\bf k}_\perp \neq {\bf k}'_\perp$. 
When taking the ensemble average of the momentum balance equation 
in Eq.~(\ref{totmomb}), 
we find that the effects of the turbulent fluctuations on the momentum 
transport are included in $\langle \mbox{\boldmath$\pi$}_\land \rangle_{\rm ens}$ 
and $\langle {\bf P}_\Psi \rangle_{\rm ens}$ through 
the correlation between the nonadiabatic part of 
the turbulent distribution 
function and the turbulent electrostatic potential which 
are described by 
$\langle \hat{h}_{a{\bf k}_\perp}^* \hat{\phi}_{{\bf k}_\perp}
\rangle_{\rm ens}$. 
Neglecting terms of higher orders in $\rho/L$, we find that 
the turbulent contribution 
$
\langle 
\mbox{\boldmath$\pi$}_{\land \Psi}
\rangle_{\rm ens}
$
to $\langle \mbox{\boldmath$\pi$}_\land \rangle_{\rm ens}$ 
is given by  
\begin{eqnarray}
\label{piens}
\langle 
\mbox{\boldmath$\pi$}_{\land \Psi}
\rangle_{\rm ens}
& = & 
\frac{c}{B}
\sum_{{\bf k}_\perp} 
 [ {\bf b} ( {\bf k}_\perp \times {\bf b}  ) 
 + ( {\bf k}_\perp \times {\bf b}  ) {\bf b} ]
\nonumber \\ & & 
\hspace*{-10mm}
\mbox{} \times
\sum_a 
\int d^3 v \, 
m_a v_\parallel
J_0(k_\perp \rho_a)
{\rm Im} [
\langle 
\hat{h}_{a{\bf k}_\perp}^* \hat{\phi}_{{\bf k}_\perp}
\rangle_{\rm ens}
]
,
\hspace*{10mm}
\end{eqnarray}
and 
$\langle {\bf P}_\Psi \rangle_{\rm ens}$ 
is written as 
\begin{eqnarray}
\label{PPsiens}
\langle 
{\bf P}_\Psi
\rangle_{\rm ens}
& = & 
\sum_{{\bf k}_\perp} 
\left[
\frac{1}{2} 
({\bf I} - {\bf b} {\bf b})
-
\frac{{\bf k}_\perp {\bf k}_\perp }{k_\perp^2}
\right]
\sum_a e_a
\nonumber \\ & & 
\hspace*{-5mm}
\mbox{} \times
\int d^3 v \; 
k_\perp \rho_a
\, 
J_1( k_\perp \rho_a )
{\rm Re} [
\langle 
\hat{h}_{a{\bf k}_\perp}^* \hat{\phi}_{{\bf k}_\perp}
\rangle_{\rm ens}
]
,
\hspace*{10mm}
\end{eqnarray}
where ${\bf I}$ denotes the unit tensor. 

We now consider the axisymmetric toroidal system,  
in which the magnetic field is given by 
\begin{equation}
{\bf B}
= I \nabla \zeta + \nabla \zeta \times \nabla \chi
, 
\end{equation}
where $\zeta$, $I$, and $\chi$ represent the toroidal angle, 
the covariant toroidal magnetic field component, and the 
poloidal flux function, respectively. 
Then, using  
$
({\bf k}_\perp \times {\bf b} ) \cdot \nabla \chi
= - B R^2 \nabla \zeta \cdot {\bf k}_\perp 
$, 
the radial transport of the toroidal angular momentum 
due to the turbulent electric field are obtained from 
Eqs.~(\ref{piens}) and (\ref{PPsiens}) as 
\begin{eqnarray}
\label{pi+PPsi}
& & 
\nabla \chi \cdot 
\langle 
\mbox{\boldmath$\pi$}_{\land \Psi}
+
{\bf P}_\Psi
\rangle_{\rm ens} 
\cdot R^2 \nabla \zeta
\nonumber \\  
& = & 
- \sum_a 
\sum_{{\bf k}_\perp}
\int d^3 v \,
\left[
\frac{c I}{B} m_a v_\parallel
J_0(k_\perp \rho_a) {\rm Im} [
\langle 
\hat{h}_{a{\bf k}_\perp}^* \hat{\phi}_{{\bf k}_\perp}
\rangle_{\rm ens}
]
\right. 
\nonumber \\ & & 
\mbox{} \times
( {\bf k}_\perp \cdot R^2 \nabla \zeta ) 
+
e_a
k_\perp \rho_a
\, 
J_1( k_\perp \rho_a )
{\rm Re} [
\langle 
\hat{h}_{a{\bf k}_\perp}^* \hat{\phi}_{{\bf k}_\perp}
\rangle_{\rm ens} 
]
\nonumber \\ & & 
\mbox{} \times
\left. 
\frac{
( {\bf k}_\perp \cdot \nabla \chi )
( {\bf k}_\perp \cdot R^2 \nabla \zeta )
}{k_\perp^2}
\right]
.
\end{eqnarray}
The flux-surface average of 
Eq.~(\ref{pi+PPsi}) agrees with 
the electrostatic and low-flow ordering limit of the 
result given in Eq.~(53) of Ref.~\cite{Sugama1998} where the 
turbulent radial transport of the toroidal angular momentum 
double-averaged over the ensemble and the flux surface
is presented for the general case allowing 
the turbulent magnetic field and 
the high-flow ordering~\cite{Abel,Sugama2017,Belli2018}. 
It should be emphasized here that Eq.~(\ref{pi+PPsi}) is 
not surface-averaged 
but it presents the spatially-local expression. 
In the axisymmetric configuration with 
up-down symmetry, the flux-surface average of Eq.~(\ref{pi+PPsi}) is 
shown to vanish in the case of the low-flow ordering~\cite{Sugama2011,Parra2011} 
although it does not imply that the local value of Eq.~(\ref{pi+PPsi}) 
itself vanishes as well.

\section{ANOTHER DERIVATION OF THE MOMENTUM BALANCE}

In Sec.~IV, the momentum balance is derived using the 
invariance of the Lagrangian of the system under the infinitesimal 
spatial coordinate transformation induced by the vector field 
$\xi^i (x)$ which has an arbitrary functional form. 
For comparison with that derivation, 
another derivation of the momentum balance is given 
in this appendix in the way closer to the conventional derivation  
of the canonical  momentum conservation law 
which generally involves the asymmetric canonical 
pressure tensor. 

Equation~(\ref{dLGKa3}) for the invariance of 
the  gyrokinetic Lagrangian $L_{GKa}$ of species $a$ can be 
rewritten without separating boundary terms from the spatial integral 
as 
\begin{equation}
\label{dLGKa6}
\overline{\delta} L_{GKa} 
= \int_V d^3 x \, J_{GKa} 
= 0
,
\end{equation}
where
\begin{eqnarray}
\label{JGKa}
J_{GKa}
& = & 
 \int_V d^3 x \int d^3 v \, 
 F_a \left( \xi^i \frac{\partial L_{GYa}}{\partial x^i}  + 
\frac{\partial L_{GYa}}{\partial u_{ax}^i} \overline{\delta} u_{ax}^i
\right.  
\nonumber \\ & & 
\left. \mbox{} 
+ \frac{\partial L_{GYa}}{\partial u_{a\vartheta}^i} 
\overline{\delta} u_{a\vartheta}^i
+ (\overline{\delta} L_{GYa})_u
\right)
\nonumber \\ 
& = & 
\xi^i  
\int d^3 v \left[
\frac{\partial}{\partial t}
\left( 
 F_a 
p_{ai}
\right)
-
{\cal K}_a
p_{ai}
\right.  
\nonumber \\ & & \mbox{} 
\left. 
+ \frac{\partial}{\partial x^j} 
\left( F_a u_{ax}^j 
p_{ai}
\right)
-  F_a  
\left( 
\frac{\partial L_{GYa}}{\partial x^i} 
\right)_u
\right]
.
\end{eqnarray}
Here, 
$p_{ai} \equiv \partial L_{GYa}/\partial u_{ax}^i$ 
is the canonical momentum [see Eq.~(\ref{pai})] and 
\begin{eqnarray}
\label{dLGYau}
(\overline{\delta} L_{GYa})_u
& \equiv & 
( \overline{\delta} p_{aj}  ) u_{ax}^j 
- \overline{\delta} H_{GYa}
\nonumber \\
& = & 
- \left( \xi^i \frac{\partial p_{aj}}{\partial x^i}
+ \frac{\partial \xi^i}{\partial x^j} p_{ai}
\right) u_{ax}^j 
+\xi^i \frac{\partial H_{GYa}}{\partial x^i}
\nonumber \\
& = & 
- \xi^i 
\left( 
\frac{\partial L_{GYa}}{\partial x^i} 
\right)_u
- \frac{\partial \xi^i}{\partial x^j} p_{ai} u_{ax}^j 
\end{eqnarray}
represents 
the variation in the functional form of 
the single gyrocenter Lagrangian $L_{GYa}$ 
under the infinitesimal spatial coordinate transformation 
with $(u_x^i, u_\vartheta)$ kept fixed in $L_{GYa}$. 
   The term $(\partial L_{GYa}/\partial x^i)_u$ in Eq.~(\ref{dLGYau}) 
is written down as 
\begin{equation}
\left( 
\frac{\partial L_{GYa}}{\partial x^i} 
\right)_u
\equiv
u_{ax}^j
\frac{\partial p_{aj}}{\partial x^i}
- \mu \frac{\partial B}{\partial x^i}
- e_a \frac{\partial \Psi_a}{\partial x^i}
.
\end{equation}
Equations~(\ref{dpidt}), (\ref{GKB}), (\ref{dupAg}), and (\ref{dLGYau}) 
are used for deriving Eq.~(\ref{JGKa}). 
Since Eq.~(\ref{dLGKa6}) holds for an arbitrary spatial integral domain $V$, 
we find $J_{GKa} = 0$ for any $\xi^i$. 
Then, the canonical momentum balance equation 
for the gyrocenters of species $a$ is derived from Eq.~(\ref{JGKa}) as 
\begin{eqnarray}
\label{canmomba}
& & 
\frac{\partial}{\partial t}
\left(
\int d^3 v \, F_a 
p_{ai}
\right)
+
 \frac{\partial}{\partial x^j} 
\left(
\int d^3 v \, F_a 
m_a u_{ax}^j 
p_{ai}
\right)
\nonumber \\ 
& & 
=
\int d^3 v \, {\cal K}_a 
p_{ai}
+ 
\int d^3 v \, 
 F_a  
\left( 
\frac{\partial L_{GYa}}{\partial x^i} 
\right)_u
. 
\end{eqnarray}
Equation~(\ref{canmomba}) also can be derived directly from 
multiplying Eq.~(\ref{GKB}) by 
$p_{ai}$, taking its $v$-space integral, and using Eq.~(\ref{dpidt}). 
It should be recalled that the general spatial coordinates $(x^i)_{i=1,2,3}$ 
are used here. 
For example, 
when $(x^i)_{i=1,2,3}$ represent the Cartesian coordinates $(x, y, z)$, 
Eq.~(\ref{canmomba}) represents the linear canonical momentum balance. 
As another interesting example, 
we can treat the canonical angular momentum balance in toroidal plasmas 
such as tokamaks and stellarator/heliotron devices, 
for which it is convenient to use the cylindrical coordinates and/or 
the magnetic flux coordinates. 
In these coordinate systems, the toroidal angle component of 
Eq.~(\ref{canmomba}) 
represents the toroidal angular momentum balance. 
It is also noted that the momentum balance equations shown in 
Eqs.~(\ref{mombal}) and (\ref{canmomba}) are equivalent to each other 
although the transformation from Eq.~(\ref{canmomba}) to Eq.~(\ref{mombal}) 
requires complicated procedures involving infinite number of 
times of partial integration. 

We next consider 
the invariance of the Lagrangian $L_{GKF}$ of the whole system 
under the infinitesimal spatial coordinate transformation 
shown in Eq.~(\ref{dLGKF2}).  
In contrast to Eq.~(\ref{dLGKF3}), 
we now rewrite Eq.~(\ref{dLGKF2})   
without separating boundary terms from the spatial integral as 
\begin{equation}
\label{dLGKF5}
\overline{\delta} L_{GKF} 
= \int_V d^3 x \, J_{GKF} 
= 0
,
\end{equation}
with
\begin{eqnarray}
\label{JGKF}
J_{GKF}
& = & 
\sum_a 
\int d^3 v \, 
\left[ \xi^i \left\{
\frac{\partial}{\partial t}
\left(  
 F_a 
p_{ai}
\right)
-
{\cal K}_a
p_{ai}
\right\}
\right. 
\nonumber \\ & & \mbox{} 
\left. 
+ \frac{\partial}{\partial x^j} 
\left( \xi^i F_a u_{ax}^j 
p_{ai}
\right)
\right] 
+ J_{GKFA} 
\nonumber \\ & & \mbox{} 
+ J_{GKFg} + J_{GKF\phi}
\nonumber \\ &  &  \mbox{} 
= 0
, 
\end{eqnarray}
where Eqs.~(\ref{dpidt}), (\ref{GKP}), (\ref{GKB}), and (\ref{dupAg}) 
are used.   
Here, $J_{GKFA}$ and $J_{GKFg}$ 
originate from the terms including $\overline{\delta} A_j$ 
and $\overline{\delta} g_{ij}$ 
in Eq.~(\ref{dLGKF2}), respectively, and 
they are written as   
\begin{eqnarray}
\label{JGKFA}
& & J_{GKFA} 
 \equiv 
\frac{1}{c} J^j \overline{\delta} A_j
+ \frac{\partial}{\partial x^j} 
\left( 
\epsilon^{jkl} 
M_l \overline{\delta} A_k 
\right)
\nonumber \\
&  & \hspace*{3mm} = 
-\frac{\sqrt{g}}{c} \epsilon_{ijk} 
\xi^i J^j B^k
- \frac{1}{c} J^j \frac{\partial (\xi^i A_i )}{\partial x^j} 
\nonumber \\
&  &  \mbox{} \hspace*{6mm}
- \epsilon^{jkl} \frac{\partial }{\partial x^j}  
\left[ M_l 
\left( \xi^i
\frac{\partial A_k}{\partial x^i}
+ \frac{\partial \xi^i }{\partial x^k} A_i
\right)
\right]
\nonumber \\
& & \hspace*{3mm} = 
-\frac{\sqrt{g}}{c} \epsilon_{ijk} 
\xi^i J^j B^k
- 
\left( \frac{1}{c} J^j 
- \epsilon^{jkl} \frac{\partial M^l}{\partial x^k} 
 \right)
\frac{\partial (\xi^i A_i )}{\partial x^j} 
\nonumber \\
&  &  \mbox{} \hspace*{6mm}
- \epsilon^{jkl} \epsilon_{ikm}
\frac{\partial }{\partial x^j}
\left(
\sqrt{g} \xi^i M_l B^m 
\right)
\nonumber \\
& & \hspace*{3mm} 
= 
-\frac{\sqrt{g}}{c} \epsilon_{ijk} 
\xi^i J^j B^k
- \left( 
\sum_a 
\frac{e_a}{c}
\int d^3 v \, F_a u_{ax}^j
\right) 
\frac{\partial (\xi^i A_i )}{\partial x^j} 
\nonumber \\
&  &  \mbox{}
\hspace*{6mm}
- 
\frac{\partial }{\partial x^i}
\left(
\sqrt{g} \xi^i M_j B^j 
\right)
+
\frac{\partial }{\partial x^j}
\left(
\sqrt{g} \xi^i M_i B^j 
\right)
,
\end{eqnarray}
and 
\begin{equation}
\label{JGKFg}
J_{GKF g} 
\equiv
\sum_a \int d^3 v \,
 F_a
\sum_J  \frac{\partial L_{GYa}}{\partial (\partial_J g_{ij})} 
\overline{\delta} (\partial_J g_{ij})
+
\frac{\partial {\cal L}_F}{\partial g_{ij}} 
\overline{\delta} g_{ij}
.
\end{equation}
The last term 
$J_{GKF\phi}$ included in Eq.~(\ref{JGKF}) is 
given in terms of the turbulent electrostatic field as 
\begin{eqnarray}
\label{JGKFphi}
& & 
J_{GKF\phi} 
 =  
\frac{\partial ( \xi^i {\cal L}_F)}{\partial x^i}
+
\sum_{n=1}^\infty
\sum_{k=1}^n
(-1)^k
\frac{\partial}{\partial x^{j_k}}
\left[
\frac{\partial^{n-k} (\xi^i \partial \phi/\partial x^i) }{\partial x^{j_{k+1}} 
\cdots \partial x^{j_n} }
\right.
\nonumber \\ & & \mbox{}
\left.
\hspace*{12mm}
\times
\frac{\partial^{k-1}}{\partial x^{j_1} \cdots \partial x^{j_{k-1}}}
\left(
\frac{\partial {\cal L}_{GKF}}{\partial 
(\partial^n \phi/\partial x^{j_1} \cdots \partial x^{j_n})}
\right)
\right]
.
\end{eqnarray}
Here, because of the symmetry with respect to permutations 
of the indices $j_1, \cdots, j_n$, 
the variables 
$\partial^n \phi/\partial x^{j_1} \cdots \partial x^{j_n}$ 
$[ \equiv - \partial^n (E_L)_{j_n}/\partial x^{j_1} \cdots \partial x^{j_{n-1}} ]$ 
are not regarded as completely independent variables and the meaning of the expression 
$\partial {\cal L}_{GKF} / \partial (\partial^n \phi/\partial x^{j_1} \cdots \partial x^{j_n})$ in Eq.~(\ref{JGKFphi}) needs to be clearly mentioned. 
It is defined such that the infinitesimal variations 
$
\delta (\partial^n \phi/\partial x^{j_1} \cdots \partial x^{j_n})
$ 
give rise to the variation 
$
\delta {\cal L}_{GKF} 
= 
[\partial {\cal L}_{GKF} / \partial (\partial^n \phi/\partial x^{j_1} \cdots \partial x^{j_n}) ] 
\delta (\partial^n \phi/\partial x^{j_1} \cdots \partial x^{j_n})
$ 
where
$
\partial {\cal L}_{GKF} / \partial (\partial^n \phi/\partial x^{j_1} \cdots \partial x^{j_n})
$
must be symmetric with respect to arbitrary permutations of 
the indices $j_1, \cdots, j_n$. 

We now suppose that $\xi^i$ represents a Killing vector field or a vector field 
which generates an isometric transformation so that 
$\overline{\delta} g_{ij} = - L_\xi g_{ij} = 0$,  
$\overline{\delta} ( \partial_J g_{ij}) 
= \partial_J  (\overline{\delta}g_{ij}) = 0$ 
and accordingly $J_{GKF g} = 0$. 
The isometry is generated by
the infinitesimal linear translation and the infinitesimal rotation, 
and they are represented by the Killing vector fields, 
$
\mbox{\boldmath$\xi$} = \epsilon {\bf n}
$
and 
$
\mbox{\boldmath$\xi$} = \epsilon {\bf n} \times {\bf r}
$,
respectively, where $\epsilon$ is an infinitesimal constant, 
${\bf n}$ a constant unit vector, and ${\bf r}$ the spatial position vector. 
Here and hereafter, the Cartesian coordinate system is used 
for $(x^i)_{i=1,2,3}$ and three-dimensional vectors are represented 
in terms of boldface  letters. 
For example, the canonical momentum of the gyrocenter of species $a$ 
is denoted by 
$
{\bf p}_{a}
\equiv
\partial L_{GYa} / \partial {\bf u}_{ax} 
=
m_a v_\parallel {\bf b} + 
(e_a/c) {\bf A}
$.

Recall again that $J_{GKF}$ defined in Eq.~(\ref{JGKF}) vanishes because 
Eq.~(\ref{dLGKF5}) holds for any $V$. 
Then, substituting $\mbox{\boldmath$\xi$} = \epsilon {\bf n}$ 
into Eq.~(\ref{JGKF}) and noting that 
 $J_{GKF} = 0$ is satisfied for any direction vector ${\bf n}$, 
we obtain the linear canonical 
momentum balance equation for the whole system, 
\begin{eqnarray}
\label{totcanmomb}
& & 
\frac{\partial}{\partial t}
\left(  
\sum_a 
\int d^3 v \, 
 F_a 
{\bf p}_{a}
\right)
+ \nabla \cdot
\left( 
\sum_a 
\int d^3 v \, 
F_a 
 {\bf u}_{ax}
{\bf p}_{a}
\right)
\nonumber \\ &  &  \mbox{} 
\hspace*{3mm}
+  \nabla \cdot \left[ - ( {\bf B} \cdot  {\bf M} ) {\bf I} +   {\bf B} {\bf M} 
+ 
\frac{|{\bf E}_L|^2}{8\pi} {\bf I} - \frac{{\bf E}_L {\bf E}_L}{4\pi} 
+ \mbox{\boldmath$\Pi$}_\Psi 
\right]
\nonumber \\ 
&  & 
=
\left( 
\sum_a 
\frac{e_a}{c}
\int d^3 v \, F_a {\bf u}_{ax}
\right) 
\cdot 
\nabla 
{\bf A}
+ 
\frac{1}{c}
{\bf J} \times {\bf B}
\nonumber \\ &  &  \mbox{} 
\hspace*{3mm}
+ \sum_a 
\int d^3 v \, 
{\cal K}_a
{\bf p}_{a}
,
\end{eqnarray}
where ${\bf M}$ is the magnetization vector defined 
in Eq.~(\ref{magnetization}) 
and $\mbox{\boldmath$\Pi$}_\Psi$ is defined using Eq.~(\ref{LPsi}) as 
\begin{eqnarray}
& & (\mbox{\boldmath$\Pi$}_\Psi)^{ij} 
= 
\sum_{n=1}^\infty
\sum_{k=1}^n
(-1)^k
\left[
\frac{\partial^{n-k} (E_L)^j }{\partial x^{j_k} 
\cdots \partial x^{j_{n-1}} }
\right.
\nonumber \\ & & \mbox{}
\left.
\times
\frac{\partial^{k-1}}{\partial x^{j_1} \cdots \partial x^{j_{k-1}}}
\left(
\frac{\partial {\cal L}_\Psi}{\partial 
(\partial^{n-1} (E_L)_i /\partial x^{j_1} \cdots \partial x^{j_{n-1}})}
\right)
\right]
.
\hspace*{10mm}
\end{eqnarray}
Like ${\bf P}_\Psi$ given in Eq.(\ref{PPsi}), 
$\mbox{\boldmath$\Pi$}_\Psi$ contains the effect of 
the turbulent electrostatic field on the momentum transport 
although ${\bf P}_\Psi$ and $\mbox{\boldmath$\Pi$}_\Psi$ 
are symmetric and asymmetric tensors, respectively. 
Using Eqs.~(\ref{eK}) and (\ref{charge_conservation}), 
Eq.~(\ref{totcanmomb}) also can be rewritten as 
\begin{eqnarray}
\label{totcanmomb2}
& & 
\frac{\partial}{\partial t}
\left(  
\sum_a 
\int d^3 v \, 
 F_a 
m_a v_\parallel {\bf b}
\right)
\nonumber \\ &  &  \mbox{} \hspace*{3mm}
+ \nabla \cdot
\left( 
\sum_a 
\int d^3 v \, 
F_a m_a v_\parallel
 {\bf u}_{ax}
 {\bf b}
\right)
\nonumber \\ &  &  \mbox{} \hspace*{3mm}
+  \nabla \cdot \left[ - ( {\bf B} \cdot  {\bf M} ) {\bf I} +   {\bf B} {\bf M} 
+ 
\frac{|{\bf E}_L|^2}{8\pi} {\bf I} - \frac{{\bf E}_L {\bf E}_L}{4\pi} 
+ \mbox{\boldmath$\Pi$}_\Psi 
\right]
\nonumber \\ 
& & =
- \left( 
\sum_a 
e_a
\int d^3 v \, F_a 
\right) 
\frac{1}{c}\frac{\partial {\bf A}}{\partial t}
+ 
\frac{1}{c}
{\bf J} \times {\bf B}
\nonumber \\ &  &  \mbox{} \hspace*{3mm}
+ \sum_a 
\int d^3 v \, 
{\cal K}_a
m_a v_\parallel {\bf b}
.
\end{eqnarray}
The momentum balance equation shown in 
Eq.~(\ref{totcanmomb2}) can be transformed into 
Eq.~(112) although again 
it is so complicated involving infinite number of times of partial integration. 

Substituting 
$\mbox{\boldmath$\xi$} = \epsilon {\bf n} \times {\bf r}$
 into Eq.~(\ref{JGKF}) and  using 
 $J_{GKF} = 0$ for any direction vector ${\bf n}$, 
the angular 
momentum balance equation for the whole system is 
derived as 
\begin{eqnarray}
& & 
\frac{\partial}{\partial t}
\left(  {\bf r} \times 
\sum_a 
\int d^3 v \, 
 F_a   m_a v_\parallel
{\bf b} 
\right)
\nonumber \\ &  &  \mbox{} \hspace*{3mm}
+ \nabla \cdot
\left(  
\sum_a 
\int d^3 v \, 
F_a m_a v_\parallel
 {\bf u}_{ax}
{\bf r} \times 
{\bf b} 
\right)
\nonumber \\ &  &  \mbox{} \hspace*{3mm}
+ 
 \nabla \cdot 
\left[  \left\{ ({\bf B} \cdot  {\bf M} ) {\bf I} - {\bf B} {\bf M} 
- 
\frac{|{\bf E}_L|^2}{8\pi} {\bf I} + \frac{{\bf E}_L {\bf E}_L}{4\pi} 
\right\} \times {\bf r}  +  {\bf T}_\Psi \right]
\nonumber \\ 
&  & = 
{\bf r} \times \left[ 
- \left( 
\sum_a 
e_a
\int d^3 v \, F_a 
\right) 
\frac{1}{c}\frac{\partial {\bf A}}{\partial t}
+ 
\frac{1}{c}
{\bf J} \times {\bf B}
\right. 
\nonumber \\ &  &  \mbox{} \hspace*{3mm}
\left. 
+ \sum_a 
\int d^3 v \, 
{\cal K}_a m_a v_\parallel
{\bf b} 
\right]
,
\end{eqnarray}
where ${\bf T}_\Psi$ is the asymmetric tensor  
representing the transport of the angular momentum due to 
the electrostatic field and defined by
\begin{eqnarray}
& & 
({\bf T}_\Psi)^{ij} 
=
\sum_{n=1}^\infty
\sum_{k=1}^n
(-1)^k
\left[
\frac{\partial^{n-k} ({\bf r} \times {\bf E}_L)^j }{\partial x^{j_k} 
\cdots \partial x^{j_{n-1}} }
\right.
\nonumber \\ & & \mbox{} \hspace*{3mm}
\left.
\times
\frac{\partial^{k-1}}{\partial x^{j_1} \cdots \partial x^{j_{k-1}}}
\left(
\frac{\partial {\cal L}_\Psi}{\partial 
(\partial^{n-1} (E_L)_i /\partial x^{j_1} \cdots \partial x^{j_{n-1}})}
\right)
\right]
.
\nonumber \\ & & 
\end{eqnarray}

\section{ENERGY BALANCE IN THE GYROKINETIC SYSTEM}

In this Appendix, the energy balance in the electrostatic gyrokinetic turbulent system 
is derived. 
In contrast to the case of in Sec.~IV where the momentum balance is derived, 
we here use only the Cartesian coordinate system and represent 
three-dimensional vectors in terms of boldface letters. 
Then, the metric tensor components are represented by the Kronecker delta, 
and they form the $3 \times 3$ unit matrix with determinant unity.

The partial time derivative of the gyrokinetic 
Lagrangian density ${\cal L}_{GKa}$ for particle species $a$ 
is written as 
\begin{eqnarray}
\label{dLGKadt}
 \frac{\partial {\cal L}_{GKa} }{\partial t}
& = & 
\frac{\partial}{\partial t}
\left( 
\int d^3 v \, F_a L_{GYa}
\right)
\nonumber \\ 
& = & 
\int d^3 v \left[
\frac{\partial F_a}{\partial t}  L_{GYa} + 
 F_a \left\{
\frac{\partial L_{GYa} }{\partial {\bf u}_{ax}} 
\cdot  \frac{\partial {\bf u}_{ax}}{\partial t}
\right.  \right.
\nonumber \\ & & \mbox{}
\left.  \left. 
+ \frac{\partial L_{GYa}}{\partial  u_{a\vartheta}}  
\frac{\partial u_{a\vartheta}}{\partial t}
+ \left( \frac{\partial L_{GYa}}{\partial t} \right)_u
\right\}
\right]
, 
\end{eqnarray}
where 
$( \partial L_{GYa}/\partial t )_u$ denotes  
the time derivative of $L_{GYa}$ with $(u_{ax}^i, u_{a\vartheta})$ 
kept fixed in $L_{GYa}$. 
Here, we consider the gyrocenter distribution function $F_a$ satisfying 
Eq.~(\ref{GKB}) which includes the term ${\cal K}_a$ representing the rate of temporal 
change in $F_a$ due to collisions and/or external sources for the species $a$.  
Substituting Eq.~(\ref{GKB}) into 
Eq.~(\ref{dLGKadt}) and using 
$\delta I_{GK}/\delta {\bf x}_{a E} =0$ [Eq.~(\ref{dpidt})], 
$\delta I_{GK}/\delta v_{a \parallel E} =0$ [Eq.~(\ref{dLdv})], 
$\delta I_{GK}/\delta \mu_{a E} =0$ [Eq.~(\ref{dIdmu})], 
and $\delta I_{GK}/\delta \vartheta_{a E} =0$ [Eq.~(\ref{dIdth})], 
we obtain the energy balance equation, 
\begin{eqnarray}
\label{energy_balance}
& & 
\frac{\partial }{\partial t} 
\left(
\int d^3 v \, F_a {\cal E}_a 
\right)
+ \nabla
\cdot 
\left(
\int d^3 v \, 
F_a {\cal E}_a {\bf u}_{ax}
 \right)
\nonumber \\
& & 
=
\int d^3 v \, \left(  F_a   \dot{\cal E}_a 
+ {\cal K}_a {\cal E}_a 
\right) 
,
\end{eqnarray}
where  
the gyrocenter velocity ${\bf u}_{ax}$ is given by  
Eq.~(\ref{uxieq}) 
and ${\cal E}_a$ represents 
the energy 
of the single particle 
(or the gyrocenter Hamiltonian $H_{GYa}$) 
defined by  
\begin{eqnarray}
{\cal E}_a 
& \equiv & 
H_{GYa} 
\equiv
\frac{\partial L_{GYa} }{\partial {\bf u}_{ax}} 
\cdot {\bf u}_{ax}
+ \frac{\partial L_{GYa}}{\partial  u_{a\vartheta}}  
 u_{a\vartheta}
-  L_{GYa}
\nonumber \\
& = & 
\frac{1}{2} m_a v_\parallel^2 
+ \mu B + e_a \Psi_a
. 
\end{eqnarray}
The rate of change in the particle's energy is 
given by 
\begin{equation}
\dot{\cal E}_a 
\equiv 
- \left( \frac{\partial L_{GYa}}{\partial t} \right)_u
= 
 e \frac{\partial \Psi_a}{\partial t} 
+ \mu \frac{\partial B}{\partial t} 
- \frac{e_a}{c} 
{\bf u}_{ax}  \cdot  \frac{\partial {\bf A}_a^* }{\partial t} 
.
\end{equation}
The energy balance equation shown in Eq.~(\ref{energy_balance}) 
agrees with Eq.~(C7) in Ref.~\cite{Sugama2018} 
except for the effects of the turbulent electrostatic potential included 
here. 
It can be seen in Eq.~(\ref{energy_balance}) 
how the energy balance is modified 
when the collision (or source) term ${\cal K}_a$ 
is added into the gyrokinetic equation. 

We now consider the energy balance in the extended system consisting of 
particles of all species and the self-consistent electrostatic field. 
The partial time derivative of the Lagrangian density ${\cal L}_{GKF}$ 
of this extended system is written as 
\begin{eqnarray}
\label{dLGKFdt}
& & 
\frac{\partial {\cal L}_{GKF} }{\partial t}
=
\frac{\partial}{\partial t}
\left( {\cal L}_{GK}  + {\cal L}_F \right)
=
\frac{\partial}{\partial t}
\left( \sum_a {\cal L}_{GKa}  + {\cal L}_F \right)
\nonumber \\ 
& & =
\frac{\partial}{\partial t}
\left( \sum_a  \int d^3 v \, F_a L_{GYa} + {\cal L}_F \right)
\nonumber \\ 
&  & =
\sum_a \int d^3 v \left\{
\frac{\partial F_a}{\partial t}  L_{GYa} + 
 F_a \left(
\frac{\partial L_{GYa} }{\partial {\bf u}_{ax}} 
\cdot  \frac{\partial {\bf u}_{ax}}{\partial t}
\right.  \right.
\nonumber \\ & & \hspace*{3mm}
\mbox{}
\left.  \left. 
+ \frac{\partial L_{GYa}}{\partial  u_{a\vartheta}}  
\frac{\partial u_{a\vartheta}}{\partial t}
\right) 
\right\}
+ \sum_J \frac{\partial {\cal L}_{GKF} }{\partial (\partial_J \phi)} 
\frac{\partial (\partial_J \phi)}{\partial t} 
\nonumber \\ 
& & \hspace*{3mm}
\left. 
\mbox{}
+ \frac{\partial {\cal L}_{GKF} }{\partial {\bf A} } 
\frac{\partial {\bf A}}{\partial t}
+  \frac{\partial {\cal L}_{GKF} }{\partial (\partial {\bf A}/\partial x^i) } 
\cdot \frac{\partial (\partial {\bf A}/\partial x^i) }{\partial t}
\right]
, 
\end{eqnarray}
In the same way as in deriving Eq.~(\ref{energy_balance}), 
we use the Euler-Lagrange equations 
[Eqs.~(\ref{dpidt}), (\ref{dLdv}),  (\ref{dIdmu}), and (\ref{dIdth})]   
for $({\bf u}_{ax}, u_{av_\parallel}, u_{a\mu}, u_{a\vartheta})$,  
Eq.~(\ref{GKB}) for $F_a$,  
and Eq.~(\ref{GKP}) for $\phi$ 
in order to rewrite 
Eq.~(\ref{dLGKFdt}) as 
\begin{eqnarray}
\label{totenergyb0}
& & 
\frac{\partial}{\partial t} 
\left( 
\sum_a \int d^3 v \, F_a H_{GYa} - {\cal L}_F
\right) 
\nonumber \\ & & \mbox{} \hspace*{3mm}
+ \nabla 
\cdot 
\left(
\sum_a \int d^3 v \, F_a H_{GYa} {\bf u}_{ax} 
\right)
+ 
\sum_{n=1}^\infty
\sum_{k=1}^n (-1)^{k-1}
\nonumber \\ & & \mbox{} \hspace*{6mm}
\times
\frac{\partial}{\partial x^{j_k}}
\left\{
\frac{\partial^{k-1}}{\partial x^{j_1} \cdots \partial x^{j_{k-1}}}
\left(
\frac{\partial {\cal L}_{GKF} }{\partial 
( \partial^n \phi /\partial x^{j_1} \cdots \partial x^{j_n})}
\right)
\right. 
\nonumber \\ & & \mbox{} \hspace*{6mm}
\left. 
\times 
\frac{\partial}{\partial t}
\left(
\frac{\partial^{n-k} \phi }{\partial x^{j_{k+1}} \cdots \partial x^{j_n}}
\right)
\right\}
\nonumber \\ & & 
= 
{\bf J} \cdot {\bf E}_T 
+ \nabla  \cdot ( c {\bf E}_T  \times {\bf M} )
+
\sum_a \int d^3 v \,  {\cal K}_a 
\nonumber \\ & & \mbox{} \hspace*{6mm} 
\times 
\left\{ \frac{1}{2} m_a v_\parallel^2 + \mu B 
+ e_a (\Psi_{E1a} + \Psi_{E2a}) \right\}
,
\end{eqnarray}
where 
$
{\bf E}_T \equiv - c^{-1} \partial {\bf A} / \partial t
$
represents the electric field induced by the temporal change in the 
background magnetic field and 
the magnetization  vector ${\bf M}$ is defined by 
Eq.~(\ref{magnetization}). 
It should be recalled that the background magnetic field is allowed 
to temporally change in this paper so that the long time evolution 
of the system over the transport time scale can be treated. 
It is also noted that  
Eq.~(\ref{eK})
is used in deriving Eq.~(\ref{totenergyb0}).

It can be shown after several analytical manipulations  
that Eq.~(\ref{totenergyb0}) is further rewritten as  
\begin{eqnarray}
\label{totenergyb1}
& & 
\frac{\partial}{\partial t} 
\left[
\sum_a \int d^3 v \, F_a 
\left(
\frac{1}{2} m_a v_\parallel^2 + \mu B
\right) 
+ \frac{|{\bf E}_L|^2}{8\pi}
\right.
\nonumber \\ & & \mbox{} \hspace*{3mm}
\left. 
+ \frac{{\bf E}_L \cdot {\bf P}_D}{2} 
+
\frac{1}{2}\sum_{n=1}^\infty
\frac{\partial^n (E_L)_i }{\partial x^{j_1} \cdots \partial x^{j_n}}
Q_E^{i j_1 \cdots j_n}
+ \frac{|{\bf B}|^2}{8\pi}
\right]
\nonumber \\ & & \mbox{} \hspace*{3mm}
+ \nabla
\cdot 
\left[
\sum_a \int d^3 v \, F_a \left\{ \frac{1}{2} m_a v_\parallel^2 + \mu B 
+ e_a (\Psi_{E1a} 
\right. 
\right.
\nonumber \\ & & \mbox{} \hspace*{3mm}
\left.
+ \Psi_{E2a}) \biggr\} {\bf u}_{ax} 
+ 
\frac{c}{4\pi} {\bf E} \times {\bf H}
- \frac{1}{4\pi} \frac{\partial \phi_D}{\partial t} {\bf E}_T
+ \phi \frac{\partial ({\bf P}_G)_T}{\partial t}
\right]
\nonumber \\ & & \mbox{} \hspace*{3mm}
+ 
\frac{\partial}{\partial x^i}
\left[
\sum_{n=1}^\infty
\sum_{k=0}^{n-1} (-1)^{n-k}
\frac{\partial^k (E_L)_{j_n}}{\partial x^{j_{n-k}} \cdots \partial x^{j_{n-1}}}
\right.
\nonumber \\ & & \mbox{} \hspace*{6mm}
\left. 
\times
\frac{\partial^{n-k} }{\partial x^{j_1} \cdots \partial x^{j_{n-k-1}}}
\left(
\frac{\partial Q^{i j_1 \cdots j_n}}{\partial t}
\right)
\right]
\nonumber \\ & & 
= 
\left( 
{\bf J}_T 
- \frac{c}{4\pi}  \nabla \times {\bf B} 
\right)
\cdot {\bf E}
+
\sum_a \int d^3 v \,  {\cal K}_a 
\nonumber \\ & & \mbox{}
\hspace*{6mm} 
\times 
\left\{ \frac{1}{2} m_a v_\parallel^2 + \mu B 
+ e_a (\Psi_{E1a} + \Psi_{E2a}) \right\}
, 
\end{eqnarray}
or 
\begin{eqnarray}
\label{totenergyb2}
& & 
\frac{\partial}{\partial t} 
\left[
\sum_a \int d^3 v \, F_a 
\left(
\frac{1}{2} m_a v_\parallel^2 + \mu B
\right) 
+ \frac{|{\bf E}_L|^2}{8\pi}
+ \frac{{\bf E}_L \cdot {\bf P}_D}{2} 
\right.
\nonumber \\ & & \mbox{} \hspace*{3mm}
\left. 
+
\frac{1}{2}\sum_{n=1}^\infty
\frac{\partial^n (E_L)_i }{\partial x^{j_1} \cdots \partial x^{j_n}}
Q_E^{i j_1 \cdots j_n}
+ \frac{{\bf E}_T \cdot {\bf D}_L }{4 \pi}
+ \frac{|{\bf B}|^2}{8\pi}
\right]
\nonumber \\ & & \mbox{} \hspace*{3mm}
+ \nabla
\cdot 
\left[
\sum_a \int d^3 v \, F_a \left\{ \frac{1}{2} m_a v_\parallel^2 + \mu B 
+ e_a (\Psi_{E1a}  
\right.
\right.
\nonumber \\ & & \mbox{} \hspace*{3mm}
\left.
+ \Psi_{E2a}) \biggr\} {\bf u}_{ax} 
+ 
\frac{c}{4\pi} {\bf E} \times {\bf H}
+ \frac{1}{4\pi} \phi_D \frac{\partial {\bf E}_T}{\partial t} 
+ \phi \frac{\partial ({\bf P}_G)_T}{\partial t}
\right]
\nonumber \\ & & \mbox{} \hspace*{3mm}
+ 
\frac{\partial}{\partial x^i}
\left[
\sum_{n=1}^\infty
\sum_{k=0}^{n-1} (-1)^{n-k}
\frac{\partial^k (E_L)_{j_n}}{\partial x^{j_{n-k}} \cdots \partial x^{j_{n-1}}}
\right.
\nonumber \\ & & \mbox{} \hspace*{6mm}
\left. 
\times
\frac{\partial^{n-k} }{\partial x^{j_1} \cdots \partial x^{j_{n-k-1}}}
\left(
\frac{\partial Q^{i j_1 \cdots j_n}}{\partial t}
\right)
\right]
\nonumber \\ & & \mbox{}
= 
\left( 
{\bf J}_T 
- \frac{c}{4\pi}  \nabla \times {\bf B} 
\right)
\cdot {\bf E}
 +
\sum_a \int d^3 v \,  {\cal K}_a 
\nonumber \\ & & \mbox{} \hspace*{6mm}
\times
\left\{ \frac{1}{2} m_a v_\parallel^2 + \mu B 
+ e_a (\Psi_{E1a} + \Psi_{E2a}) \right\}
, 
\end{eqnarray}
where the longitudinal (irrotational) part of the electric displacement vector 
[see Eq.~(\ref{Di})] is denoted by 
${\bf D}_L \equiv ({\bf E} + 4\pi {\bf P}_G)_L \equiv - \nabla \phi_D$ and the magnetic intensity field ${\bf H}$ is defined
by 
$
{\bf H} \equiv {\bf B} - 4\pi {\bf M}
$
with the magnetic induction field ${\bf B}$ and
the magnetization vector field ${\bf M}$ [see Eq.~(\ref{magnetization})].
Equations~(\ref{totenergyb1}) and (\ref{totenergyb2}) 
take the conservative form including no other terms than   
the time derivative of the total energy density and the divergence of 
the total energy flux 
when the right-hand sides vanish.  
The effect of the collision (or source) term on the total energy 
balance is shown by the integral term which appears on the right-hand sides of 
Eqs.~(\ref{totenergyb1}) and (\ref{totenergyb2}). 
This term vanishes when 
${\cal K}_a$ represents the collision operator 
which conserves the summation of the gyrocenter kinetic and 
polarization energies 
$\sum_a \int d^3 v \{ \frac{1}{2} m_a v_\parallel^2 + \mu B 
+ e_a (\Psi_{E1a} + \Psi_{E2a}) \}$.  
In addition, the right-hand sides of 
Eqs.~(\ref{totenergyb1}) and (\ref{totenergyb2})  
contain the difference between the transverse current density ${\bf J}_T$ 
and $(c/4\pi) \nabla \times {\bf B}$ which vanishes if 
the self-consistency condition 
$\nabla \times {\bf B} = (4\pi/c){\bf J}_T$ is imposed 
as is done in the Darwin model~\cite{Kaufman}
and in our previous work by including the magnetic 
energy with the minus sign into the Lagrangian 
for the drift kinetic system with self-consistent fields.~\cite{Sugama2018} 
We also find from Eqs.~(\ref{totenergyb1}) and (\ref{totenergyb2})  
that the terms including ${\bf E}_T$ appear 
on their left-hand sides 
in the way consistent with the case of 
the energy conservation law for the 
Vlasov-Poisson-Amp\`{e}re (or Vlasov-Darwin) system 
shown in Eq.(22) of Ref.~\cite{Sugama2013}. 

All the kinetic energy, the electric energy, and the magnetic energy 
with the polarization including the dipole and other 
multipole moment effects are described in 
Eqs.~(\ref{totenergyb1}) and (\ref{totenergyb2}), 
where  
the energy flux contains the kinetic energy flow 
due to the gyrocenter motion,   
the Poynting vector, and the extra energy flux due to 
electrostatic turbulent 
fluctuations with wavelengths of the order of gyroradius.

\section{THE LOCAL CONSERVATION LAW OF THE CANONICAL MOMENTUM 
IN THE DIRECTION OF SYMMETRY}

It is shown in this Appendix how the local conservation law of 
the canonical momentum in 
the direction of symmetry can be derived for the gyrokinetic system 
considered in the present paper. 
We start from Eq.~(\ref{dLGKF3}) for which the gyrokinetic equation including the 
collision (or source) term ${\cal K}_a$ [see Eq.~(\ref{GKB})] is used. 
Equation~(\ref{dLGKF3}) is rewritten as
\begin{eqnarray}
\label{G1}
\overline{\delta}
L_{GKF}
& = & 
 \int_V d^3 x \, 
\left[ 
\xi^i  \sum_a \int d^3 v \left\{
\frac{\partial}{\partial t}
\left( 
 F_a 
p_{ai}
\right)
-
{\cal K}_a
p_{ai}
\right\}
\right.  
\nonumber \\ & &
\hspace*{-5mm}
\left.  \mbox{} 
+ \frac{1}{c} J^i
\overline{\delta} A_i
+ \frac{\delta L_{GKF}}{\delta \phi} 
\overline{\delta} \phi
+ \frac{1}{2} \Theta^{ij}
\overline{\delta} g_{ij}
\right]
+  \mbox{B.T.}
,
\hspace*{10mm}
\end{eqnarray}
where we have used 
the canonical momentum of the single gyrocenter, 
the electric current density, and 
the symmetric pressure tensor
defined by 
$p_{ai}\equiv \partial L_{GYa}/\partial u_{ax}^i$, 
$J^i \equiv \sum_a e_a \Gamma_a^i = c \sum_a \delta L_{GKa}/\delta A_i 
= c \, \delta L_{GKF}/\delta A_i$, 
and 
$\Theta^{ij}\equiv 
2 \, \delta L_{GKF}/\delta g_{ij}$, 
respectively, 
as shown in Eqs.~(\ref{pai}), 
(\ref{dLGKadA}), (\ref{current}), and (\ref{Theta}).  
Next, 
we use
$\delta L_{GKF}/\delta \phi = 0$, which is equivalent to 
the gyrokinetic Poisson's equation [see Eq.~(\ref{GKP})], and 
substitute 
$\overline{\delta} A_i = - \xi^j (\partial_j A_i) - 
(\partial_i \xi^j) A_j$ [Eq.~(\ref{dEgg})], 
$\overline{\delta} g_{ij} = - \nabla_i \xi_j - \nabla_i \xi_i $ 
[Eq.~(\ref{dg})]
into Eq.~(\ref{G1}). 
Then, Eq.~(\ref{G1}) is rewritten after performing partial integrals as 
\begin{eqnarray}
\label{G2}
\overline{\delta}
L_{GKF}
& = & 
 \int_V d^3 x \, 
\left[ 
\xi^j  \sum_a \int d^3 v \left\{
\frac{\partial}{\partial t}
\left( 
 F_a 
p_{aj}
\right)
-
{\cal K}_a
p_{aj}
\right\}
\right.  
\nonumber \\ & &
\left.  \mbox{} 
+ \xi^j \frac{1}{c}
\left\{ 
\partial_i (J^i A_j)
- J^i (\partial_j A_i )
\right\}
+ \xi_j \nabla_i \Theta^{ij}
\right]
\nonumber \\ & &
 \mbox{} +  \mbox{B.T.}
,
\end{eqnarray}
where we can also write the last term of the integrand 
as  $\xi_j \nabla_i \Theta^{ij} = \xi^j \nabla_i \Theta^i_j$ 
by using $\xi_j \equiv g_{jk} \xi^k$, 
$\Theta^i_j \equiv g_{jk} \Theta^{ik}$, and Eq.~(A8). 
We now recall the invariance of the Lagrangian $L_{GKF}$ 
under an arbitrary 
spatial coordinate transformation which implies that 
$\overline{\delta} L_{GKF} = 0$ for any $\xi^j$. 
Consequently, the canonical momentum balance equation is obtained from 
Eq.~(\ref{G2}) as
\begin{eqnarray}
\label{G3}
& & 
\frac{\partial}{\partial t}
\left( 
\sum_a \int d^3 v \,
 F_a 
p_{aj}
\right)
- \sum_a \int d^3 v \,
{\cal K}_a
p_{aj}
\nonumber \\ & &
\mbox{} 
+ \frac{1}{c}
\left\{ 
\partial_i (J^i A_j)
- J^i (\partial_j A_i )
\right\}
+ \nabla_i \Theta^i_j
 = 0
.
\end{eqnarray}
It is emphasized here that Eq.~(\ref{G3}) is valid for arbitrary spatial coordinates 
$(x^i)_{i=1,2,3}$ in which the spatial position vector in the Cartesian coordinate
system is represented as ${\bf r} = {\bf r}(x^1, x^2, x^3)$. 

Hereafter, we assume that the background magnetic 
field has symmetry with respect to continuous isometric transformations 
(spatial translations, rotations, or screw motions generated by the combination of 
translations and rotations) in a certain direction. 
Owing to the use of general spatial coordinates, 
translation, rotation, and screw (or helical) symmetries can be treated in 
a unified manner.  
Under the symmetry assumption for the background magnetic field, 
there exists a certain coordinate system $(x^i)_{i=1,2,3}$ where 
all components of the magnetic field, the vector potential, and the 
metric tensor become independent of one coordinate for which $x^3$ 
is chosen here without loss of generality, 
\begin{equation}
\label{G4}
\partial_3 B^i = 0,
\hspace*{3mm}
\partial_3 A_i = 0,
\hspace*{3mm}
\mbox{and}
\hspace*{3mm}
\partial_3 g_{ij} = 0
.
\end{equation}
For example, 
such coordinate systems are given by 
Cartesian, cylindrical (or spherical), and 
helical coordinate systems for the cases of translation, rotation, 
and screw (or helical) symmetries, respectively. 
Then, from Eq.~(\ref{G3}) with $j=3$, we have
\begin{eqnarray}
\label{G5}
& & 
\frac{\partial}{\partial t}
\left( 
\sum_a \int d^3 v \,
 F_a 
p_{a3}
\right)
+ \frac{1}{c}
\partial_i (J^i A_3)
+ \nabla_i \Theta^i_3
\nonumber \\ & &
\hspace*{3mm} 
= 
\sum_a \int d^3 v \,
{\cal K}_a
p_{a3}
.
\end{eqnarray}
Here, we define the basis vector 
 ${\bf e} \equiv \partial {\bf r}/ \partial x^3$ associated with 
the coordinate $x^3$. 
For translation symmetry,  ${\bf e}$ represents the constant vector parallel 
to the direction of symmetry while, for rotation symmetry, we have 
${\bf e} = {\bf n} \times {\bf r}$ where ${\bf n}$ is the direction vector of 
the rotation axis. 
For helical symmetry, ${\bf e}$ is given by the combination of 
the two types of the vectors described above. 
Then, the boldface notation ${\bf p}_a$ is used to represent the 
vector with the covariant components $(p_{aj})_{j=1,2,3}$,  
and the inner product of the vectors  ${\bf p}_a$ and ${\bf e}$ 
is represented by 
${\bf p}_a \cdot {\bf e} = p_{aj} e^j = p_{a3}$ 
where the $j$th contravariant component 
of ${\bf e}\equiv \partial {\bf r}/\partial x^3$ is 
given by $e^j = \partial x^j/ \partial x^3 = \delta^j_3$. 
Thus, $p_{a3}$ is hereafter treated as a scalar field produced 
by the inner product of the vectors. 
In the same way, using ${\bf A}$ to represent the vector 
with the covariant components $(A_j)_{j=1,2,3}$, 
we can write $A_3 \equiv {\bf A} \cdot {\bf e}$ which 
is treated as a scalar field produced by the inner product of 
${\bf A}$ and ${\bf e}$.  

It is noted that, with respect to the spatial coordinate transformation, 
the distribution function $F_a$ is transformed as a scalar density field 
[see Eq.~(B12)], which means $F_a/\sqrt{g}$ is regarded as a scalar field. 
Since $F_a$ enters the definitions of $J^i$ [Eq.~(113)] 
and $\Theta^{ij}$ [Eq.~(\ref{Theta})], $J^i$ and $\Theta^{ij}$ are the components 
of the contravariant vector density field and the symmetric contravariant 
tensor density field, respectively. 
Correspondingly, $J^i/\sqrt{g}$ and $\Theta^{ij}/\sqrt{g}$ represent 
the vector field and the tensor field. 
In the Cartesian coordinates where $\sqrt{g}=1$, 
the scalar, vector, and tensor density fields 
have the same components as the corresponding 
scalar, vector, and tensor fields so that we do not need to 
distinguish these density fields from the corresponding fields.    
Using the notation ${\bf J}$ 
to represent the vector field with the contravariant components 
$(J^i/\sqrt{g})_{i=1,2,3}$, 
we can regard 
$J^i A_3/\sqrt{g}$ as the $i$th contravariant 
component of the vector field ${\bf J} ({\bf A} \cdot {\bf e})$. 
We also use Eqs.~(A2), (A8), and the formula
$\Gamma^i_{ij} = (\partial_j \sqrt{g})/\sqrt{g}$
to write 
\begin{eqnarray}
\label{G6}
 \partial_i ( J^i  A_3 )
& = & 
\sqrt{g} \left\{
\partial_i[( J^i/\sqrt{g}) A_3]
+ \Gamma^i_{ik} (J^i/\sqrt{g}) A_3
\right\}
\nonumber \\ 
& = & 
\sqrt{g} 
\nabla \cdot [ {\bf J} ({\bf A} \cdot {\bf e}) ]
. 
\end{eqnarray} 
From Eq.~(A4) and $\partial_3 g_{ij} = 0$ in Eq.~(\ref{G4}), 
we find that 
$\Gamma_{i,j3} = \frac{1}{2} (\partial_j g_{i3} - \partial_i g_{j3}) 
= - \Gamma_{j,i3}$, 
which is combined with $\Theta^{ij} = \Theta^{ji}$ to 
obtain $\Gamma^j_{i3} \Theta^i_j = \Gamma_{j,i3} \Theta^{ij} = 0$.  
We now use the notation $\mbox{\boldmath$\Theta$}$ to represent 
the tensor field with the contravariant 
components $\Theta^{ij}/\sqrt{g}$.  
Also, the tensor-vector contraction 
 $\mbox{\boldmath$\Theta$}\cdot {\bf e}$ is used  
to represent the vector field,   
the $i$th component of which is written as 
$(\Theta^{ij}/\sqrt{g}) e_j = (\Theta^i_j/\sqrt{g}) e^j 
= \Theta^i_3/\sqrt{g}$. 
Then, using Eq.~(A7) and (A8), 
we obtain
\begin{eqnarray}
\label{G7}
\nabla_i \Theta^i_3 
& = & 
\sqrt{g} \nabla_i (\Theta^i_3 /\sqrt{g} ) 
 \nonumber  \\
& = & 
\sqrt{g} 
[\partial_i (\Theta^i_3 /\sqrt{g} )
+ \Gamma^i_{ij} (\Theta^j_3 /\sqrt{g} )
- \Gamma^j_{i3} (\Theta^i_j /\sqrt{g} ) ]
 \nonumber  \\
& = &  
\sqrt{g} 
[
\partial_j (\Theta^j_3 /\sqrt{g} )
+ \Gamma^i_{ij} (\Theta^j_3 /\sqrt{g} )] 
]
 \nonumber  \\
& = &  
\sqrt{g} \nabla \cdot ( \mbox{\boldmath$\Theta$} \cdot {\bf e} )
= \partial_i \Theta^i_3
, 
\end{eqnarray}
where $\Gamma^i_{ij} = (\partial_j \sqrt{g})/\sqrt{g}$ 
and $\Gamma^j_{i3} \Theta^i_j = 0$ are used.  

Using Eq.~(\ref{G5})--(\ref{G7}), the local canonical momentum balance equation 
in the direction of symmetry is written as 
\begin{eqnarray}
\label{G8}
& & 
\frac{\partial}{\partial t}
\left( 
\sum_a \int d^3 v \,
 F_a 
p_{a3}
\right)
+ 
\partial_i 
\left( \frac{1}{c}J^i A_3
+  \Theta^i_3 \right) 
\nonumber \\ 
& &  = 
\sum_a \int d^3 v \,
{\cal K}_a
p_{a3}
,
\end{eqnarray}
in the spatial coordinates where $x^3$ is the coordinate in the 
direction of symmetry. 
Equation~(\ref{G8}) 
is also written using the conventional vector and tensor notations in the Cartesian coordinates (with $\sqrt{g}=1$) as 
\begin{eqnarray}
\label{G9}
& & 
\frac{\partial}{\partial t}
\left( 
\sum_a \int d^3 v \,
F_a \, 
({\bf p}_a  \cdot {\bf e} )
\right)
+
\nabla \cdot 
\left[\frac{1}{c} {\bf J} \, ({\bf A} \cdot {\bf e}) 
+  \mbox{\boldmath$\Theta$} \cdot {\bf e}\right] 
\nonumber 
\\ & & 
\hspace*{3mm}
 = 
\sum_a \int d^3 v \, 
{\cal K}_a   \,
( {\bf p}_a  \cdot {\bf e} )
.
\end{eqnarray}
Finally, we clearly see that Eqs.~(\ref{G8}) and (\ref{G9})
represent 
the local conservation law of the canonical momentum conjugate to 
the coordinate in the direction of symmetry 
when the collision (or source) term ${\cal K}_a$ satisfies  
$\sum_a \int d^3 v \, {\cal K}_a \,
( {\bf p}_a  \cdot {\bf e} ) =0$. 

Let us compare Eqs.~(\ref{G8})--(\ref{G9})
with the result derived 
from the conventional method explained in Appendix~E. 
In deriving Eqs.~(\ref{G8}) and (\ref{G9}), 
after the invariance of the Lagrangian under arbitrary spatial coordinate 
transformations is used to obtain the local momentum balance equation, Eq.~(\ref{G1}),  
in the general spatial coordinates, Eq.~(\ref{G1}) 
is represented in the special coordinates, one of which 
is the coordinate in the direction of symmetry. 
On the other hand, in Appendix~E, 
the invariance of the Lagrangian under 
not arbitrary but only translational and rotational coordinate 
transformations are used to derive the local linear and angular 
momentum equations in Eqs.~(E11) and (E14). 
In the case where the background magnetic field has translation symmetry along the
constant direction vector ${\bf e}$, 
the inner product of Eq.~(E11) and ${\bf e}$ results in 
the local canonical momentum conservation law, 
\begin{eqnarray}
\label{G10}
& & 
\frac{\partial}{\partial t}
\left( 
\sum_a \int d^3 v \,
F_a \, 
({\bf p}_a  \cdot {\bf e} )
\right)
+
\nabla \cdot 
\left[\frac{1}{c} {\bf J} \, ({\bf A} \cdot {\bf e}) 
+  \mbox{\boldmath$\Pi$}_c \cdot {\bf e}\right] 
 \nonumber 
\\ & & 
\hspace*{3mm}
= 
\sum_a \int d^3 v \, 
{\cal K}_a   \,
( {\bf p}_a  \cdot {\bf e} )
, 
\end{eqnarray}
with the asymmetric pressure tensor $\mbox{\boldmath$\Pi$}_c$ given by 
\begin{eqnarray}
\label{G11}
\mbox{\boldmath$\Pi$}_c
& \equiv & 
\sum_a 
\int d^3 v \, 
F_a 
 {\bf u}_{ax}
{\bf u}_{ax}
- ( {\bf B} \cdot  {\bf M} ) {\bf I} +   {\bf B} {\bf M} 
 \nonumber 
\\ & & 
\mbox{} + 
\frac{|{\bf E}_L|^2}{8\pi} {\bf I} - \frac{{\bf E}_L {\bf E}_L}{4\pi} 
+ \mbox{\boldmath$\Pi$}_\Psi 
, 
\end{eqnarray}
where Eqs.(\ref{pai}), (\ref{current}), and ${\bf e} \cdot \nabla {\bf A} = 0$ are used, 
and the term including ${\cal K}_a$ is still retained 
for comparison with Eq.~(\ref{G9}). 
When the background field has rotational symmetry around the axis  
which is parallel to the direction vector ${\bf n}$ and passes through 
the origin of the position vector ${\bf r}$, 
${\bf A}$ satisfies 
$({\bf n} \times {\bf r}) \cdot \nabla {\bf A} 
= {\bf n} \times {\bf A}$. 
Then, we can use the inner product of Eq.~(E14) and ${\bf n}$ 
to derive the local canonical angular momentum conservation law which has the 
same form as shown in Eqs.~(\ref{G10})--(\ref{G11}) 
where ${\bf e}$ is regarded as given by  
${\bf e}= {\bf n} \times {\bf r}$. 
In the same way, it can be shown that, 
for the case of helical symmetry, 
the local conservation law  of the canonical momentum 
in the direction of symmetry is written again by 
Eqs.~(\ref{G10}) and (\ref{G11}), where we put 
${\bf e} = {\bf n} + \alpha {\bf n} \times {\bf r}$ with a 
constant $\alpha$ to represent the direction of helical symmetry. 
Now, it is clear that the difference between 
Eqs.~(\ref{G9}) and (\ref{G10}) is 
expressed by that between the symmetric and asymmetric 
pressure tensors denoted as 
 $\mbox{\boldmath$\Theta$}$
and $\mbox{\boldmath$\Pi$}_c$, respectively.
It is not easy to find out the well-known CGL tensor 
${\bf P}_{CGL}$ [Eq.~(\ref{CGL})] 
in $\mbox{\boldmath$\Pi$}_c$ while ${\bf P}_{CGL}$ naturally appears 
in  $\mbox{\boldmath$\Theta$}$ [Eq.~(\ref{Theta})]. 
Incidentally, 
the procedure of 
the Belinfante-Rosenfeld type is known 
for derivation of the symmetric pressure tensor from the asymmetric canonical 
pressure tensor~\cite{Sugama2013,Dixon} 
although the symmetric tensor is more directly derived by 
the present method using the variational derivative of 
the Lagrangian with the metric tensor.  



\end{document}